\documentclass[letterpaper,english,twocolumn,pra]{revtex4-1}
\usepackage[T1]{fontenc}
\usepackage[latin9]{inputenc}
\usepackage{verbatim}
\usepackage{prettyref}
\usepackage{amsmath}
\usepackage{amssymb}
\usepackage{graphicx}

\makeatletter

\pdfpageheight\paperheight
\pdfpagewidth\paperwidth

\providecommand{\tabularnewline}{\\}

\usepackage{stackengine}
\stackMath
\usepackage{graphicx}
\usepackage{graphics}
\usepackage{slashbox}
\def\shrinkage{2.1mu}
\def\vecsign{\mathchar"017E}
\def\dvecsign{\smash{\stackon[-1.95pt]{\mkern-\shrinkage\vecsign}{\rotatebox{180}{$\mkern-\shrinkage\vecsign$}}}}

\@ifundefined{showcaptionsetup}{}{%
 \PassOptionsToPackage{caption=false}{subfig}}
\usepackage{subfig}
\makeatother

\usepackage{babel}
\begin{document}
\noindent\begin{minipage}[t]{1\columnwidth}%
\global\long\def\anticommutator#1#2{\left\{  #1,#2\right\}  }

\global\long\def\commutator#1#2{\left[#1,#2\right]}

\global\long\def\braket#1#2{\langle#1|#2\rangle}

\global\long\def\bra#1{\langle#1|}

\global\long\def\ket#1{|#1\rangle}

\global\long\def\Tr{\operatorname{Tr}}

\global\long\def\sgn{\operatorname{sgn}}

\global\long\def\diag{\operatorname{diag}}

\global\long\def\dvec#1{\global\long\def\useanchorwidth{T}
\stackon[-4.2pt]{#1}{\,\dvecsign}}

\global\long\def\scaledmatrix#1{\scalebox{.7}{\global\long\def\arraystretch{.7}
\ensuremath{#1}}}

\begin{comment}
Potential Changes:

Presentation:

-add projection lines for Bloch vectors and right angles between correlation
axes to make 3d apparent

-consistent terminology: sometimes qubit, sometimes system, subsystem
... be consistent

-Make LMM sv diagram larger than the others?

Research:

-seek a representation where the convex hull of product states is
separable, while those outside are entangled.

-linear combination of irreducible unitaries ... a suitable linear
combo chosen, different for each rho, such that effect has easy geometric
presentation

-A better geometric representation of matrix R (9 d.f.), such that
the \prettyref{eq:U1effect} is a simple geometric operation ... breaking
it up to a 3 column, or 3 row vectors doesn't work well, and SVD too
complicated here

- plot 3d space of the values of the 3 invariants, write positivity
equations in terms of them to find the range .... plot plane of pure
states in this 3d space.

- What do Krauss operations and POVMs do to the Bloch Matrix?

-unitary on qubit is rotation of Bloch vector, what kind on qubit
is reflection of bloch vector?
\end{comment}
%
\end{minipage}

\title{Entangled Bloch spheres: Bloch matrix and two-qubit state space}

\author{Omar Gamel}
\email{ogamel@berkeley.edu}

\affiliation{Department of Chemistry, University of California, Berkeley, California
94720, USA~\\
 Physical Biosciences Division, Lawrence Berkeley National Laboratory,
Berkeley, California 94720, USA }

\date{\today}
\begin{abstract}
We represent a two-qubit density matrix in the basis of Pauli matrix
tensor products, with the coefficients constituting a Bloch matrix,
analogous to the single qubit Bloch vector. We find the quantum state
positivity requirements on the Bloch matrix components, leading to
three important inequalities, allowing us to parametrize and visualize
the two-qubit state space. Applying the singular value decomposition
naturally separates the degrees of freedom to local anr nonlocal,
and simplifies the positivity inequalities. It also allows us to geometrically
represent a state as two entangled Bloch spheres with superimposed
correlation axes. It is shown that unitary transformations, local
or nonlocal, have simple interpretations as axis rotations or mixing
of certain degrees of freedom. The nonlocal unitary invariants of
the state are then derived in terms of local unitary invariants. The
positive partial transpose criterion for entanglement is generalized,
and interpreted as a reflection, or a change of a single sign. The
formalism is used to characterize maximally entangled states, and
generalize two-qubit isotropic and Werner states.

\begin{description}
\item [{PACS~numbers}] 03.65.Ud, 03.67.Mn, 03.67.Bg, 03.67.Lx. %
\begin{comment}
Entanglement and quantum nonlocality 03.65.Ud., Quantum Mechanics
03.65.-w, Quantum computation, 03.67.Lx. Entanglement and quantum
nonlocality in quantum information 03.67.Mn, 03.67.Bg, 
\end{comment}
{\small \par}
\end{description}
\end{abstract}


\maketitle

\section{Introduction}

Probabilistic mixtures of single qubit quantum states can be represented
by a density matrix \citep{VonNeumann1927}. The density matrix may
be written in the Pauli Matrix basis, with the coefficients making
up the Bloch vector \citep{Bloch1946}. The latter has the simple
geometry of a vector inside a unit Bloch sphere, whose magnitude indicates
the state's purity, and whose rotations are unitary transformations.

The simplicity of this representation motivated many authors to generalize
it to quantum systems of higher dimensions. In three dimensions, the
basis of Gell-Mann matrices \citep{GellMann1964} led to an irregularly
shaped Bloch vector space \citep{Kimura2003,Kimura2005,Byrd2003}.
Generalized Gell-Mann matrices have been used as the basis in the
four-dimensional (two-qubit) case \citep{Jakobczyk2001,Tilma2002,Bertlmann2008},
again leading to a space without much symmetry. Two-qubit state space
has also been analyzed through Hopf fibrations \citep{Mosseri2001},
and steering ellipsoids \citep{Jevtic2014,Milne2014}.

In this work, we make use of tensor products of Pauli matrices as
our four-dimensional system basis, with the coefficients representing
entries of a \emph{Bloch matrix. }Numerous authors have use a similar
approach \citep{Horodecki1995,Horodecki1996a,Linden1999,Makhlin2000,Barrett2001,Kus2001,Abouraddy2002,Spengler2011,Jevtic2014,Milne2014,Aerts2015,James2001,Kagalwala2015,Rajagopal2001}.
We go further by studying the properties of this representation, and
in particular, deriving the positivity conditions. 

The positivity of the quantum states leads to three inequalities that
allow us to parametrize and visualize the state space. The inequalities
suggest a singular value decomposition, which simplifies the positivity
conditions and reproduces known unitary invariants \citep{Makhlin2000}
with additional insights. The conditions also allow us to generalize
the positive partial transpose criterion for entanglement \citep{Peres1996,Horodecki1996},
and strikingly interpret it as a reflection, or a change of a single
sign. We also find that the most basic nonlocal transformations \citep{Zhang2002a}
reduce to a family of two-dimensional rotation matrices which mix
various degrees of freedom of the Bloch matrix representation.

The paper is organized as follows; Secs. \ref{sec:oneLevelBloch}
and \ref{sec:Two-Qubit-System} review the Bloch vector representation,
with the former on qubits and qutrits, and the latter on two-qubit
systems. The positivity inequalities, a key result of this paper,
are derived in Sec. \ref{sec:The-Positivity-Condition}. The singular
value decomposition, along with its simplification of the positivity
conditions and representation of a quantum state as a pair of entangled
Bloch spheres are presented in Sec. \ref{sec:Singular-Value-decomposition}.
The actions of unitary operations, local and nonlocal, and their invariants
are expressed in the Bloch representation in Sec. \ref{sec:Unitary-Operations}.
Section \ref{sec:Entanglement-Criteria} provides a novel geometric
interpretation and generalization of the positive partial transpose
entanglement criterion. Section \ref{sec:Special-Classes-of-States}
applies the formalism to the characterization of maximally entangled,
pure states, and generalized isotropic/Werner states. Geometric visualization
of the quantum state space, indicating separability and entanglement,
takes place in Sec. \ref{sec:Geometry-of-State-Space}. Finally, we
recapitulate and propose future extensions in Sec. \ref{sec:Summary}.

We make use of Einstein summation notation where repeated indices
in the subscript are summed over, unless otherwise indicated. Greek
indices $\alpha,\beta,\gamma,\delta,\mu,\nu$ run from 0 to 3, and
Roman indices $i,j,k,$ run from $1$ to $3$, unless otherwise indicated.
Column vectors are denoted with an over right arrow ($\vec{u}$),
while row vectors are given a conjugate transpose dagger ($\vec{v}^{\dagger}$).
The Bloch matrix is denoted $\dvec r$, with the two sided over-arrow
indicating its two-dimensional tensorial nature. The identity matrix
is denoted $I$, with the context implying dimensionality.

We take a thorough approach, reproducing some known results to keep
this work reasonably self-contained, and relegating some detail to
the appendices.

\section{\label{sec:oneLevelBloch}Bloch Representations of Single Systems}

A quantum state may be represented by a density matrix $\rho$ containing
all its observable information \citep{VonNeumann1927}. The expectation
value of any observable $O$ is given by $\langle O\rangle=Tr[\rho O]$,
where the latter is the trace operator. The time evolution of a quantum
system, governed by the Schr\"{o}dinger equation for a pure state
\citep{Schrodinger1926}, is given by a unitary transformation on
the density matrix for a mixed state, $\rho\rightarrow U\rho U^{\dagger}$,
with $U$ a unitary matrix.

However, it is insightful to complement the density matrix with an
alternative representation of the quantum state space. To this end,
we examine the Bloch vector and its generalized representation.

\subsection{Pauli spin matrices}

For two-level systems, we study the\emph{ extended Pauli matrices};
with the identity matrix added, 
\begin{align}
\sigma_{0} & =\begin{pmatrix}1 & 0\\
0 & 1
\end{pmatrix}=I, &  & \sigma_{1}=\begin{pmatrix}0 & 1\\
1 & 0
\end{pmatrix},\nonumber \\
\sigma_{2} & =\begin{pmatrix}0 & -i\\
i & 0
\end{pmatrix}, &  & \sigma_{3}=\begin{pmatrix}1 & 0\\
0 & -1
\end{pmatrix}.\label{eq:paulisId}
\end{align}
The Pauli matrices form a set of generators for the group of 2$\times$2
special unitary matrices SU(2). Along with the identity, they constitute
a complete basis of the space of 2$\times$2 Hermitian matrices over
the real numbers. 

Pauli matrices satisfy the following well-known product, commutation,
and anticommutation relations
\begin{eqnarray}
\sigma_{i}\sigma_{j\;} & = & \delta_{ij}I+i\varepsilon_{ijk}\sigma_{k},\label{eq:pauliProduct}\\
\commutator{\sigma_{i}}{\sigma_{j}} & = & 2i\varepsilon_{ijk}\sigma_{k},\label{eq:pauliCommutator}\\
\anticommutator{\sigma_{i}}{\sigma_{j}} & = & 2\delta_{ij}I,\label{eq:pauliAntiCommutator}
\end{eqnarray}
respectively, where $\delta_{ij}$ is the Kronecker delta and $\varepsilon_{ijk}$
is the Levi-Civita symbol. To generalize to higher dimensions, we
wish to extend \prettyref{eq:pauliProduct}, \prettyref{eq:pauliCommutator},
and \prettyref{eq:pauliAntiCommutator} to include $\sigma_{0}$.
One can verify by trial that the four matrices in \prettyref{eq:paulisId}
satisfy the following product identity
\begin{equation}
\boxed{\sigma_{\alpha}\sigma_{\beta}=(\theta_{\alpha\beta\gamma}+i\varepsilon_{\alpha\beta\gamma})\sigma_{\gamma}},\label{eq:pauliIdProduct}
\end{equation}
where the third order tensors $\theta_{\alpha\beta\gamma}$ and $\varepsilon_{\alpha\beta\gamma}$
are defined
\begin{equation}
\theta_{\alpha\beta\gamma}\equiv\begin{cases}
1 & \text{\small one index is 0, the other two equal}\\
0 & \text{\small otherwise,}
\end{cases}\label{eq:thetaDef}
\end{equation}
and
\begin{equation}
\varepsilon_{\alpha\beta\gamma}\equiv\begin{cases}
\hphantom{-}1 & \text{\small\ensuremath{\alpha\beta\gamma\in\left\{  123,231,312\right\} } }\\
-1 & \text{\small\ensuremath{\alpha\beta\gamma\in\left\{  321,213,132\right\} } }\\
\hphantom{-}0 & \text{\small repeated indices, or any index is 0.}
\end{cases}\label{eq:epsDef}
\end{equation}

Of the $4^{3}=64$ entries in each of the two tensors, $\theta_{\alpha\beta\gamma}$
takes the nonzero value of 1 for 10 entries, $\alpha\beta\gamma\in\left\{ 000,011,101,110,022,202,220,033,303,330\right\} $,
and $\varepsilon_{\alpha\beta\gamma}$ takes a nonzero value for the
six entries defined in \prettyref{eq:epsDef}. Note that $\varepsilon_{\alpha\beta\gamma}$
is just the Levi-Civita symbol extended to take the value zero if
any index is zero. The tensor $\theta_{\alpha\beta\gamma}$ is symmetric
under the exchange of any two indices, while $\varepsilon_{\alpha\beta\gamma}$
is antisymmetric. 

Also note that $\theta_{\alpha\beta\gamma}$ satisfies
\begin{flalign}
\theta_{\alpha\beta0} & =\delta_{\alpha\beta},\label{eq:thetaKronecker1}\\
\theta_{\alpha\beta i} & =\delta_{\alpha0}\delta_{\beta i}+\delta_{0\beta}\delta_{\alpha i},\label{eq:thetaKronecker2}
\end{flalign}
where $\delta_{\alpha\beta}$ is the Kronecker delta extended to zero
index value. Equation \prettyref{eq:pauliIdProduct} implies the commutation
and anticommutations relations
\begin{align}
\commutator{\sigma_{\alpha}}{\sigma_{\beta}} & =2i\varepsilon_{\alpha\beta\gamma}\sigma_{\gamma},\label{eq:pauliIdCommutator}\\
\anticommutator{\sigma_{\alpha}}{\sigma_{\beta}} & =2\theta_{\alpha\beta\gamma}\sigma_{\gamma}.\label{eq:pauliIdAntiCommutator}
\end{align}

Taking the trace of \prettyref{eq:pauliIdProduct} we can also derive
the orthogonality relation
\begin{eqnarray}
\Tr(\sigma_{\alpha}\sigma_{\beta}) & = & (\theta_{\alpha\beta\gamma}+i\varepsilon_{\alpha\beta\gamma})2\delta_{\gamma0}\nonumber \\
 & = & 2(\theta_{\alpha\beta0}+i\varepsilon_{\alpha\beta0})\nonumber \\
 & = & 2\delta_{\alpha\beta},\label{eq:pauliOrthogonality}
\end{eqnarray}
where we made use of \prettyref{eq:thetaKronecker1} in the last line.

\subsection{Single qubit }

After characterizing the matrices in \prettyref{eq:paulisId}, we
can now express the 2$\times$2 density matrix in the basis they create,
\begin{eqnarray}
\rho & = & \frac{1}{2}\big(I+r_{i}\sigma_{i}\big)=\frac{1}{2}r_{\mu}\sigma_{\mu},\label{eq:bloch2d}
\end{eqnarray}
where the scalar $r_{0}$ is always unity to ensure $\Tr\rho=1$,
and scalars $r_{1},\,r_{2},$ and $r_{3}$ are the components of the
Bloch vector \citep{Bloch1946}, denoted $\vec{r}=(r_{1},r_{2},r_{3})$.
Since $\rho$ is Hermitian, $r_{i}$ are always real. Because of the
orthogonality relation \prettyref{eq:pauliOrthogonality}, the Bloch
vector is given by
\begin{equation}
r_{\mu}=\Tr(\rho\sigma_{\mu}).\label{eq:blochvecDef}
\end{equation}

As an alternative representation of the quantum state, the Bloch vector
has some advantages over the density matrix. For one, it is easier
to visualize the quantum state space in which Bloch vector exists.
To see this, recall that the purity of the density matrix, $\Tr\rho^{2}$,
is at most unity. Using \prettyref{eq:bloch2d} we have
\begin{equation}
1\geq\Tr\rho^{2}=r_{\mu}r_{\nu}\Tr(\sigma_{\mu}\sigma_{\nu})/4=\big(1+\|\vec{r}\|^{2}\big)/2,\label{eq:purityCondition}
\end{equation}
implying $\|\vec{r}\|\leq1$. Hence, the Bloch vector lies inside
a sphere of unit radius, known as the \emph{Bloch sphere}. 

Unitary transformations on the density matrix are interpreted as rotations
in the Bloch vector picture. Any unitary operator $U$ in two dimensions
can we written 
\begin{equation}
U_{\hat{a},\alpha}=\cos\frac{\alpha}{2}I-i\sin\frac{\alpha}{2}a_{i}\sigma_{i},\label{eq:unitaryDef}
\end{equation}
where $\alpha$ is an angle and $\hat{a}=(a_{1},a_{2},a_{3})$ is
a unit vector. 

A unitary transformation on the density matrix in \prettyref{eq:bloch2d}
leaves $I$ unchanged, but modifies the Bloch vector term $r_{i}\sigma_{i}$.
Making use of \prettyref{eq:unitaryDef}, writing $c=\cos\frac{\alpha}{2}$,
$s=\sin\frac{\alpha}{2}$, and suppressing subscripts on $U$, we
find the effect of a unitary transformation on the Bloch vector term
to be
\begin{align}
 & Ur_{j}\sigma_{j}U^{\dagger}=(cI-isa_{i}\sigma_{i})(r_{j}\sigma_{j})(cI+isa_{k}\sigma_{k})/2\nonumber \\
 & =\:r_{j}\big(c^{2}\sigma_{j}-icsa_{i}\commutator{\sigma_{i}}{\sigma_{j}}+s^{2}a_{i}a_{k}\sigma_{i}\sigma_{j}\sigma_{k}\big)\nonumber \\
 & =\:r_{j}\big(c^{2}\sigma_{j}+2cs\varepsilon_{ijk}a_{i}\sigma_{k}+2s^{2}a_{j}a_{i}\sigma_{i}-s^{2}\sigma_{j}\big)\nonumber \\
 & =\:r_{j}\big(\cos\alpha\sigma_{j}+(1{-}\cos\alpha)a_{j}a_{i}\sigma_{i}+\sin\alpha\varepsilon_{ijk}a_{i}\sigma_{k}\big)\nonumber \\
 & =\:\big(\cos\alpha\delta_{ij}+(1{-}\cos\alpha)a_{i}a_{j}+\sin\alpha\varepsilon_{kji}a_{k}\big)r_{j}\sigma_{i},\label{eq:unitaryRotation1}
\end{align}
where in the third line, we used \prettyref{eq:pauliCommutator} and
the identity $\sigma_{i}\sigma_{j}\sigma_{k}\equiv\delta_{ij}\sigma_{k}-\delta_{ik}\sigma_{j}+\delta_{jk}\sigma_{i}+i\varepsilon_{ijk}I$.
Setting $r_{i}^{\prime}\sigma_{i}\equiv Ur_{j}\sigma_{j}U^{\dagger}$,
where $r_{i}^{\prime}$ are the entries of the transformed Bloch vector,
we see that $r_{i}^{\prime}$ is equal to the terms that multiply
$\sigma_{i}$ in the last line. In vector notation
\begin{eqnarray}
\vec{r}^{\,\prime} & = & \big(\cos\alpha I+(1-\cos\alpha)\hat{a}\hat{a}^{\dagger}+\sin\alpha\lfloor\hat{a}\rfloor_{\times}\big)\vec{r}\nonumber \\
 & = & Q(\hat{a},\alpha)\vec{r},\label{eq:unitaryRotation2}
\end{eqnarray}
where $\hat{a}\hat{a}^{\dagger}$ is an outer product, and $\lfloor\hat{a}\rfloor_{\times}=\left[\begin{smallmatrix}0 & -a_{3} & a_{2}\\
a_{3} & 0 & -a_{1}\\
-a_{2} & a_{1} & 0
\end{smallmatrix}\right]$ is the cross product matrix of $\hat{a}$ (i.e. $\lfloor\hat{a}\rfloor_{\times}\vec{b}=\hat{a}\times\vec{b}$,
$\forall\vec{b}$). We identified the bracketed terms as the rotation
matrix $Q(\hat{a},\alpha)$, which rotates vectors by the angle $\alpha$
around $\hat{a}$. 

The rotation is more evident if we rewrite \prettyref{eq:unitaryRotation2}
as
\begin{eqnarray}
\vec{r}^{\,\prime} & = & (\vec{r}\cdot\hat{a})\hat{a}+\cos\alpha\big(\vec{r}-(\vec{r}\cdot\hat{a})\hat{a}\big)+\sin\alpha\hat{a}\times\vec{r}.\label{eq:unitaryRotation3}
\end{eqnarray}
The first term is the projection of $\vec{r}$ onto $\hat{a}$, left
unchanged by the rotation. The two other terms are of equal magnitude,
perpendicular to each other and to the first. They constitute the
rotated component of $\vec{r}$.

A final interesting property of the Bloch vector is that expectation
values become inner products. A generic qubit observable can be written
$O=sI+\vec{c}\cdot\vec{\sigma},$ for some scalar $s$ and vector
$\vec{c}$. Its expectation value is
\begin{equation}
\langle O\rangle=\Tr\big[\rho O\big]=\frac{1}{2}\big(2s+r_{i}c_{j}\Tr[\sigma_{i}\sigma_{j}]\big)=s+\vec{r}\cdot\vec{c}.\label{eq:expInnerProduct}
\end{equation}

\subsection{Single qutrit }

Given the usefulness of the Bloch vector representation, some authors
have generalized it to a qutrit system \citep{Kimura2003,Byrd2003,Kimura2005}.
They write the 3$\times$3 density matrix $\rho$ as
\begin{equation}
\rho=\frac{1}{3}\big(I+\sum_{m=1}^{8}r_{m}G_{m}\big),\label{eq:bloch3d}
\end{equation}
where $G_{m}$ are the Gell-Mann matrices \citep{GellMann1964} in
Appendix \prettyref{subsec:Gell-Mann-Matrices}, and the real coefficients
$r_{m}$ are the components of the generalized Bloch vector, still
denoted $\vec{r}$. The $G_{m}$ are Hermitian, traceless, and satisfy
the orthogonality relation $\Tr{[G_{m}G_{n}]}=2\delta_{mn}$. However
they are not unitary like the Pauli matrices. More fundamentally,
writing their commutation relations 
\begin{equation}
\commutator{G_{m}}{G_{n}}=2i\sum_{l=1}^{8}f_{lmn}G_{l},\qquad m,n=1,...,8.\label{eq:gellMannCommutator}
\end{equation}

The antisymmetric tensor $f_{ijk}$ takes the nonzero values $f_{123}=1$,
$f_{458}=f_{678}=\frac{\sqrt{3}}{2},$ $f_{147}=f_{165}=f_{246}=f_{257}=f_{345}=f_{376}=\frac{1}{2}$
\citep{Georgi1999}. The $f_{lmn}$ are the \emph{structure constants}
of the Lie algebra induced by the Gell-Mann matrices \citep{Lie1874,Serre2005}.
Comparing \prettyref{eq:gellMannCommutator} with \prettyref{eq:pauliCommutator},
the structure constants induced by Pauli matrices are given simply
by the Levi-Civita tensor, which up to antisymmetry, takes only a
single nonzero value of 1. This simplicity creates the symmetry underlying
the Bloch sphere. Conversely, the complexity of the $f_{ijk}$ implies
a lower level of symmetry, and a more complex qutrit Bloch vector
space.

\begin{figure}[h]
\includegraphics[width=1\columnwidth,height=1\columnwidth,keepaspectratio]{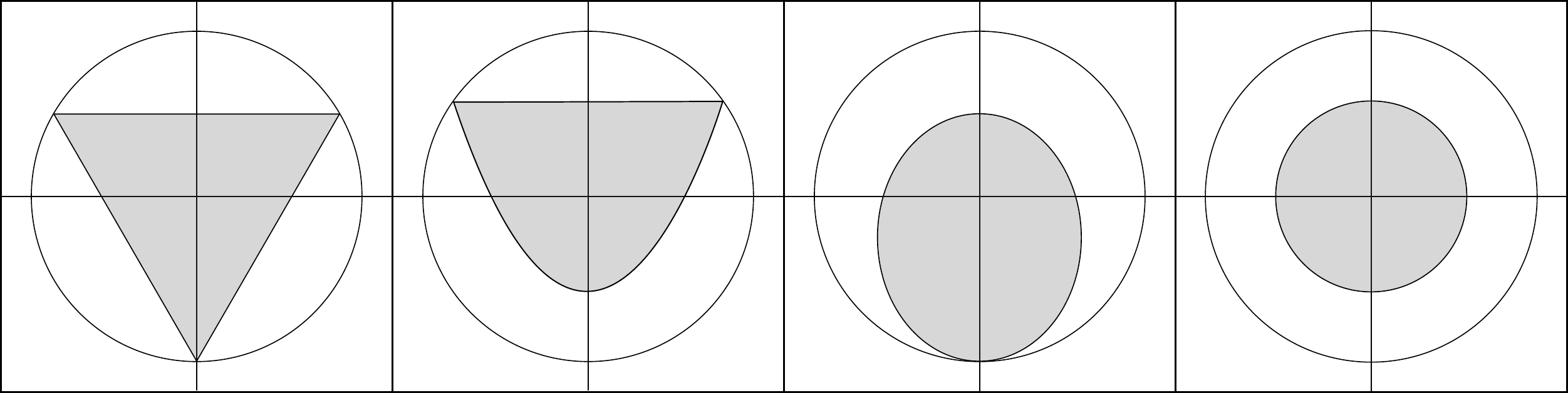}

\caption{\label{fig:kimuraCrossSections} Cross-sections of the qutrit Bloch
vector space. Allowed regions in the hypersphere are shaded. }
\end{figure}

Indeed, the space of allowable three-level Bloch vectors is a complicated
region lying inside an eight-dimensional hypersphere without filling
it. Representative cross sections of this complex eight-dimensional
space are shown in Fig. \ref{fig:kimuraCrossSections}, simplified
from Kimura \citep{Kimura2003}.

In addition, three-level unitary operators do not have a simple decomposition
as in the two-level case in \prettyref{eq:unitaryDef}, and the equivalence
between unitary transformations and rotations does not hold. Though
the Bloch vector representation of qutrits helps quantify purity and
polarization \citep{Gamel2012,Gamel2014}, the lack of symmetry limits
its usefulness. As we demonstrate in the remainder of the paper, much
symmetry and utility can be recovered in a four-level system.

\section{\label{sec:Two-Qubit-System}Two-Qubit System}

\subsection{Dirac matrices}

A 4$\times$4 density matrix may represent a single four-level system,
or more commonly, a pair of coupled two-level systems; two qubits.
Several authors analyzed the Bloch vector space of this system \citep{Jakobczyk2001,Tilma2002,Bertlmann2008}.
However, the basis they used is a generalization of the Gell-Mann
matrices with complicated structure constants, resulting in a 15-dimensional
space of allowable Bloch vectors with little useful symmetry. 

We investigate the same system using the \emph{Dirac matrices, }denoted\emph{
$D_{\mu\nu}$,} as our basis. They are defined as
\begin{equation}
D_{\mu\nu}=\sigma_{\mu}\otimes\sigma_{\nu}.\label{eq:diracMatrices}
\end{equation}
We have named them after Dirac as he used several of them in his eponymous
equation on the theory of relativistic electrons \citep{Dirac1928,Dirac1930}.
The 16 matrices are explicitly shown in Appendix \prettyref{subsec:Dirac-Matrices}.
The Dirac matrices satisfy the orthogonality relation
\begin{equation}
\Tr\big(D_{\alpha\beta}D_{\gamma\delta}\big)=4\delta_{\alpha\gamma}\delta_{\beta\delta}.\label{eq:diracOrthogonality}
\end{equation}

From \prettyref{eq:pauliIdProduct}, one can calculate the product,
the commutator, and the anticommutator, respectively given by 
\begin{align}
D_{\alpha\beta}D_{\gamma\delta} & =(\theta_{\alpha\gamma\mu}+i\varepsilon_{\alpha\gamma\mu})(\theta_{\beta\delta\nu}+i\varepsilon_{\beta\delta\nu})D_{\mu\nu},\label{eq:diracProduct}\\
\commutator{D_{\alpha\beta}}{D_{\gamma\delta}} & =2i\big(\theta_{\alpha\gamma\mu}\varepsilon_{\beta\delta\nu}+\varepsilon_{\alpha\gamma\mu}\theta_{\beta\delta\nu}\big)D_{\mu\nu},\label{eq:diracCommutator}\\
\anticommutator{D_{\alpha\beta}}{D_{\gamma\delta}} & =2\big(\theta_{\alpha\gamma\mu}\theta_{\beta\delta\nu}-\varepsilon_{\alpha\gamma\mu}\varepsilon_{\beta\delta\nu}\big)D_{\mu\nu}.\label{eq:diracAntiCommutator}
\end{align}

In the right hand sides of \prettyref{eq:diracCommutator} and \prettyref{eq:diracAntiCommutator},
at most one of the two bracketed terms is nonzero for any index values.
Since the tensors $\theta$ and $\varepsilon$ are either zero or
have absolute value 1, the bracketed terms themselves, up to a sign,
can take a single nonzero value, unity. That is, the structure constants
of the Dirac matrices are simple, since they are derived from the
Pauli matrices. We then expect the representation of two-qubit density
matrices in the Dirac basis to yield useful symmetries in the Bloch
vector space.

\subsection{The Bloch matrix}

Writing the density matrix in the Dirac basis, 
\begin{equation}
\rho=\frac{1}{4}r_{\mu\nu}D_{\mu\nu},\label{eq:bloch4d}
\end{equation}
where the scalar coefficients $r_{\mu\nu}$ constitute the 16 entries
of the \emph{Bloch matrix} $\dvec r$. 

The orthogonality relation \prettyref{eq:diracOrthogonality} implies
the Bloch matrix entries are accessible as the expectation values
of tensor products of local observables, as per
\begin{equation}
r_{\mu\nu}=\Tr\big(\rho D_{\mu\nu}\big)=\langle\sigma_{\mu}\otimes\sigma_{\nu}\rangle.\label{eq:blochMatrixDef}
\end{equation}

For $\rho$ to be a density matrix, it is necessary and sufficient
that it be Hermitian, of unit trace, and positive semidefinite. The
first two conditions imply $\dvec r$ is real and $r_{00}=1$. Translating
positivity to a condition on $\dvec r$ is involved, and we defer
it to Sec. \ref{sec:The-Positivity-Condition}. %
\begin{comment}
The reality of $\dvec r$changes the field to which linear combination
coefficients belong from $\mathbb{\mathbb{C}}$ to $\mathbb{\mathbb{R}}$,
which changes the dimensionality of bases, particularly when it comes
to pure states.
\end{comment}

It is instructive to split the Bloch matrix $\dvec r$ into four components;
a scalar of value unity, two three-dimensional vectors, and a $3\times3$
matrix. We write
\begin{equation}
\dvec r=\left[\begin{array}{c|ccc}
1 & r_{01} & r_{02} & r_{03}\\
\hline r_{10} & r_{11} & r_{12} & r_{13}\\
r_{20} & r_{21} & r_{22} & r_{23}\\
r_{30} & r_{31} & r_{32} & r_{33}
\end{array}\right]\equiv\begin{bmatrix}1 & \vec{v}^{\dagger}\\
\vec{u} & R
\end{bmatrix},\label{eq:p2uvP}
\end{equation}
where $u_{i}=r_{i0}$, $v_{j}=r_{0j}$, and $R_{ij}=r_{ij}$. 

The vector $\vec{u}=\Tr_{2}[\rho]$ ($\vec{v}=\Tr_{1}[\rho]$) is
the \emph{local Bloch vector of the first (second) subsystem }once
the other subsystem is traced out, while $R$ is the \emph{correlation
matrix} between the two subsystems. Bloch matrix components are used
by many authors \citep{Horodecki1995,Horodecki1996a,Linden1999,Makhlin2000,Barrett2001,Kus2001,Abouraddy2002,Byrd2003,Spengler2011,Jevtic2014,Milne2014,Aerts2015}.
However, we go further in our analysis and characterization.

\begin{comment}

\subsection{Explicit Representation}
\end{comment}

\begin{widetext}
For a density matrix with entries $\rho_{ij}$, its Bloch matrix $\dvec r$
is, explicitly, 
\[
\dvec r=\begin{bmatrix}1 & \hphantom{{-}}2\Re(\rho_{12}{+}\rho_{34}) & {-}2\Im(\rho_{12}{+}\rho_{34}) & \rho_{11}-\rho_{22}+\rho_{33}-\rho_{44}\\
\hphantom{{-}}2\Re(\rho_{13}{+}\rho_{24}) & \hphantom{{-}}2\Re(\rho_{23}{+}\rho_{14}) & \hphantom{{-}}2\Im(\rho_{23}{-}\rho_{14}) & \hphantom{{-}}2\Re(\rho_{13}{-}\rho_{24})\\
{-}2\Im(\rho_{13}{+}\rho_{24}) & {-}2\Im(\rho_{23}{+}\rho_{14}) & \hphantom{{-}}2\Re(\rho_{23}{-}\rho_{14}) & {-}2\Im(\rho_{13}{-}\rho_{24})\\
\rho_{11}+\rho_{22}-\rho_{33}-\rho_{44} & \hphantom{{-}}2\Re(\rho_{12}{-}\rho_{34}) & {-}2\Im(\rho_{12}{-}\rho_{34}) & \rho_{11}-\rho_{22}-\rho_{33}+\rho_{44}
\end{bmatrix},
\]
where $\Re$ and $\Im$ are, respectively, the real and imaginary
components of what follow. Conversely, given a Bloch matrix $\dvec r$
with components \textbf{$\vec{u}$, $\vec{v}$}, and $R$, defined
in \prettyref{eq:p2uvP}, the density matrix $\rho$ it constructs
is given by\textbf{ }
\[
\rho=\frac{1}{4}\begin{bmatrix}1{+}R_{33}{+}u_{3}{+}v_{3} & R_{31}{-}iR_{32}{+}v_{1}{-}iv_{2} & R_{13}{-}iR_{23}{+}u_{1}{-}iu_{2} & R_{11}{-}iR_{12}{-}iR_{21}{-}R_{22}\\
R_{31}{+}iR_{32}{+}v_{1}{+}iv_{2} & 1{-}R_{33}{+}u_{3}{-}v_{3} & R_{11}{+}iR_{12}{-}iR_{21}{+}R_{22} & {-}R_{13}{+}iR_{23}{+}u_{1}{-}iu_{2}\\
R_{13}{+}iR_{23}{+}u_{1}{+}iu_{2} & R_{11}{-}iR_{12}{+}iR_{21}{+}R_{22} & 1{-}R_{33}{-}u_{3}{+}v_{3} & {-}R_{31}{+}iR_{32}{+}v_{1}{-}iv_{2}\\
R_{11}{+}iR_{12}{+}iR_{21}{-}R_{22} & {-}R_{13}{-}iR_{23}{+}u_{1}{+}iu_{2} & {-}R_{31}{-}iR_{32}{+}v_{1}{+}iv_{2} & 1{+}R_{33}{-}u_{3}{-}v_{3}
\end{bmatrix}.
\]
\end{widetext}

\subsection{Example States}

Here we consider the Bloch matrices of common quantum states. The
\emph{maximally mixed} state, $\rho=I/4$, has a Bloch matrix where
all the entries except $r_{00}$ are zero. 

For a \emph{product state,} $\rho=\rho_{1}\otimes\rho_{2}$, there
are no classical or quantum correlation between the two subsystems.
Supposing the single-qubit density matrices $\rho_{1}$ and $\rho_{2}$
have the Bloch vectors $\vec{u}$ and $\vec{v}$ respectively, then
the Bloch matrix of the product state is given by 
\begin{equation}
\dvec r_{prod}=\begin{bmatrix}1 & \vec{v}^{\dagger}\\
\vec{u} & \vec{u}\vec{v}^{\dagger}
\end{bmatrix}=\begin{bmatrix}1\\
\vec{u}
\end{bmatrix}\begin{bmatrix}1 & \vec{v}^{\dagger}\end{bmatrix}.
\end{equation}

That is, the correlation matrix is equal to the outer product of the
two Bloch vectors, $R=\vec{u}\vec{v}^{\dagger}$, and the Bloch matrix
$\dvec r_{prod}$ itself is an outer product of two 4-vectors. Hence
the interesting algebraic property of the Bloch matrix representation:
\emph{tensor products of operators become outer products of vectors}.

A \emph{separable state} is one that can be written as a convex sum
of product states, and therefore exhibits classical correlations,
but no quantum correlations. A state that is not separable is said
to be \emph{entangled}. Given an arbitrary state, it is not practical
to judge whether it is separable or entangled by attempting to write
it as a convex sum of product states. In practice, one uses the powerful
entanglement criterion discussed in Sec. \ref{sec:Entanglement-Criteria}.

The four maximally entangled Bell states are given by $\ket{\Phi^{\pm}}=\frac{1}{\sqrt{2}}\big(\ket{00}\pm\ket{11}\big),\ket{\Psi^{\pm}}=\frac{1}{\sqrt{2}}\big(\ket{01}\pm\ket{10}\big)$
\citep{Bell1964,Nielsen2011}. Their density matrices are $\rho_{\Phi^{\pm}}=\ket{\Phi^{\pm}}\bra{\Phi^{\pm}}$
and $\rho_{\Psi^{\pm}}=\ket{\Psi^{\pm}}\bra{\Psi^{\pm}}.$ We find
their Bloch matrices to be
\begin{eqnarray}
\dvec r_{\Phi^{+}} & {=}{\begin{bmatrix}\begin{array}{r|rrr}
1 & 0 & 0 & 0\\
\hline 0 & \hphantom{-}1 & 0 & 0\\
0 & 0 & -1 & 0\\
0 & 0 & 0 & \hphantom{-}1
\end{array}\end{bmatrix}}\!, & \;\dvec r_{\Phi^{-}}{=}{\begin{bmatrix}\begin{array}{r|rrr}
1 & 0 & 0 & 0\\
\hline 0 & -1 & 0 & 0\\
0 & 0 & \hphantom{-}1 & 0\\
0 & 0 & 0 & \hphantom{-}1
\end{array}\end{bmatrix}}\!,\hspace{1em}\nonumber \\
\dvec r_{\Psi^{+}} & {=}{\begin{bmatrix}\begin{array}{r|rrr}
1 & 0 & 0 & 0\\
\hline 0 & \hphantom{-}1 & 0 & 0\\
0 & 0 & \hphantom{-}1 & 0\\
0 & 0 & 0 & -1
\end{array}\end{bmatrix}}\!, & \;\dvec r_{\Psi^{-}}{=}{\begin{bmatrix}\begin{array}{r|rrr}
1 & 0 & 0 & 0\\
\hline 0 & -1 & 0 & 0\\
0 & 0 & -1 & 0\\
0 & 0 & 0 & -1
\end{array}\end{bmatrix}}\!.\hspace{1em}\label{eq:bellBlochMatrices}
\end{eqnarray}
As expected for the Bell states, the local Bloch vectors for the individual
systems are always zero, since the partial trace of a maximally entangled
state yields a maximally mixed state on the subsystem. The singlet
state, $\ket{\Psi^{-}}$, has a correlation matrix that is the negative
identity, often making it simpler to deal with algebraically than
the other Bell states. However, this is a superficial distinction;
the singlet state has no fundamental properties not shared by other
maximally entangled states.

We can additionally find the Bloch matrices of generalized Bell states,
$\frac{1}{\sqrt{2}}\big(\ket{00}+e^{i\theta}\ket{11}\big)$ and $\frac{1}{\sqrt{2}}\big(\ket{01}+e^{i\theta}\ket{10}\big),$
also maximally entangled. Recall that an orthogonal matrix $Q$ is
real matrix that satisfies $Q^{\dagger}Q=QQ^{\dagger}=I$, and hence
has determinant $\pm1$. Interestingly, one finds the correlation
matrices of all the aforementioned maximally entangled states to be
orthogonal with determinant $-1$. We shall see in Sec. \ref{subsec:Maximally-Entangled-states}
that these are in fact defining properties of maximally entangled
states. 

\subsection{\label{subsec:Observables}Observables}

To complete our understanding of the Bloch matrix representation,
it is instructive to represent observables in the Dirac basis as well.
We write an observable $A$ as 
\begin{equation}
A=[\dvec A]_{\alpha\beta}D_{\alpha\beta},\label{eq:observable4d}
\end{equation}
with $[\dvec A]$ the Dirac basis representation of $A$. Note that
\prettyref{eq:observable4d} lacks the factor of $\frac{1}{4}$ present
in the Bloch matrix definition \prettyref{eq:bloch4d}. Since $A$
is Hermitian, $[\dvec A]$ is real. 

The expectation value of $A$ is
\begin{eqnarray}
\langle A\rangle & = & r_{\mu\nu}[\dvec A]_{\mu\nu}\equiv\dvec r\cdot[\dvec A].\label{eq:expInnerProduct4d}
\end{eqnarray}
The result is reminiscent of the qubit inner product expectation value
in \prettyref{eq:expInnerProduct}. As an example, suppose we seek
the expectation value of local spins measured in the singlet state.
The observable $B=(\hat{c}\cdot\vec{\sigma})\otimes(\hat{d}\cdot\vec{\sigma})$
is represented in the Dirac basis as $[\dvec B]=\left[\begin{smallmatrix}0 & 0\\
0 & \hat{c}\hat{d}^{\dagger}
\end{smallmatrix}\right]$. The expectation value is given by 
\[
\langle B\rangle_{\Psi^{-}}=\dvec r_{\Psi^{-}}\cdot[\dvec B]=\Tr\big(-I\hat{d}\hat{c}^{\dagger}\big)=-\hat{c}\cdot\hat{d}.
\]
For the singlet state, expectation values of local observables reduce
to inner products because its correlation matrix is the negative identity.
This algebraic convenience is the reason it is more common than other
Bell states.

It is sometimes useful to take the inner product of operators, which
can be shown to yield
\begin{equation}
\Tr[AB]=4[\dvec A]\cdot[\dvec B].\label{eq:observableInnerProduct}
\end{equation}

We also examine the representation of the square of an observable,
which will later help us derive the positivity inequalities. The square
of $A$ is
\begin{eqnarray}
A^{2} & = & \frac{1}{2}\anticommutator AA=\frac{1}{2}[\dvec A]_{\alpha\beta}[\dvec A]_{\gamma\delta}\anticommutator{D_{\alpha\beta}}{D_{\gamma\delta}}\nonumber \\
 & = & [\dvec A]_{\alpha\beta}[\dvec A]_{\gamma\delta}\big(\theta_{\alpha\gamma\mu}\theta_{\beta\delta\nu}-\varepsilon_{\alpha\gamma\mu}\varepsilon_{\beta\delta\nu}\big)D_{\mu\nu},\nonumber \\
 & \equiv & [\dvec{A^{2}}]_{\mu\nu}D_{\mu\nu},\label{eq:ASquaredDirac}
\end{eqnarray}
where we substituted \prettyref{eq:diracAntiCommutator} in the second
line, which also serves as a definition of $[\dvec{A^{2}}]_{\mu\nu}$.
Applying \prettyref{eq:thetaKronecker1} and \prettyref{eq:thetaKronecker2}
to the definition, we find the components of $[\dvec{A^{2}}]$ to
be 
\begin{align}
[\dvec{A^{2}}]_{00} & =[\dvec A]\cdot[\dvec A],\nonumber \\
{}[\dvec{A^{2}}]_{i0} & =2([\dvec A]_{00}[\dvec A]_{i0}+[\dvec A]_{ij}[\dvec A]_{0j}),\nonumber \\
{}[\dvec{A^{2}}]_{0j} & =2([\dvec A]_{00}[\dvec A]_{0j}+[\dvec A]_{i0}[\dvec A]_{ij}),\nonumber \\
{}[\dvec{A^{2}}]_{ij} & =2([\dvec A]_{00}[\dvec A]_{ij}+[\dvec A]_{i0}[\dvec A]_{0j})\nonumber \\
 & \hphantom{===}-\varepsilon_{i_{1}i_{2}i}\varepsilon_{j_{1}j_{2}j}[\dvec A]_{i_{1}j_{1}}[\dvec A]_{i_{2}j_{2}}.\label{eq:observableProduct}
\end{align}

\section{\label{sec:The-Positivity-Condition}The Positivity Condition}

In this key section, we translate the positivity condition on $\rho$,
to a set of conditions on $\dvec r$, or more precisely, on its components
$\vec{u}$, $\vec{v}$ and $R$. 

\subsection{\label{subsec:The-Characteristic-Polynomial}The characteristic polynomial
and Newton's identities}

We begin with the general procedure employed by Kimura \citep{Kimura2003}
for the derivation of positivity conditions. For a 4$\times$4 density
matrix $\rho$ with eigenvalues $\lambda_{l}$ to be positive, it
must satisfy
\begin{equation}
\lambda_{1},\,\lambda_{2},\,\lambda_{3},\,\lambda_{4}\ge0.\label{eq:eigenvaluesPositive}
\end{equation}

We consider the characteristic polynomial of $\rho$, defined as $c(\lambda)\equiv\det\big(\rho-\lambda I\big)$.
We can write this polynomial as a factorized product of terms involving
its roots (the eigenvalues of $\rho$), or as a sum of powers of $\lambda$,
as per
\begin{equation}
c(\lambda)=\prod_{l=1}^{4}(\lambda-\lambda_{l})=\sum_{m=0}^{4}(-1)^{m}a_{m}\lambda^{4-m},\label{eq:charPolynomial}
\end{equation}
where the coefficients $a_{m}$ are themselves functions of the roots
$\lambda_{l}$. If one expands \prettyref{eq:charPolynomial} and
compares coefficients of $\lambda,$ one finds the $a_{m}$ are the
elementary symmetric polynomials, given by Vieta's formulas \citep{Macdonald1995},
\begin{eqnarray*}
a_{0} & = & 1,\\
a_{1} & = & \lambda_{1}+\lambda_{2}+\lambda_{3}+\lambda_{4},\\
a_{2} & = & \lambda_{1}\lambda_{2}+\lambda_{1}\lambda_{3}+\lambda_{1}\lambda_{4}+\lambda_{2}\lambda_{3}+\lambda_{2}\lambda_{4}+\lambda_{3}\lambda_{4},\\
a_{3} & = & \lambda_{1}\lambda_{2}\lambda_{3}+\lambda_{1}\lambda_{2}\lambda_{4}+\lambda_{1}\lambda_{3}\lambda_{4}+\lambda_{2}\lambda_{3}\lambda_{4},\\
a_{4} & = & \lambda_{1}\lambda_{2}\lambda_{3}\lambda_{4}.
\end{eqnarray*}

Descartes' rule of signs colloquially states that the roots of a polynomial
are all positive if and only if its coefficients alternate signs.
More precisely, and given the manner in which the $a_{m}$ were defined
in \prettyref{eq:charPolynomial}, we have
\begin{equation}
\lambda_{l}\ge0,\:\forall l\quad\Leftrightarrow\quad a_{m}\ge0,\;\forall m.\label{eq:posCondition}
\end{equation}

Next, we note that the power sums of the eigenvalues are equivalent
to the trace of the power of the density matrix, as per
\begin{equation}
s_{n}\equiv\lambda_{1}^{n}+\lambda_{2}^{n}+\lambda_{3}^{n}+\lambda_{4}^{n}=\Tr\rho^{n},\quad n=1,2,3,4.\label{eq:tracePower}
\end{equation}

The elementary symmetric polynomials $a_{m}$ and the power sums $s_{n}$
are related by Newton's identities \citep{Macdonald1995}
\begin{eqnarray}
1!a_{1} & = & s_{1},\nonumber \\
2!a_{2} & = & s_{1}^{2}-s_{2},\nonumber \\
3!a_{3} & = & s_{1}^{3}-3s_{1}s_{2}+2s_{3},\nonumber \\
4!a_{4} & = & s_{1}^{4}-6s_{1}^{2}s_{2}+8s_{1}s_{3}+3s_{2}^{2}-6s_{4}.\label{eq:newtonsIdentities}
\end{eqnarray}

Making the substitutions $s_{k}=\Tr\rho^{k},\:s_{1}=1$ in the identities
\prettyref{eq:newtonsIdentities}, and making use of \prettyref{eq:posCondition},
we find that the positivity of $\rho$, defined by \prettyref{eq:eigenvaluesPositive},
is equivalent to the truth of the following four inequalities \begin{subequations}\label{eq:traceIneq}
\begin{eqnarray}
0 & \le & 1,\nonumber \\
0 & \le & 1-\Tr\rho^{2},\label{eq:traceIneq1}\\
0 & \le & 1-3\Tr\rho^{2}+2\Tr\rho^{3},\label{eq:traceIneq2}\\
0 & \le & 1-6\Tr\rho^{2}+8\Tr\rho^{3}+3(\Tr\rho^{2})^{2}-6\Tr\rho^{4}.\label{eq:traceIneq3}
\end{eqnarray}
\end{subequations} 

The three nontrivial inequalities above depend on the trace of the
powers of $\rho$, which we now need to write in terms of the Bloch
matrix and its components. 

\subsection{The density matrix as an observable}

We proceed to calculate $\Tr\rho^{2}$, $\Tr\rho^{3}$, and $\Tr\rho^{4}$
in terms of $\dvec r$ and its components $\vec{u}$, $\vec{v}$ and
$R$. 

To this end, we define $\tilde{R}$ as the cofactor matrix\emph{ }of
$R$. It is the $3\times3$ matrix whose $(i,j)$ element is $(-1)^{i+j}$
times the $(i,j)$ minor of $R$. Recall the minor is the determinant
of the $2\times2$ submatrix obtained from $R$ once the $i^{\text{th}}$
row and $j^{\text{th}}$ column have been removed. The cofactor matrix
satisfies the following identity
\begin{equation}
R\tilde{R}^{\dagger}=\tilde{R}^{\dagger}R=(\det R)I.\label{eq:cofactorDet}
\end{equation}

The above implies $\tilde{R}^{\dagger}$ is proportional to the inverse
of $R$, if the latter is invertible. Explicitly, the entries of $\tilde{R}$
are given by \citep{Grinfeld2013}
\begin{equation}
\tilde{R}_{ij}\equiv\frac{1}{2}\varepsilon_{i_{1}i_{2}i}\varepsilon_{j_{1}j_{2}j}r_{i_{1}j_{1}}r_{i_{2}j_{2}}.\label{eq:cofactor}
\end{equation}

We now define an observable proportional to the density matrix,
\begin{equation}
A\equiv4\rho,\label{eq:ADef}
\end{equation}
which along with \prettyref{eq:bloch4d} and \prettyref{eq:observable4d}
implies that in the Dirac basis representation 
\begin{equation}
[\dvec A]_{\alpha\beta}=r_{\alpha\beta},\qquad\quad[\dvec A]=\begin{bmatrix}1 & \vec{v}^{\dagger}\\
\vec{u} & R
\end{bmatrix}\!.\label{eq:Aequalsr}
\end{equation}
That is, $A$ is the observable whose Dirac basis representation is
equivalent to the Bloch matrix of $\rho$. 

We also find the Dirac basis representation of $A^{2}$. Substituting
the components from \prettyref{eq:Aequalsr} into \prettyref{eq:observableProduct}
and simplifying, we have
\begin{equation}
[\dvec{A^{2}}]=\begin{bmatrix}\|\dvec r\|^{2} & 2\big(\vec{v}^{\dagger}+\vec{u}^{\dagger}R\big)\\
2\big(\vec{u}+R\vec{v}\big) & 2\big(\vec{u}\vec{v}^{\dagger}+R-\tilde{R}\big)
\end{bmatrix}\!,\label{eq:Asquared}
\end{equation}
where we have used the cofactor matrix definition \prettyref{eq:cofactor},
and $\|\dvec r\|^{2}$ is the square magnitude of the Bloch matrix
$\dvec r$. The latter satisfies
\begin{equation}
\|\dvec r\|^{2}=1+\|\vec{u}\|^{2}+\|\vec{v}\|^{2}+\|R\|^{2},\label{eq:blochMatrixMagnitude}
\end{equation}
and $\|R\|^{2}=\Tr\big(R^{\dagger}R\big)$, using the Hilbert-Schmidt
inner product.

Additionally, note that \prettyref{eq:ADef} implies
\begin{equation}
\Tr\rho^{n}=\frac{1}{4^{n}}\Tr A^{n}.\label{eq:traceRhoPower}
\end{equation}

\subsection{The trace of the powers of $\rho$}

We now have the tools we need to calculate the trace of the powers
of $\rho$. Starting with \prettyref{eq:traceRhoPower} for $n=2,3,4$,
we split $A^{n}$ to products of $A$ and $A^{2}$, apply \prettyref{eq:observableInnerProduct}
to find the result in terms of inner products of $[\dvec A]$ and
$[\dvec{A^{2}}]$. Then we use the expressions \prettyref{eq:Aequalsr}
and \prettyref{eq:Asquared} to find the trace in terms of $\vec{u}$,
$\vec{v}$ and $R$. 

Proceeding in this manner we have for $n=2,$ 
\begin{eqnarray}
\Tr\rho^{2} & {=} & \frac{1}{4^{2}}\Tr A^{2}{=}\frac{1}{4^{2}}\Tr(AA){=}\frac{1}{4}[\dvec A]{\cdot}[\dvec A]{=}\frac{1}{4}\|\dvec r\|^{2}.\label{eq:trRho^2}
\end{eqnarray}

\begin{widetext}
For $n=3,$
\begin{eqnarray}
\Tr\rho^{3} & = & \frac{1}{4^{3}}\Tr A^{3}=\frac{1}{4^{3}}\Tr(A^{2}A)=\frac{1}{4^{2}}[\dvec{A^{2}}]\cdot[\dvec A]\nonumber \\
 & = & \frac{1}{16}\Big(\|\dvec r\|^{2}+2\big(\vec{v}^{\dagger}+\vec{u}^{\dagger}R\big)\vec{v}+2\vec{u}^{\dagger}\big(\vec{u}+R\vec{v}\big)+2\Tr\big[R(\vec{u}\vec{v}^{\dagger}+R-\tilde{R})^{\dagger}\big]\Big)\nonumber \\
 & = & \frac{1}{16}\Big(3\|\dvec r\|^{2}-2+6\vec{u}^{\dagger}R\vec{v}-6\det R\Big),\label{eq:trRho^3}
\end{eqnarray}
where in the last line we used \prettyref{eq:cofactorDet} and \prettyref{eq:blochMatrixMagnitude}.
Finally, for $n=4$ we have,
\begin{eqnarray}
\Tr\rho^{4} & = & \frac{1}{4^{4}}\Tr A^{4}=\frac{1}{4^{4}}\Tr(A^{2}A^{2})=\frac{1}{4^{3}}[\dvec{A^{2}}]\cdot[\dvec{A^{2}}]\nonumber \\
 & = & \frac{1}{64}\Big(\|\dvec r\|^{4}+4\big(\vec{v}^{\dagger}+\vec{u}^{\dagger}R\big)\big(\vec{v}^{\dagger}+\vec{u}^{\dagger}R\big)^{\dagger}+4\big(\vec{u}+R\vec{v}\big)^{\dagger}\big(\vec{u}+R\vec{v}\big)+4\Tr\big[(\vec{u}\vec{v}^{\dagger}+R-\tilde{R})(\vec{u}\vec{v}^{\dagger}+R-\tilde{R})^{\dagger}\big]\Big)\nonumber \\
 & = & \frac{1}{64}\Big(\|\dvec r\|^{4}+4\big(\|\dvec r\|^{2}-1+\|\vec{u}\|^{2}\|\vec{v}\|^{2}+\|\vec{u}^{\dagger}R\|^{2}+\|R\vec{v}\|^{2}+\|\tilde{R}\|^{2}+6\vec{u}^{\dagger}R\vec{v}-2\vec{u}^{\dagger}\tilde{R}\vec{v}-6\det R\big)\Big),\label{eq:trRho^4}
\end{eqnarray}
and again we made use of \prettyref{eq:cofactorDet}, \prettyref{eq:blochMatrixMagnitude},
as well as $\vec{v}^{\dagger}R^{\dagger}\vec{u}=\vec{u}^{\dagger}R\vec{v}$,
and $\|\tilde{R}\|^{2}\equiv\Tr\big(\tilde{R}\tilde{R}^{\dagger}\big)$.
\end{widetext}

\subsection{Final positivity conditions}

\begin{comment}
So far in \prettyref{sec:The-Positivity-Condition}, we have derived
the conditions on the Bloch matrix components for the underlying quantum
state to be positive. We started by translating the positivity of
the eigenvalues of $\rho$ to some conditions on the coefficients
of its characteristic polynomial. We then used Newton's identities
to convert the latter to a set of inequalities on $\Tr\rho^{n}$.
We then defined an observable based on $\rho$ , and used the formula
for multiplying observables from Section \prettyref{subsec:Observables}
to calculate $\Tr\rho^{n}$.
\end{comment}

To conclude this section, we plug the expressions for $\Tr\rho^{n}$
from \prettyref{eq:trRho^2}, \prettyref{eq:trRho^3}, and \prettyref{eq:trRho^4}
into \prettyref{eq:traceIneq}. Doing so yields three inequalities,
which constitute necessary and sufficient conditions for the positivity
(i.e. physicality) of the underlying quantum state. These inequalities
are the first of the principal results of this paper, and are given
by 
\begin{widetext}
\begin{subequations} \label{eq:bounds}
\begin{gather}
4-\|\dvec r\|^{2}\geq0,\label{eq:bound1}\\
2\big(\vec{u}{}^{\dagger}R\vec{v}-\det R\big)-\big(\|\dvec r\|^{2}-2\big)\geq0,\label{eq:bound2}\\
8\big(\vec{u}{}^{\dagger}R\vec{v}-\det R\big)+\big(\|\dvec r\|^{2}-2\big)^{2}+8\vec{u}{}^{\dagger}\tilde{R}\vec{v}-4\big(\|\vec{u}\|^{2}\|\vec{v}\|^{2}+\|\vec{u}{}^{\dagger}R\|^{2}+\|R\vec{v}\|^{2}+\|\tilde{R}\|^{2}\big)\geq0.\label{eq:bound3}
\end{gather}

\end{subequations} 
\end{widetext}

To recap, $\dvec r$ is the Bloch matrix, $\vec{u}$,$\vec{v}$, the
local Bloch vectors of the two subsystems, $R$ the correlation matrix
between them, and $\tilde{R}$ the cofactor matrix of $R$. The positivity
inequalities \prettyref{eq:bounds} are equivalent to those independently
derived in Ref. \citep{Englert2001}.

The inequality \prettyref{eq:bound1} is analogous to \prettyref{eq:purityCondition},
setting a limit on the magnitude of the Bloch matrix. The matrix entries
lie inside a $15$-dimensional hypersphere, but don't fill it due
to the other inequalities \prettyref{eq:bound2} and \prettyref{eq:bound3}.
The vector $\vec{u}$ ($\vec{v}$ ) of the first (second) subsystem
always multiplies $R$ and $\tilde{R}$ from the left (right). This
will facilitate an important simplification in Sec. \ref{sec:Singular-Value-decomposition}.

It is instructive to operationally interpret some terms. We write
the three Cartesian canonical (column) unit vectors as $\hat{e}_{1},$
$\hat{e}_{2},$ $\hat{e}_{3}$. The $i^{\text{th}}$ row of $R$ (i.e.
$\hat{e}_{i}^{\dagger}R$) can be thought of as a pseudo-Bloch vector
of the second subsystem provided we simultaneously measure the operator
$\hat{e}_{i}\cdot\vec{\sigma}=\sigma_{i}$ on the first subsystem.
Hence, measurements along $\hat{e}_{i}$ in the first subsystem are
correlated with those along $\hat{e}_{i}^{\dagger}R$ in the second.
The $j^{\text{th}}$ column of $R$ (i.e. $R\hat{e}_{j}$) has the
analogous interpretation as a pseudo-Bloch vector of the first subsystem.

Therefore, $\det R$ can be thought of as the triple product of the
three pseudo-Bloch vectors for either subsystem, equivalent to the
signed volume of the parallelepiped they subtend. This volume can
be contrasted with the volume of the unit cube subtended by $\hat{e}_{1},$
$\hat{e}_{2},$ $\hat{e}_{3},$ since the latter volume in one subsystem
is in some sense correlated with the former volume in the other. Even
though these volumes do not correspond to actual regions of space,
the ratio between them quantifies the overall correlation of the subsystems
in three-dimensional space. This is particularly true when dealing
with a spin-$\frac{1}{2}$ system and the $\hat{e}_{k}$ correspond
to directions along which spin is measured.

The term $\vec{u}{}^{\dagger}R\vec{v}$ is the expectation value if
each subsystem is simultaneously measured along its local Bloch vector.
In the case of an uncorrelated product state $(R=\vec{u}\vec{v}^{\dagger})$,
this reduces to $\|\vec{u}\|^{2}\|\vec{v}\|^{2}$. Hence its departure
from this latter quantity is a gauge to what extent the two subsystems
are correlated.

Similar to the interpretation of the rows of $R,$ the term $\vec{u}{}^{\dagger}R$
is the pseudo-Bloch vector of the second subsystem, provided we simultaneously
measure the operator $\vec{u}\cdot\vec{\sigma}$ on the first subsystem.
That is, the local Bloch vector $\vec{u}$ in the first subsystem
is correlated with $\vec{u}{}^{\dagger}R$ in the second. Also the
Bloch vector $\vec{v}$ in the second subsystem is correlated with
$R\vec{v}$ in the first. 

\begin{comment}
Create a Table to summarize the terms?

What is the intuitive interpretation of the various terms in bound2
and bound3? What is $\tilde{R}$?

-could these terms be better understood in light of entangling unitaries?
\end{comment}

\section{\label{sec:Singular-Value-decomposition}Singular Value Decomposition}

\subsection{Definitions}

To further simplify the representation of the two-qubit quantum state,
we apply the singular value decomposition (SVD) to the correlation
matrix $R$ \citep{Noble1988}. Any real matrix $R$ can be written
as the following matrix product
\begin{equation}
R=M\Sigma N^{\dagger},\label{eq:svd}
\end{equation}
where $M$ and $N$ are orthogonal corresponding to the two subsystems,
and $\Sigma$ is non-negative diagonal. The diagonal entries of $\Sigma=\diag(x_{1},x_{2},x_{3})$
are the \emph{singular values }of $R$. The rank of $R$ is the number
of nonzero $x_{i}$.

One may write the matrices in terms of their column vectors, $M=[\,\hat{m}_{1}\:\hat{m}_{2}\:\hat{m}_{3}\,]$,
$N=[\,\hat{n}_{1}\:\hat{n}_{2}\:\hat{n}_{3}\,]$. Each set of columns
vectors is an orthonormal basis for three-dimensional space. The unit
vector $\hat{m}_{i}$ is the \emph{left singular vector} and $\hat{n}_{i}$
is the \emph{right singular vector }of the singular value $x_{i}$.
We may write \prettyref{eq:svd} as a sum of singular vector outer
products weighted by the singular values, 
\begin{equation}
R=x_{1}\hat{m}_{1}\hat{n}_{1}^{\dagger}+x_{2}\hat{m}_{2}\hat{n}_{2}^{\dagger}+x_{3}\hat{m}_{3}\hat{n}_{3}^{\dagger}.\label{eq:svdouterproducts}
\end{equation}

The three singular values are uniquely defined for a given $R$, however
they may always be reordered arbitrarily as long as the columns of
$M$ and $N$ (i.e. the singular vectors) are reordered in the same
manner. There is additional freedom in defining $M$ and $N$, in
that we may always flip the signs of both the left and right singular
vectors for a given singular value. If some singular values are degenerate
(i.e. equivalent), then an arbitrary orthogonal transformation may
be applied to both the subspaces spanned by the degenerate right and
left singular vectors. If a singular value is zero (implying $R$
has rank $2$ or less), then the sign of either one of its singular
vectors may be flipped. 

The SVD splits the $9$ degrees of freedom in $R$ to $3$ each for
$\Sigma,$ $M,$ and $N$. The left and right singular vectors are
the \emph{primary correlation axes} for their respective subsystems.
This means measurements along the vector $\hat{m}_{i}$ in the first
subsystem have a correlation coefficient, defined as the joint expectation
value, of $x_{i}$ with measurements along $\hat{n}_{i}$ in the second
subsystem, and zero correlation with measurements orthogonal to $\hat{n}_{i}$%
\begin{comment}
definition of correlation here E(XY) is not the one used in statistics
E(XY)/E(X)E(Y)
\end{comment}
. More compactly, 
\begin{equation}
\langle\hat{m}_{i}\cdot\vec{\sigma}\otimes\hat{n}_{j}\cdot\vec{\sigma}\rangle=\hat{m}_{i}^{\dagger}R\hat{n}_{j}=\delta_{ij}x_{i}.\label{eq:svdMeasureExpectation}
\end{equation}

Orthogonal matrices have determinant $\pm1$. A positive (negative)
determinant is equivalent to the matrix representing a rotation (rotoreflection),
and its columns constituting a \emph{right (left)-handed basis}. We
define the \emph{correlation orientation} or \emph{orientation }of
$R$\emph{, }denoted $d$, 
\begin{equation}
d\equiv\det(M)\det(N),\label{eq:ddef}
\end{equation}
which takes on values $\pm1$. The orientation is $+1$ if the two
bases created by the right and left singular vectors have the same
handedness, and $-1$ if they have the opposite handedness. Note that
$d$ is uniquely defined for any $R$ of rank 3, since the freedom
of flipping the signs of a right and left singular vector simultaneously
leaves $d$ unchanged. In this case $d=\sgn(\det R)$. 

For $R$ of rank $2$ or less, $d$ is not uniquely defined, since
one may flip the sign of a single singular vector. Ambiguity can be
mitigated by choosing a particular $M$ and $N$ consistently for
the decomposition of a given $R$. This can be done, for example,
by choosing them such that $d=-1$ whenever there is an ambiguity,
a preference motivated by the negative orientation of Bell states.

Since the quantum state depends also on local Bloch vectors, we define
the \emph{relative Bloch vectors }$\vec{g}$, $\vec{h}$ as
\begin{equation}
\vec{g}\equiv M^{\dagger}\vec{u},\quad\vec{h}\equiv N^{\dagger}\vec{v}.\label{eq:relBlochVectors}
\end{equation}
These are simply the Bloch vectors expressed in the bases set by the
columns of $M$ and $N$. Any ambiguity in defining $M$ and $N$
discussed above translates to ambiguity in $\vec{g}$ and $\vec{h}$,
and can be mitigated the same way. 

Therefore, the $15$ degrees of freedom in the quantum state $\dvec r$
may be split to $3$, $3$, $9$ for $\vec{u}$, $\vec{v}$, $R$
in the Bloch matrix picture, or to five sets of $3$ for $\Sigma$,
$M$, $N$, $\vec{g}$, $\vec{h}$ in the SVD picture. The two pictures
yield complementary insights and we use both in the remainder of the
paper.

\subsection{Positivity inequalities}

We can now simplify the positivity inequalities in \prettyref{eq:bounds}
by making use of \prettyref{eq:svd}, \prettyref{eq:ddef}, and \prettyref{eq:relBlochVectors}
to write them in terms of $\Sigma$, $M$, $N$, $\vec{g}$, $\vec{h}$,
and $d$. To this end, it is straightforward to show that 
\begin{align}
\vec{u}{}^{\dagger}R\vec{v} & =\vec{g}{}^{\dagger}\Sigma\vec{h}, & \det R & =d\det\Sigma, & \|R\|^{2} & =\|\Sigma\|^{2},\nonumber \\
\|\vec{u}\|^{2} & =\|\vec{g}\|^{2}, & \|\vec{v}\|^{2} & =\|\vec{h}\|^{2}, & \|\vec{u}{}^{\dagger}R\|^{2} & =\|\Sigma\vec{g}\|^{2},\nonumber \\
\|R\vec{v}\|^{2} & =\|\Sigma\vec{h}\|^{2}, & \|\dvec r\|^{2} & =1+\|\vec{g}\|^{2}+\|\vec{h}\|^{2}+\|\Sigma\|^{2}.\span\span\label{eq:svdsubstitutions}
\end{align}

\begin{comment}
det is a signed volume, d is the sign, det sigma the magnitude
\end{comment}

We also need to express the cofactor matrix $\tilde{R}$ in terms
of the SVD. If $R$ is an invertible matrix, then the cofactor identity
\prettyref{eq:cofactorDet} implies
\begin{eqnarray}
\tilde{R} & = & (\det R)R^{-\dagger}\nonumber \\
 & = & d\det\Sigma(N\Sigma^{-1}M^{\dagger})^{\dagger}\nonumber \\
 & = & dM\tilde{\Sigma}N^{\dagger},\label{eq:sigmaCofactor}
\end{eqnarray}
where $\tilde{\Sigma}\equiv\diag(x_{2}x_{3},x_{3}x_{1},x_{1}x_{2})=(\det\Sigma)\Sigma^{-1}$
is the cofactor matrix of $\Sigma$. Appendix \ref{sec:svd-Cofactor}
shows that the result of \prettyref{eq:sigmaCofactor} holds even
if $R$ is not invertible.

We can now plug \prettyref{eq:svdsubstitutions} and \prettyref{eq:sigmaCofactor}
into the positivity inequalities \prettyref{eq:bounds}, and find
the\emph{ reduced positivity inequalities} 

\begin{comment}
$\Sigma=\left[\begin{smallmatrix}x_{1} & 0 & 0\\
0 & x_{2} & 0\\
0 & 0 & x_{3}
\end{smallmatrix}\right]$

$\tilde{\Sigma}=\left[\begin{smallmatrix}x_{2}x_{3} & 0 & 0\\
0 & x_{1}x_{3} & 0\\
0 & 0 & x_{1}x_{2}
\end{smallmatrix}\right]$
\end{comment}

\begin{widetext}
\begin{subequations} \label{eq:svbounds}

\begin{gather}
4-\|\dvec r\|^{2}\geq0,\label{eq:svbound1}\\
2\big(\vec{g}{}^{\dagger}\Sigma\vec{h}-d\det\Sigma\big)-\big(\|\dvec r\|^{2}-2\big)\geq0,\label{eq:svbound2}\\
8\big(\vec{g}{}^{\dagger}\Sigma\vec{h}-d\det\Sigma\big)+\big(\|\dvec r\|^{2}-2\big)^{2}+8d\vec{g}{}^{\dagger}\tilde{\Sigma}\vec{h}-4\big(\|\vec{g}\|^{2}\|\vec{h}\|^{2}+\|\Sigma\vec{g}\|^{2}+\|\Sigma\vec{h}\|^{2}+\|\tilde{\Sigma}\|^{2}\big)\geq0.\label{eq:svbound3}
\end{gather}

\end{subequations} 
\end{widetext}

Note that the reduced positivity inequalities above have no direct
dependence on $M$ and $N$, but only indirectly through the orientation
$d$. Of the $15$ degrees of freedom in the quantum state, only $9$
matter for positivity; $3$ each for $\vec{g}$, $\vec{h}$, and $\Sigma$.
Since $d=\pm1$, it is not a continuous degree of freedom, but rather
can be thought of as a binary flag determined from some continuous
degrees of freedom. The inequalities' left hand sides resemble the
characteristic polynomial coefficients in \citep{Kus2001}. 

\subsection{Entangled Bloch spheres}

The singular value decomposition allows us to visualize a two-qubit
state through a pair of Bloch spheres, one per subsystem. The Bloch
vectors $\vec{u}$ and $\vec{v}$ are inscribed in their respective
spheres, representing $6$ degrees of freedom detectable through local
measurements. The 9 degrees of freedom that can only be detected nonlocally
are contained in $\Sigma$, $M$, and $N$, or equivalently, in the
two matrix products $M\Sigma$ and $N\Sigma$. The columns of these
two products are the \emph{scaled correlation axes}, given by $x_{i}\hat{m}_{i}$
and $x_{i}\hat{n}_{i}$ respectively.  

To complete the geometric representation of the quantum state, the
three scaled correlation axes for each system can be added to their
respective Bloch sphere, where they represent the magnitude and direction
of the correlation. The scaled correlation axes in the two systems
are paired off by a shared index $i$. 

As per \prettyref{eq:svdMeasureExpectation}, spin in the directions
of two such axes with the same index are correlated, proportional
to their shared length $x_{i}$, while spin along axes with different
indices are uncorrelated. That is, simultaneously measuring the two
spins on multiple copies of the system, each along the direction of
its scaled correlation axis $i$, yields an expectation value equal
to the axis length. Measuring the two spins simultaneously along correlation
axes with different indices, $i\neq j$, yields zero expectation value.

\begin{figure}
\subfloat[\label{fig:Generic-State}Generic State]{\centering{}\includegraphics[width=0.47\columnwidth]{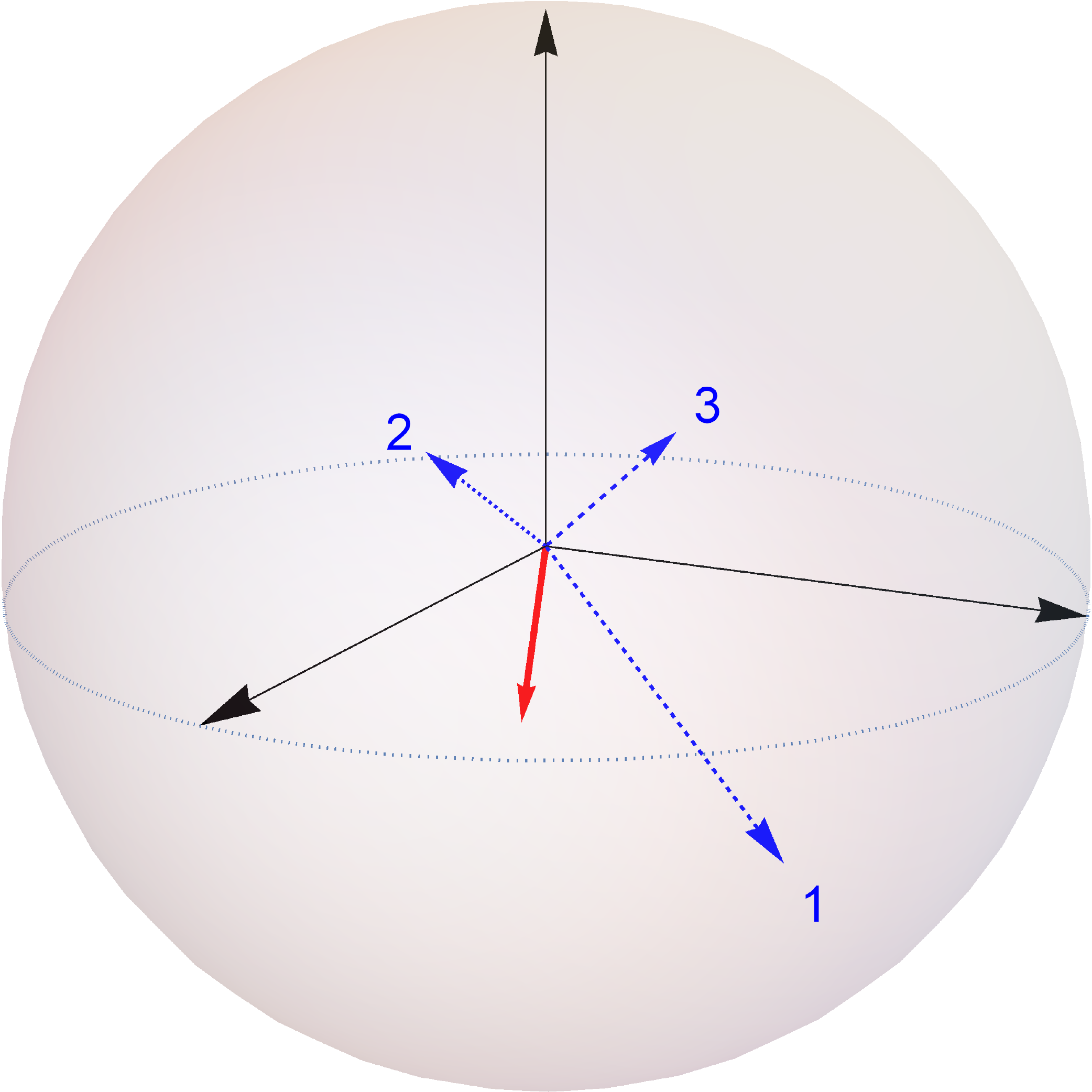}\quad{}\includegraphics[width=0.47\columnwidth]{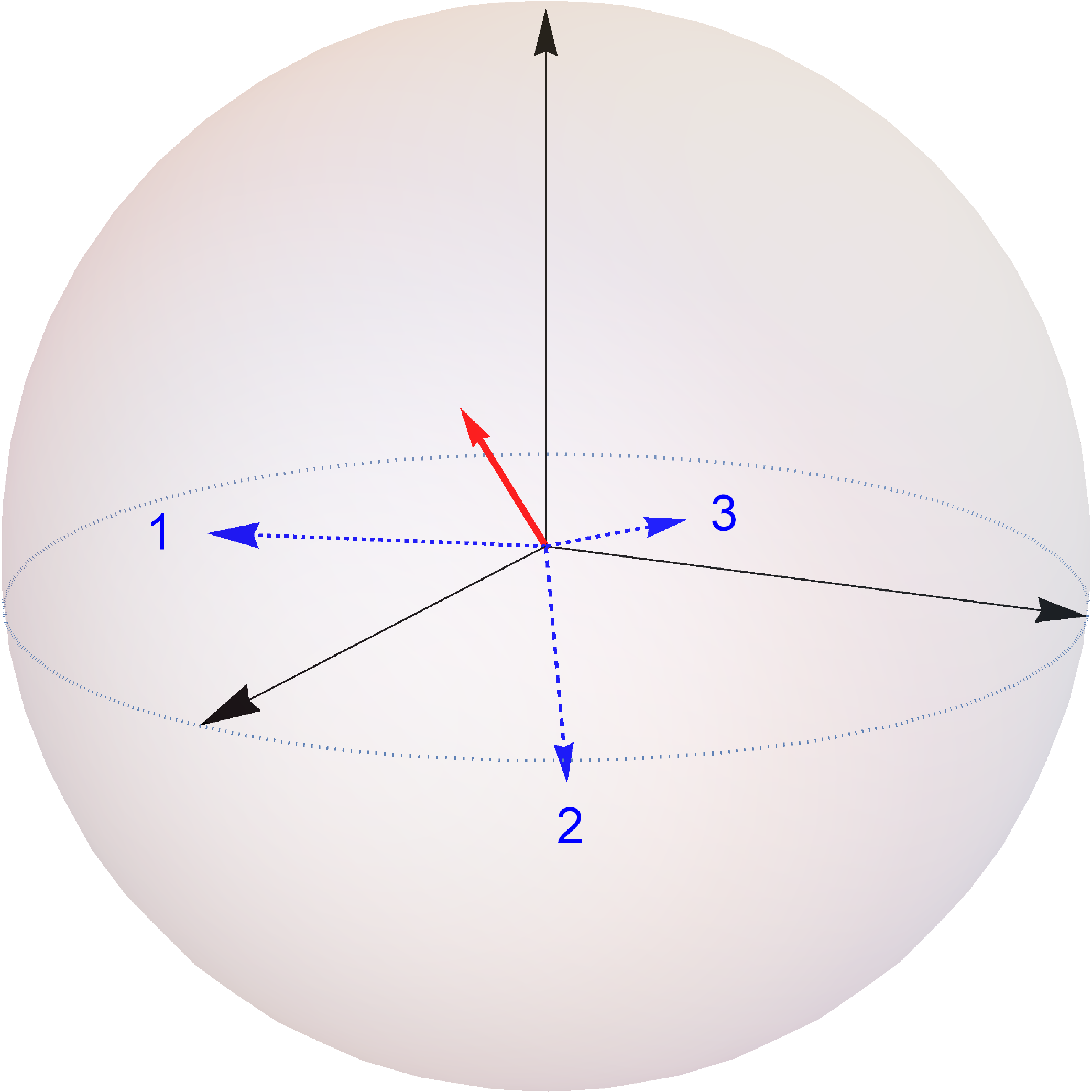}}

\subfloat[\label{fig:Pure-State}Pure State]{\begin{centering}
\includegraphics[width=0.47\columnwidth]{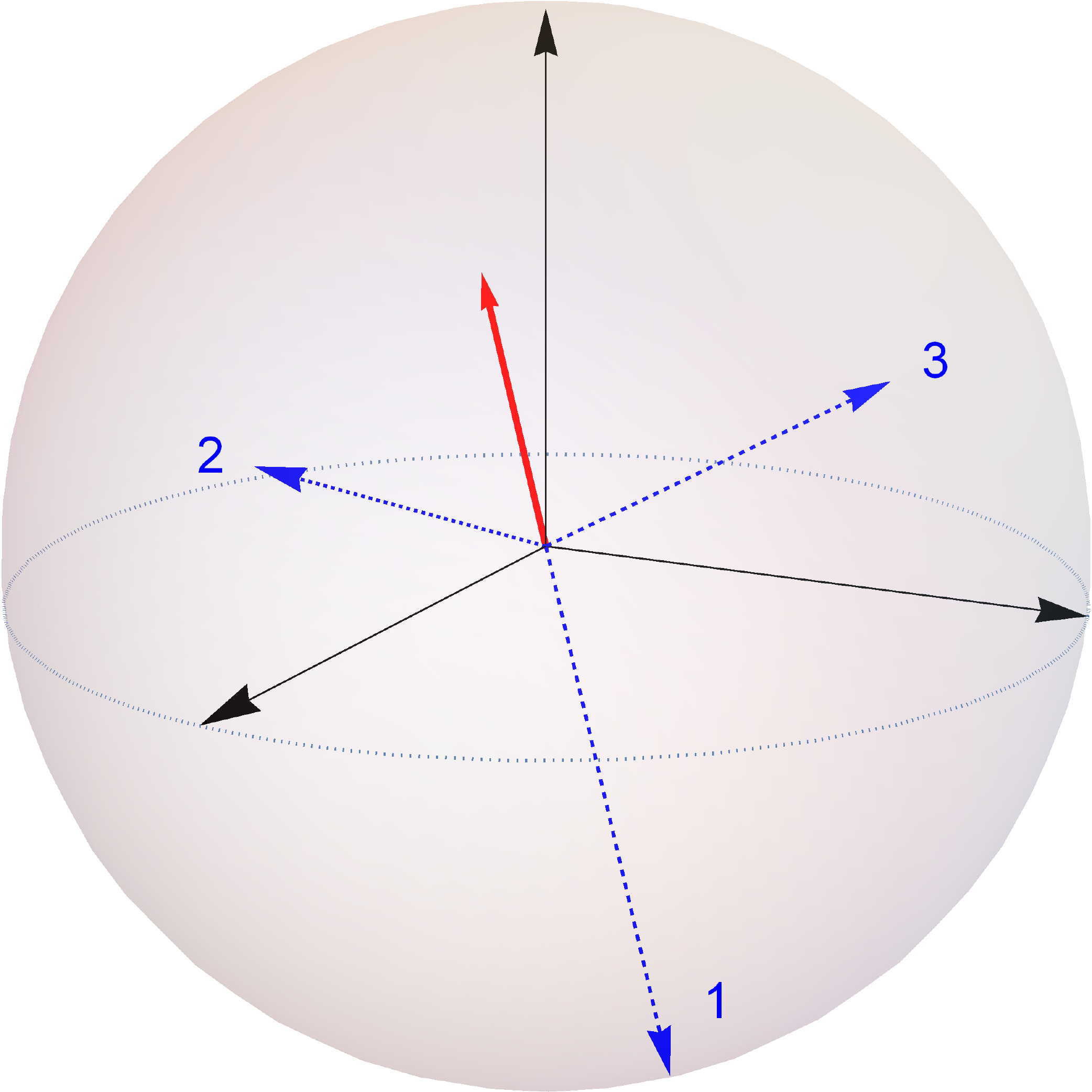}\quad{}\includegraphics[width=0.47\columnwidth]{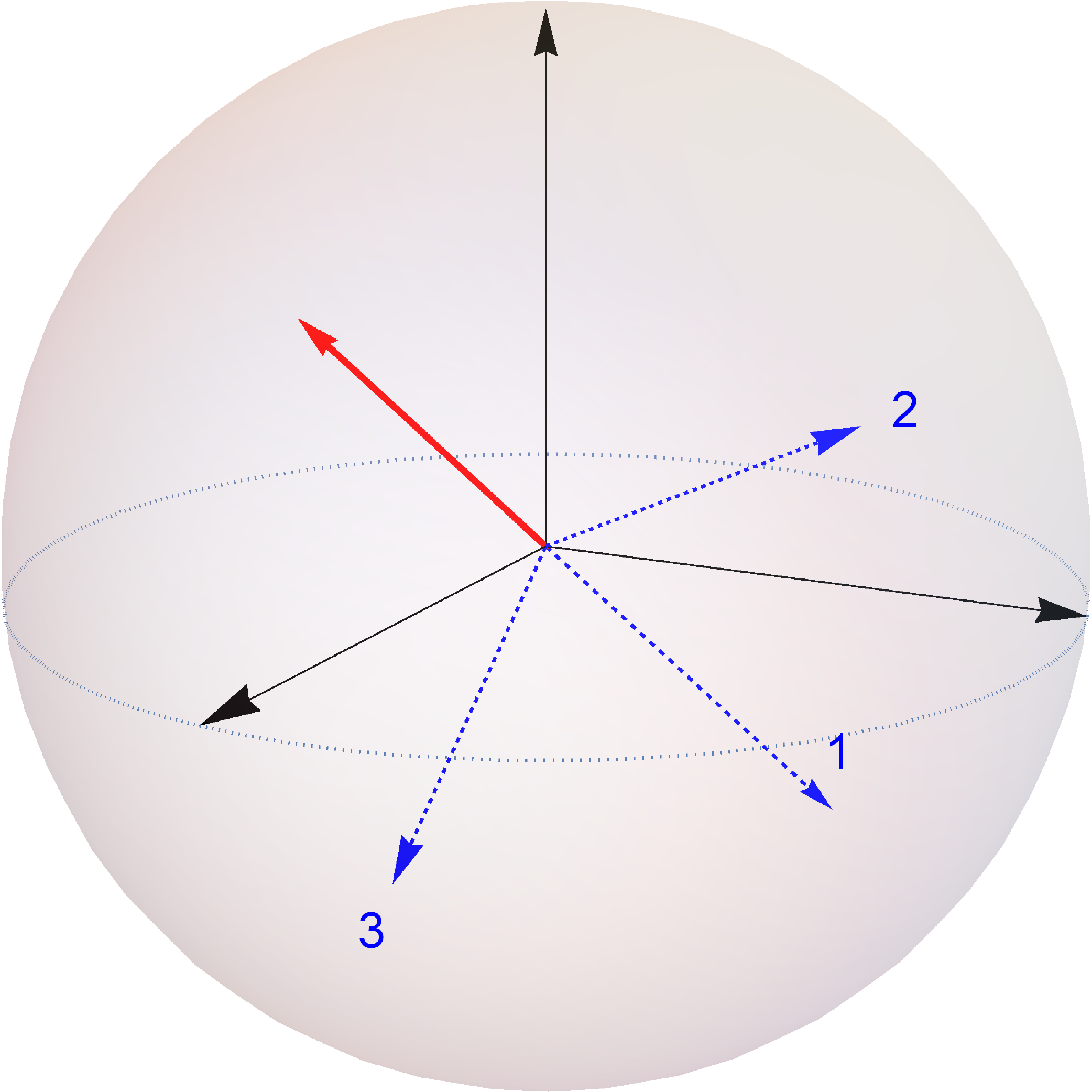}
\par\end{centering}
}

\subfloat[\label{fig:Product-State}Product State]{\centering{}\includegraphics[width=0.47\columnwidth]{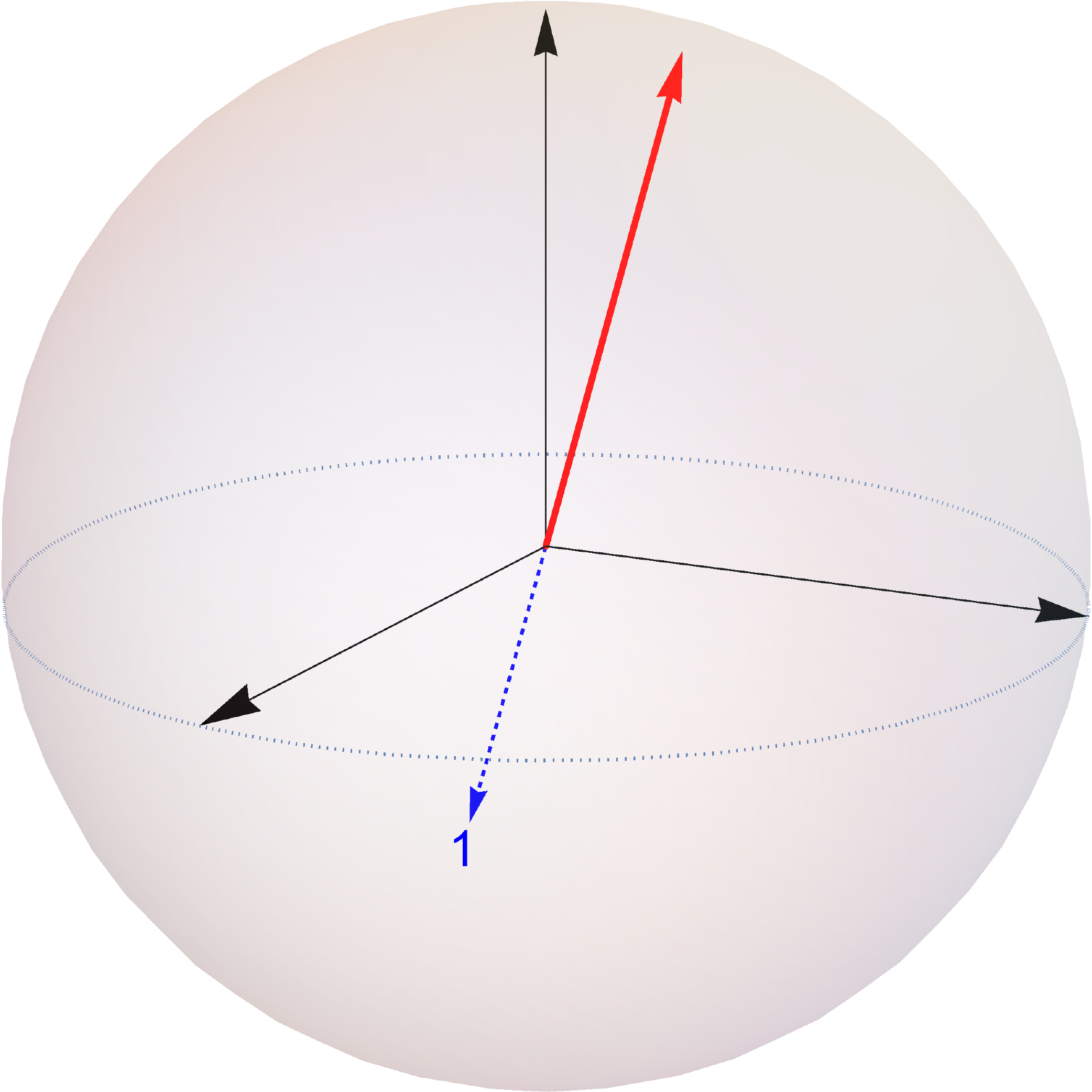}\quad{}\includegraphics[width=0.47\columnwidth]{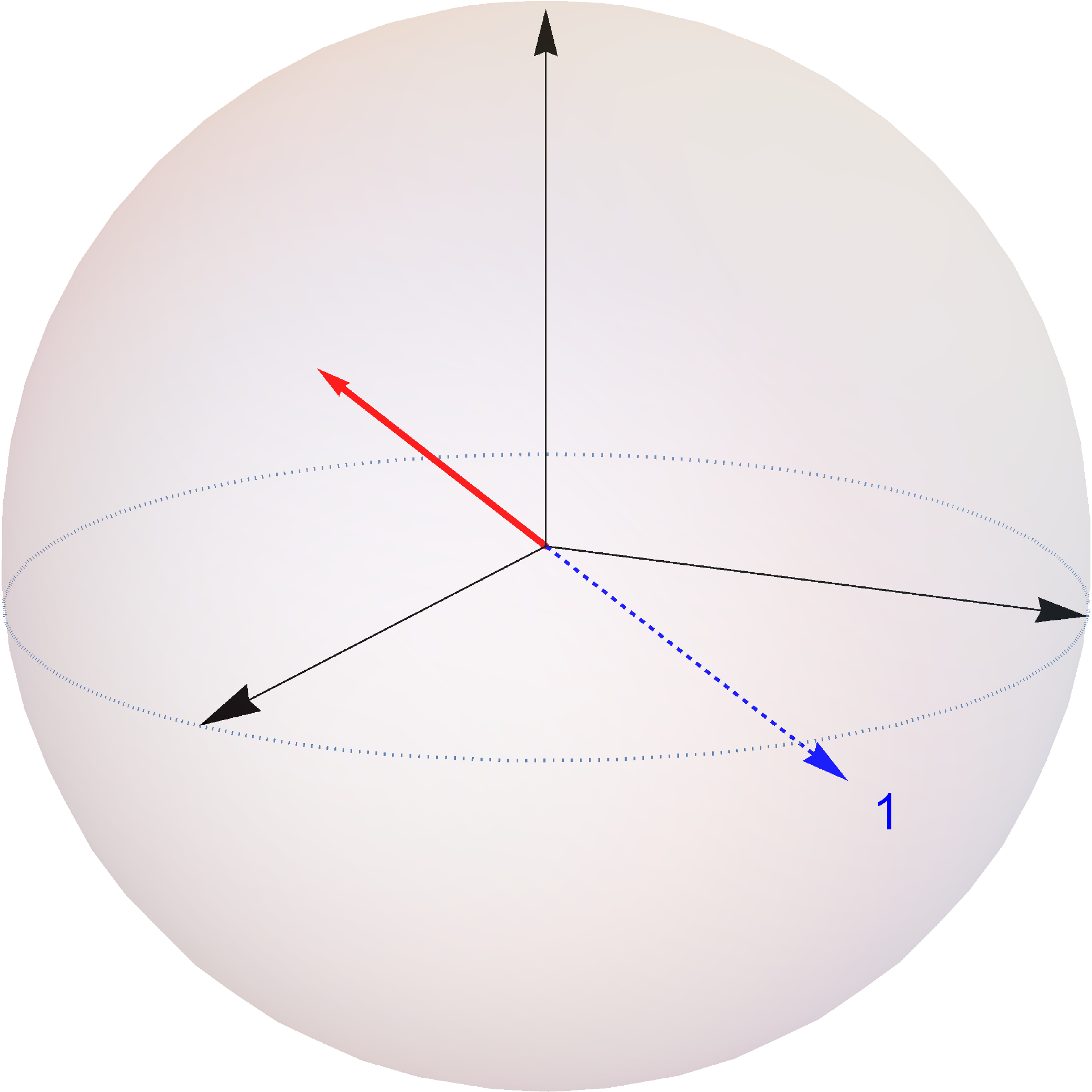}}

\subfloat[\label{fig:Singlet-State}Singlet State]{\centering{}\includegraphics[clip,width=0.47\columnwidth]{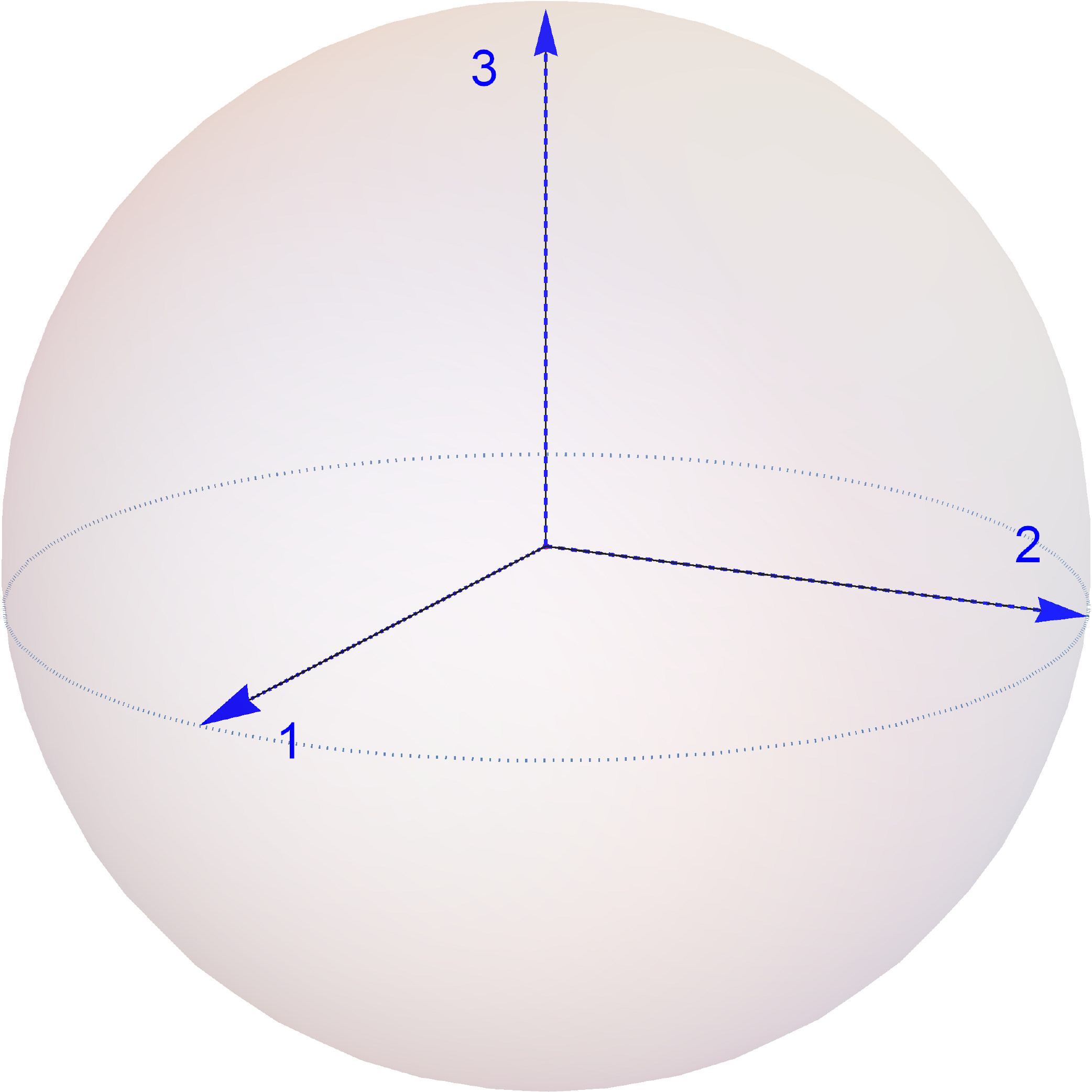}\quad{}\includegraphics[clip,width=0.47\columnwidth]{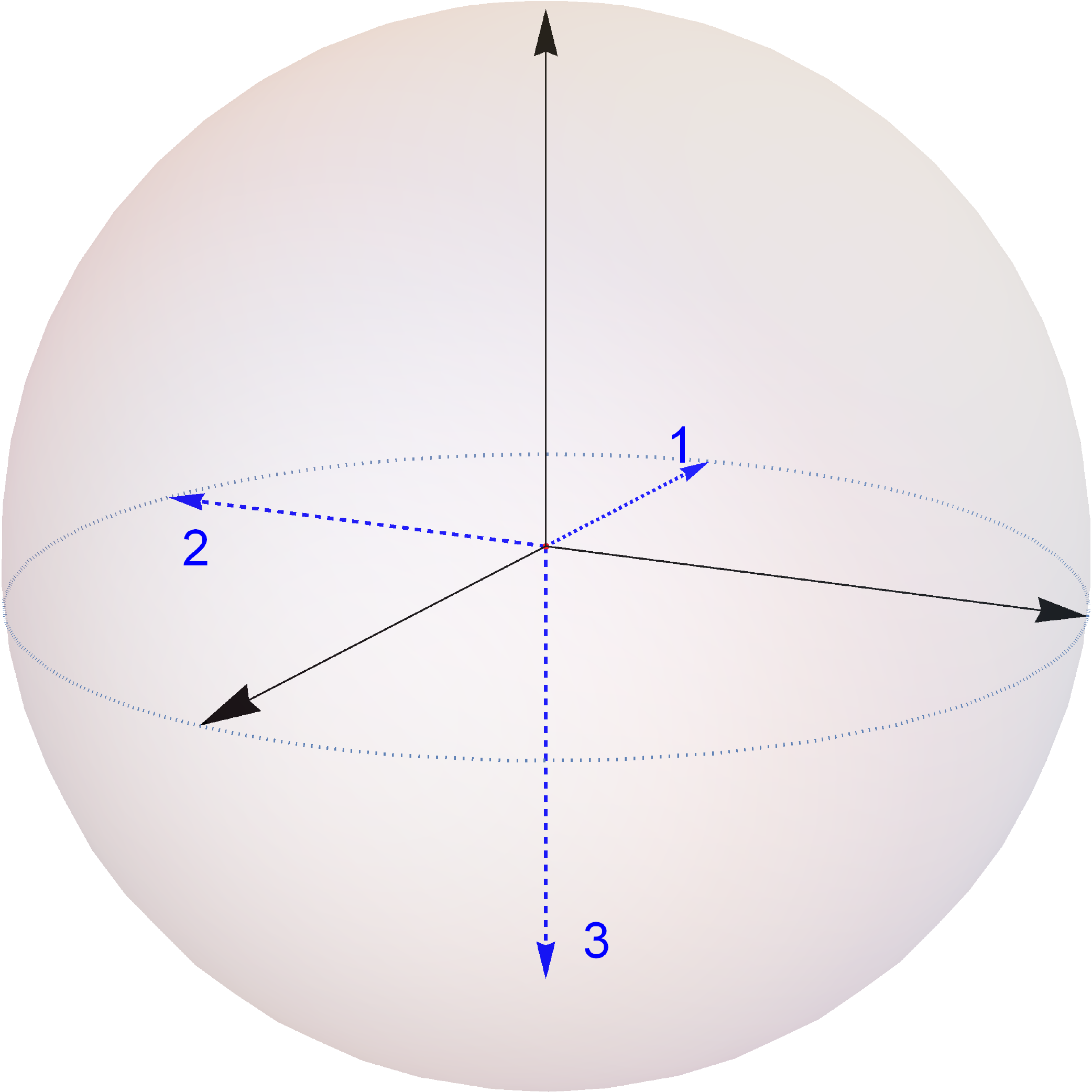}}

\caption{\label{fig:EntangledBlochSpheres}The Bloch spheres of the two subsystems
for: (a) generic state, (b) pure state, (c) product state, and (d)
singlet state. In each Bloch sphere, the local Cartesian axes vectors
are in black, the subsystem's Bloch vector is in red, and the scaled
correlation axes ($x_{i}\hat{m}_{i}$ or $x_{i}\hat{n}_{i}$) dashed
in blue. The scaled correlation axes are mutually orthogonal in each
Bloch sphere, and are labeled with their index $i$ to indicate the
correlation pairing between the two Bloch spheres.}
\end{figure}

Figure \ref{fig:EntangledBlochSpheres} includes the described Bloch
sphere pair diagrams for each of four representative quantum states;
a randomly generated generic state, a pure state, a product state,
and the maximally entangled singlet state. All but the product state
are entangled and have a negative orientation ($d=-1$). The dotted
correlation axes in each Bloch sphere are mutually orthogonal, and
axes with the same label in the two spheres have equal magnitude,
though the projection of three-dimensional vectors onto a two-dimensional
diagram may obscure these facts. 

Figure \ref{fig:Generic-State} represents an arbitrary state generated
from a randomly selected $4\times4$ density matrix. 

In the pure state Fig. \ref{fig:Pure-State}, the first scaled correlation
axis has unit magnitude, while the magnitudes of the second and third
are equivalent. The Bloch vector is colinear with the first correlation
axis. Section \ref{subsec:Pure-States} shows that these are always
properties of pure states. 

The product state in Fig. \ref{fig:Product-State} has only one scaled
correlation axis, which is colinear with the Bloch vector. The second
and third scaled correlation axes vanish as $x_{2}=x_{3}=0$. The
magnitude of the non-vanishing correlation axis is equivalent to the
product of the magnitudes of the two Bloch vectors. The SVD of product
states demonstrating these properties can be discerned by comparing
$R=\vec{u}\vec{v}^{\dagger}=\|\vec{u}\|\|\vec{v}\|\hat{u}\hat{v}^{\dagger}$
with \prettyref{eq:svdouterproducts}.

Finally in the singlet state in Fig. \ref{fig:Singlet-State} the
Bloch vectors vanish, the scaled correlation axes all have unit magnitude,
with an opposite handedness in each Bloch sphere. We see in Sec. \ref{subsec:Maximally-Entangled-states}
that these are properties of all maximally entangled states. The unique
additional feature of the singlet state lies in the fact that all
paired correlation axes between the two spheres differ only by a sign.
This follows from its correlation matrix $R$ being the negative identity
matrix, as shown in \prettyref{eq:bellBlochMatrices}.

The ambiguities of the SVD are better understood in the diagrammatic
representations above. Reordering the singular values and singular
vectors corresponds to a simple relabeling of the scaled correlation
axes. One may freely flip the signs of any two paired correlation
axes, since measurements in the negative direction of both subsystems
will still be positively correlated. If two scaled correlation axes
have the same length, an identical rotation about the third axis may
be applied to them in both Bloch spheres. For example, axes $2$ and
$3$ in both spheres for the pure state may be rotated by the same
arbitrary angle about axis $1$, leaving the underlying quantum correlations
unaffected.

Although any two-qubit quantum state can be represented as a pair
of correlated Bloch spheres, not every possible configuration of Bloch
vectors and scaled correlation axes represents a physically allowed
quantum state. For a state to be physically allowed, it must satisfy
the positivity inequalities \prettyref{eq:svbounds}. Since the latter
depend on the relative not absolute Bloch vectors, one may arbitrarily
rotate a Bloch sphere as a single unit (along with its Bloch vector
and correlation axes) without affecting the physicality of a state. 

While the Bloch sphere pairs help us visualize individual quantum
states, Sec. \ref{sec:Geometry-of-State-Space} is dedicated to visualizing
the entire quantum state space.

\section{\label{sec:Unitary-Operations}Unitary Operations}

\subsection{Local unitary transformations}

In this section we investigate the effect of unitary operations on
the Bloch matrix components as well as the singular value decomposition.
We begin with local unitary operations. 

We showed in \prettyref{eq:unitaryRotation2} that a single qubit
unitary transformation is equivalent to a rotation of the Bloch vector.
Let $U_{1}$ and $U_{2}$ be $2\times2$ unitary matrices with unitary
transformations corresponding to rotation matrices $Q_{1}$ and $Q_{2}$
respectively. It is straightforward to show that applying a local
unitary transformation to the quantum state, $\rho\rightarrow\rho^{\prime}=(U_{1}\otimes U_{2})\rho(U_{1}^{\dagger}\otimes U_{2}^{\dagger})$,
is equivalent to the following transformations on the Bloch matrix
components \citep{Horodecki1996a,Makhlin2000}:
\begin{align}
\vec{u} & \rightarrow\vec{u}^{\prime}=Q_{1}\vec{u},\nonumber \\
\vec{v} & \rightarrow\vec{v}^{\prime}=Q_{2}\vec{v},\nonumber \\
R & \rightarrow R^{\prime}=Q_{1}RQ_{2}^{\dagger},\label{eq:localUnitaryonBloch}
\end{align}
where the primed symbols indicate the value after the transformation.
The local Bloch vectors are rotated as expected, while the first rotation
is applied to the rows of the correlation matrix, and the second to
its columns.

Since $Q_{i}$ are rotations, they satisfy $\det Q_{i}=1$ and $Q_{i}^{\dagger}Q_{i}=I$.
With this in mind, it easy to show that the transformations \prettyref{eq:localUnitaryonBloch}
leave every term in the positivity inequalities \prettyref{eq:bounds}
unchanged. It is to be expected of course that local unitary transformations
do not affect positivity. Nonetheless, it is interesting that even
the individual terms in the inequalities are unaffected.

It becomes clear why this is the case when we examine the effect of
local unitary transformations in the SVD picture. The modified correlation
matrix can be expressed in its own SVD, $R^{\prime}=Q_{1}M\Sigma N^{\dagger}Q_{2}^{\dagger}\equiv M^{\prime}\Sigma N^{\prime\dagger},$
with $M^{\prime}\equiv Q_{1}M$ and $N^{\prime}\equiv Q_{2}N$ themselves
orthogonal matrices. The relative Bloch vectors are left unaffected
by the transformation, as per $\vec{g}^{\prime}=M^{\prime\dagger}\vec{u}^{\prime}=M^{\dagger}Q_{1}^{\dagger}Q_{1}\vec{u}=M^{\dagger}\vec{u}=\vec{g},$
with a similar result for $\vec{h}$. The orientation is likewise
unaffected with $d^{\prime}=\det(Q_{1})\det(Q_{2})d=d$. 

The effect of the local unitary transformation is then
\begin{align}
M & \rightarrow M^{\prime}=Q_{1}M,\nonumber \\
N & \rightarrow N^{\prime}=Q_{2}N,\label{eq:localUnitaryonSVD}
\end{align}
with $\Sigma,\vec{g},\vec{h},d$ left unchanged. Only the unaffected
degrees of freedom are present in the positivity inequalities \prettyref{eq:svbounds},
explaining why even their individual terms are left unchanged. This
representation describes in a simpler manner the local unitary invariants
derived in Ref. \citep{Makhlin2000}.

In the paired Bloch sphere diagrams, a local unitary transformation
rotates the Bloch vector and correlation axes in each sphere together,
leaving the relative Bloch vectors unchanged. Equivalently, the reverse
rotation may be applied to the absolute axes in each sphere.

There are two senses in which we speak of ``local degrees of freedom''.
We may mean the degrees of freedom that are \emph{locally measurable}.
These are simply the Bloch vectors $\vec{u},\vec{v}$. We may also
mean the degrees of freedom that are \emph{free to vary via local
unitary transformations}. That is, the orthogonal matrices $M,N$,
with the product of their determinants, the orientation $d$, left
unchanged.

\subsection{General unitary transformations}

We now consider the effect of general unitary operations on the composite
quantum state. Ideally, we would like to represent an arbitrary unitary
operator $U\in\text{SU}(4)$, as a combination of local and nonlocal
unitary transformations. A powerful result by Zhang et al. \citep{Zhang2002a}
fulfills this requirement, stating that any such $U$ can be written
as
\begin{equation}
U=\left(U_{1}\otimes U_{2}\right)\mathring{U}\big(\theta_{1},\theta_{2},\theta_{3}\big)\left(U_{3}\otimes U_{4}\right),\label{eq:generalUreduction}
\end{equation}
where the $U_{k}$ are single-qubit unitary operators, and $\mathring{U}\big(\theta_{1},\theta_{2},\theta_{3}\big)$,
which we call \emph{a basic nonlocal operator}, is given by
\begin{equation}
\mathring{U}=\exp\left[\frac{i}{2}\left(\theta_{1}\sigma_{1}{\otimes}\sigma_{1}+\theta_{2}\sigma_{2}{\otimes}\sigma_{2}+\theta_{3}\sigma_{3}{\otimes}\sigma_{3}\right)\right].\label{eq:irreduciblenonlocal}
\end{equation}

In other words, a generic unitary transformation can be reduced to
a local transformation, followed by a basic nonlocal transformation,
followed by another unitary transformation, with $6$, $3$, and $6$
degrees of freedom respectively. In the previous section we examined
the effect of local unitary transformations, and therefore only need
to consider the effect of a basic nonlocal operator $\mathring{U}$.
The above representation is not necessarily unique \citep{Zhang2002a},
however this is of no consequence for our purposes.

Since the three $\sigma_{j}\otimes\sigma_{j}$ commute, the matrix
exponential of their sum is simply the product of their matrix exponentials,
in any order. It is therefore possible to factorize $\mathring{U}$
to the product of three exponentials,
\begin{equation}
\mathring{U}\big(\theta_{1},\theta_{2},\theta_{3}\big)=\mathring{U}_{1}\big(\theta_{1}\big)\mathring{U}_{2}\big(\theta_{2}\big)\mathring{U}_{3}\big(\theta_{3}\big),\label{eq:UBasicU1U2U3}
\end{equation}
where the $\mathring{U}_{j}$, called \emph{irreducible nonlocal operators},
are given by
\begin{equation}
\mathring{U}_{j}\big(\theta_{j}\big)=\exp\left[\frac{i}{2}\theta_{j}\sigma_{j}{\otimes}\sigma_{j}\right],\quad j=1,2,3.\label{eq:UjIrreducible}
\end{equation}

To understand the action of nonlocal operations, we examine the effect
of one of the irreducible nonlocal transformations, say $\mathring{U}_{1}$,
with the understanding that $\mathring{U}_{2}$ and $\mathring{U}_{3}$
will be of similar effect. With much algebra, some of which is shown
in Appendix \ref{sec:BasicNonlocal}, the transformation $\rho\rightarrow\rho^{\prime}=\mathring{U}_{1}\rho\mathring{U}_{1}^{\dagger}$
can be shown to transform the Bloch matrix $\dvec r$ in the following
manner:
\begin{widetext}
\begin{equation}
\dvec r\rightarrow\dvec r^{\prime}=\begin{bmatrix}1 & v_{1} & \hphantom{-}v_{2}\cos\theta_{1}+R_{13}\sin\theta_{1} & \hphantom{-}v_{3}\cos\theta_{1}-R_{12}\sin\theta_{1}\\
u_{1} & R_{11} & \hphantom{-}v_{3}\sin\theta_{1}+R_{12}\cos\theta_{1} & -v_{2}\sin\theta_{1}+R_{13}\cos\theta_{1}\\
u_{2}\cos\theta_{1}+R_{31}\sin\theta_{1} & \hphantom{-}u_{3}\sin\theta_{1}+R_{21}\cos\theta_{1} & R_{22} & R_{23}\\
u_{3}\cos\theta_{1}-R_{21}\sin\theta_{1} & -u_{2}\sin\theta_{1}+R_{31}\cos\theta_{1} & R_{32} & R_{33}
\end{bmatrix}.\label{eq:U1effect}
\end{equation}
\end{widetext}

One can interpret the operation $\mathring{U}_{1}$ as resulting in
four two-variable ``mixing'' operations, where each mixture is the
mathematical application of the two-dimensional rotation matrix $\left[\begin{smallmatrix}\cos\theta & -\sin\theta\\
\sin\theta & \cos\theta
\end{smallmatrix}\right]$, with $\theta$ the mixing angle, to a vector of the two mixed variables.
The operation mixes $u_{2}$ with $R_{31}$ and $v_{2}$ with $R_{13}$
with a mixing angle $-\theta_{1}$, and $u_{3}$ with $R_{21}$, $v_{3}$
with $R_{12}$ with a mixing angle $\theta_{1}$. 

More generally, supposing $\{i,j,k\}$ to be a cyclic permutation
of $\{1,2,3\}$, the operation $\mathring{U}_{j}$ mixes $u_{k}$
with $R_{ij}$, $v_{k}$ with $R_{ji}$ with a mixing angle $-\theta_{j}$,
and $u_{i}$ with $R_{kj}$, $v_{i}$ with $R_{jk}$ with a mixing
angle $\theta_{j}$. 

This mixing action is precisely what generates entanglement. If we
start with a product state ($R_{ij}=u_{i}v_{j}$), then the modified
Bloch vector for each subsystem in \prettyref{eq:U1effect} will depend
on the other system's Bloch vector. That is, correlation was created
between the two subsystems.

One may combine the effects of the three $\mathring{U}_{j}$ to find
the action of $\mathring{U}$, as per \prettyref{eq:UBasicU1U2U3}.
The effect of the basic nonlocal transformation $\rho\rightarrow\rho^{\prime}=\mathring{U}\rho\mathring{U}^{\dagger}$
on the Bloch matrix $\dvec r$ is given in Appendix \ref{sec:BasicNonlocal}.

A question that naturally arises at this point is the effect of irreducible
nonlocal transformations on the SVD picture, i.e. its effect on $\Sigma,\vec{g},\vec{h},M,N$
and $d$. Since local operations only act on $M$ and $N$, one may
naively hope that an irreducible nonlocal operator only acts on $\Sigma,\vec{g},\vec{h},$
and $d$. However, this cannot be the case, since it would imply that
irreducible nonlocal operators commute with local operations. Given
the action of $\mathring{U}_{1}$ on $\vec{u},\vec{v},R$ shown in
\prettyref{eq:U1effect}, there is no simple way to represent its
effect on the SVD components.

\begin{figure}[h]
\includegraphics[width=0.95\columnwidth]{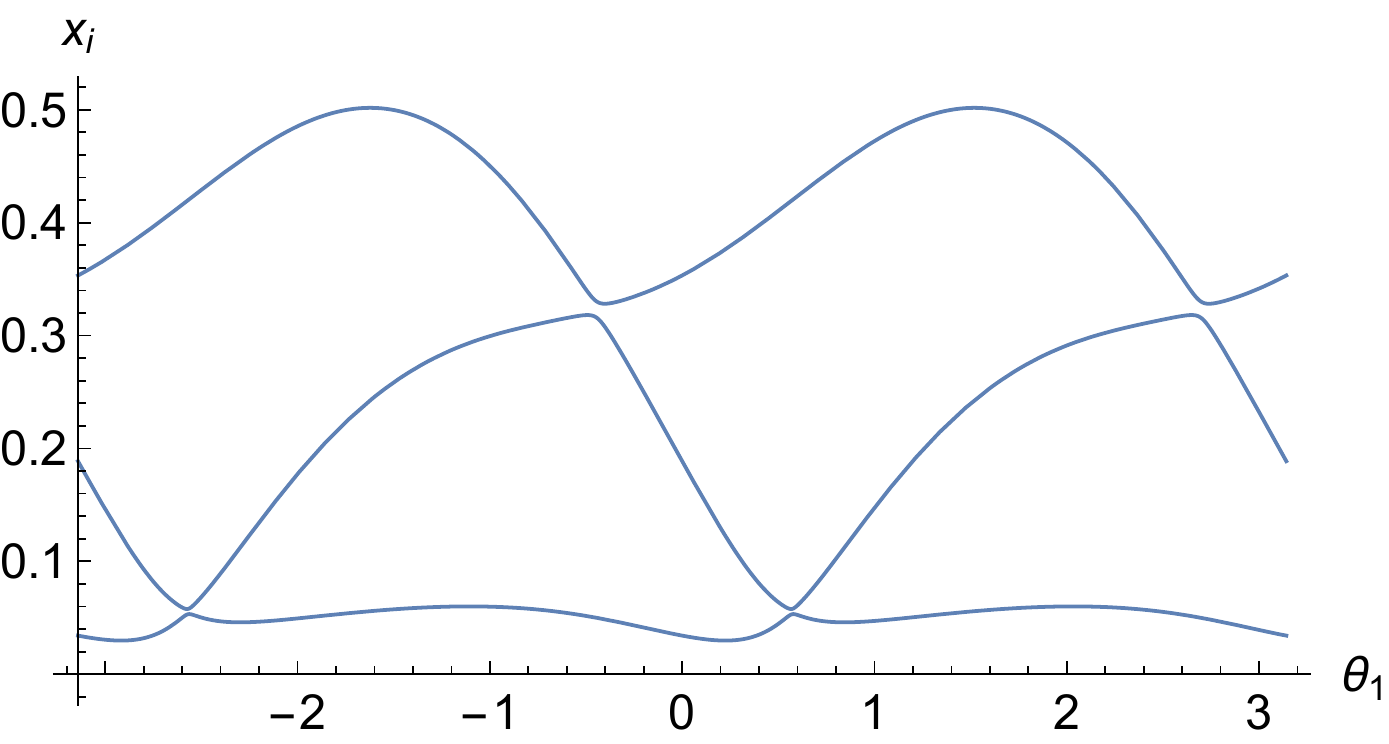}

\caption{\label{fig:svaluesU1}The effect of the irreducible nonlocal unitary
transformation due to $\mathring{U}_{1}(\theta_{1})$ on a generic
quantum state. The three singular values $x_{i}$ are plotted against
the parameter angle $\theta_{1}$ in the domain $[-\pi,\pi]$.}
\end{figure}

We demonstrate this by plotting the effect of $\mathring{U}_{1}(\theta_{1})$
on the singular values of a randomly generated quantum state in Fig.
\ref{fig:svaluesU1}. The singular values change with a period $\pi$.
We also see that $\mathring{U}_{1}$ has an effect on the singular
values akin to avoided crossings of Hermitian operator eigenvalues
\citep{VonNeumann1929}. In the region in parameter-space where the
avoided crossing between singular values $x_{i}$ and $x_{j}$ takes
place, one can show that their respective primary correlation axes
undergo a rapid but continuous transformation roughly with the net
effect that they switch places; $\hat{m}_{i}\longleftrightarrow\hat{m}_{j},$
and $\hat{n}_{i}\longleftrightarrow\hat{n}_{j}$. For some special
choices of initial state, (or with a simultaneous application of $\mathring{U}_{2}$
and/or $\mathring{U}_{3}$) one can get actual crossings.

The above figure simplifies somewhat for pure states, and more so
for maximally entangled states. However, the action of $\mathring{U}_{1}$
on the SVD components cannot, in general, be given in a form that
is simpler than its action on the Bloch matrix components in \prettyref{eq:U1effect}.

\subsection{Unitary invariants}

For a $4\times4$ matrix there are exactly four invariant quantities
unchanged by unitary transformations. The invariants of a density
matrix $\rho$ may be taken as its eigenvalues $\lambda_{1},\lambda_{2},\lambda_{3},\lambda_{4}$.
Alternatively, since functions of invariants are themselves invariant,
one may take $\Tr\rho,\Tr\rho^{2},\Tr\rho^{3},\Tr\rho^{4}$ as the
\emph{trace unitary invariants}. Since $\Tr\rho=1$, the values of
the three other traces define an equivalence class of density matrices.
Unitary transformations can take a density matrix to any other in
its equivalence class, but not to one in another class.

One may also find three invariants in terms of the Bloch matrix components.
Given their derivation from $\Tr\rho^{n}$, it is clear that the left
hand sides of the positivity inequalities \prettyref{eq:bounds} or
\prettyref{eq:svbounds}, are unitarily invariant. We call these the
\emph{positivity unitary invariants, }as their values indicate how
far the state is from violating positivity. 

We can further simplify these by extracting from them three independent
invariants, similar to those in \citep{Englert2001}, which we call
the \emph{Bloch invariants}, given by
\begin{align}
B_{1} & \equiv\|\dvec r\|^{2},\nonumber \\
B_{2} & \equiv\vec{u}{}^{\dagger}R\vec{v}-\det R=\vec{g}{}^{\dagger}\Sigma\vec{h}-d\det\Sigma,\nonumber \\
B_{3} & \equiv\|\vec{u}\|^{2}\|\vec{v}\|^{2}+\|\vec{u}{}^{\dagger}R\|^{2}+\|R\vec{v}\|^{2}+\|\tilde{R}\|^{2}-2\vec{u}{}^{\dagger}\tilde{R}\vec{v}\nonumber \\
 & =\|\vec{g}\|^{2}\|\vec{h}\|^{2}+\|\Sigma\vec{g}\|^{2}+\|\Sigma\vec{h}\|^{2}+\|\tilde{\Sigma}\|^{2}-2d\vec{g}{}^{\dagger}\tilde{\Sigma}\vec{h}.\label{eq:blochInvariants}
\end{align}

Thus, there are different levels of invariance. Local unitary transformations
will leave nine continuous degrees of freedom $\Sigma,\vec{g},\vec{h}$
as well as the discrete $d$ invariant \citep{Makhlin2000,Grassl1998}.
A general (nonlocal) unitary transformation will leave the three degrees
of freedom in the Bloch invariants \textbf{$B_{1},B_{2},B_{3}$ }unchanged.
The noteworthy feature of the expressions in \prettyref{eq:blochInvariants}
is that \emph{they express the general unitary invariants in terms
of the local unitary invariants}.

If the quantum state undergoes non-unitary evolution, as in open system
dynamics \citep{Breuer2007} or depolarizing noise channels, then
even the Bloch invariants \textbf{$B_{i}$ }will change.

Interestingly, $B_{i}$ is of order $i+1$ in the Bloch matrix terms.
If the quantum state is acted upon by a depolarizing noise channel
$\rho\rightarrow\rho^{\prime}=p\rho+(1-p)\frac{I}{4}$, with $(1-p)$
the noise ratio, then the Bloch invariants change as $B_{i}\rightarrow p^{i+1}B_{i}$.

\section{\label{sec:Entanglement-Criteria}Entanglement Criteria}

A quantum state $\rho$ is defined as \emph{separable} if it can be
written as a convex combination of product states,
\begin{equation}
\rho=\sum_{i}p_{i}\rho_{i}^{(1)}\otimes\rho_{i}^{(2)},\label{eq:rhoseparable}
\end{equation}
where $p_{i}$ are non-negative probabilities that sum to unity. A
state that is not separable is defined as entangled. 

Given a quantum state, it is important to find out whether or not
it is entangled. To this end, one can use the \emph{positive partial
transpose} (PPT)\emph{ }criterion, also known as the Peres-Horodecki
criterion. It was first stated by Woronowicz \citep{Woronowicz1976}
based on previous work by St�rmer \citep{Stormer1963}, and extended
for use in quantum systems by Peres and Horodecki \citep{Peres1996,Horodecki1996}.
The PPT criterion states that if one takes the transpose of one subsystem
(i.e. partial transpose) of the density matrix $\rho$, and the resulting
matrix is not positive (i.e. has a negative eigenvalue), then $\rho$
was entangled. This criterion is necessary and sufficient for entanglement
in the two-qubit systems addressed in this paper, and sufficient for
higher dimensions.

We apply the PPT criterion in the Bloch matrix picture. First we note
that taking the transpose of the extended Pauli matrices leaves $\sigma_{0},\sigma_{1},\sigma_{3}$
unchanged, and flips the sign of $\sigma_{2}$. Therefore, transposing
a single-qubit density matrix is equivalent to flipping the sign of
the second entry of the Bloch vector in \prettyref{eq:bloch2d}. That
is, the Bloch vector transforms as $\vec{r}\rightarrow Q_{t}\vec{r}$,
where 
\begin{equation}
Q_{t}=\left[\begin{array}{rrr}
1 & 0 & 0\\
0 & -1 & 0\\
0 & 0 & \hphantom{-}1
\end{array}\right].\label{eq:QtDef}
\end{equation}

Based on this, the partial transpose of the quantum state $\rho$
with Bloch matrix components $\vec{u},\vec{v},R$ is equivalent to
the transformations 
\begin{align}
\vec{u} & \rightarrow Q_{t}\vec{u},\text{ and }R\rightarrow Q_{t}R,\text{ or}\nonumber \\
\vec{v} & \rightarrow Q_{t}\vec{v},\text{ and }R\rightarrow RQ_{t}^{\dagger},\label{eq:PPTonBloch}
\end{align}
where the transformations in the first (second) line signify a transpose
of the first (second) subsystem. In terms of their effects on the
positivity inequalities \prettyref{eq:bounds}, the preceding transformations
only reverse the signs of the $\det R$ and $\vec{u}{}^{\dagger}\tilde{R}\vec{v}$
terms. 

To interpret this result, it is more instructive to examine the partial
transpose operation in the SVD picture. Following the example of local
unitary transformations in \prettyref{eq:localUnitaryonSVD}, it is
easy to show that the partial transpose transformations in \prettyref{eq:PPTonBloch}
are equivalent to
\begin{align}
M & \rightarrow Q_{t}M,\text{ or}\nonumber \\
N & \rightarrow Q_{t}N,\label{eq:PPTonSVD}
\end{align}
with $\Sigma,\vec{g},\vec{h},$ left unchanged. The effect of either
of the above is to flip the orientation $d\rightarrow\det Q_{t}d=-d$. 

The next step is to examine whether the partially transposed state
violates positivity. Given the above, the only change to the positivity
inequalities \prettyref{eq:svbounds} by the partial transpose operation
is to flip the sign of $d$. If after the orientation $d$ is reversed,
all the positivity inequalities remain satisfied, then the initial
state was separable, otherwise it was entangled. 

Therefore, \emph{the only meaningful effect of the partial transpose
is to flip the sign of the orientation $d$}. This may be alternatively
achieved if $Q_{t}$ is replaced by any orthogonal matrix $Q_{-}$
such that $\det Q_{-}=-1$. Given some quantum state, we call the
quantum state with identical $\Sigma,\vec{g},\vec{h},$ but the reverse
orientation $d$ its \emph{conjugate }state\emph{. Testing the entanglement
of a quantum state is equivalent to testing the positivity of its
conjugate state.}

The partial transpose may be replaced by other testing operations
with the same effect. For example, a partial anti-diagonal transpose,
corresponding to $Q_{-}=\diag(1,1,-1)$, would work just as well.
Every choice of $Q_{-}$ corresponds to a new criterion. Though the
standard PPT criterion and its anti-diagonal version are the simplest
to apply to $\rho$, others may possibly be more convenient under
some assumptions. 

If we think of our qubits as spin-$\frac{1}{2}$ systems, the axes
in the Bloch sphere correspond to three spatial dimensions. In this
case, the entanglement criterion corresponds to applying the rotoreflection
$Q_{-}$ to one subsystem's spin and testing the physicality of the
result, similar to the mirror quantum theory of Ref. \citep{Hoehn2014}.
Separable states may be interpreted as ones whose spin mirror image
in one subsystem are physical, while for entangled states the single
subsystem spin mirror images are unphysical. This makes sense when
one recalls the spin of a member of an entangled pair is not simply
an isolated vector in space, but rather a spatial distribution of
correlations.

Reflecting the spin can be thought of as a combination of spatial
parity (P) inversion and time (T) inversion, common in quantum field
theory \citep{Peskin1995}. However, it is important to note that
this PT inversion is applied to a single subsystem of the two, not
the combined state as is usually the case. Reflections become more
difficult to intuit if our qubits are not spin-$\frac{1}{2}$ systems,
but for example, two-level atoms where the Bloch vectors don't correspond
to spatial directions. In this case, reflections are simply taken
abstractly over the Bloch vector space.

Applying an entangling unitary operation will leave the invariants
in \prettyref{eq:blochInvariants} unchanged, but the individual terms
in the positivity equation will change such that entanglement criteria
are satisfied. For example, an entangling unitary transformation will
change the two quantities $\vec{u}{}^{\dagger}R\vec{v}=\vec{g}{}^{\dagger}\Sigma\vec{h}$
and $\det R=d\det\Sigma$ by the same amount such that their difference,
the invariant $B_{2}$, remains unchanged. However, the change may
be such that reversing the sign of the second expression will lead
to a violation of the positivity inequalities, and hence the transformed
state is entangled.

Quantum states where $R$ (or $\Sigma$) is of rank $1$ or $0$ cannot
be entangled, since the two terms whose sign is flipped will be zero,
and the satisfied positivity inequalities remain unchanged. Even rank
$2$ states where $\vec{u}{}^{\dagger}\tilde{R}\vec{v}=d\vec{g}{}^{\dagger}\tilde{\Sigma}\vec{h}=0$
cannot be entangled.

For a maximally entangled Bell state, the values of the left hand
sides of the positivity inequalities \prettyref{eq:svbounds}, after
the reversal of the orientation, are $0,-4,-16$ respectively. These
are ``the most negative'' values these quantities can attain for
any quantum state. The first has no $d$ dependence and of course
is never negative for any state. It is also quite common for only
the third quantity to be negative for an entangled state (e.g. $\vec{u}=\vec{v}=0,\,R=-0.4I$).
Although it remains to be rigorously verified, there do not seem to
be physical quantum states where the second is negative but the third
is not. One may therefore consider the degree of negativity of the
left hand side of the third inequality \prettyref{eq:svbound3}, after
orientation reversal $d\rightarrow-d$, as a possible candidate for
degree of entanglement. 

There remains the important question of whether unphysicality under
reflection is fundamental to entanglement, or just an artifact of
the two-qubit system. In bipartite systems larger than a qubit-qutrit
pair, the PPT criterion is sufficient but not necessary. For such
systems, a subsystem's Bloch vector space may have eight or more dimensions.
Perhaps more feasibly, one can also speculate about multipartite entanglement
between $n$ qubits. The Bloch matrix will then become a tensor with
$4^{n}$ entries. Unfortunately, useful analysis will be complicated
by the lack of a simple singular value decomposition in higher dimensions
\citep{Tucker1966,Kolda2009a}. Despite this, one may hypothesize
multiple generalized orientation parameters $d$ in higher dimensions.
If they exist, perhaps inverting them will provide workable entanglement
criteria.

\begin{comment}
Can I derive PH criterion in this picture easily? or explain it? this
would give edge to the formalism.

like improper rotation, generalized reflection, e.g. Parity inversion
(CPT), is this fundamental on entanglement level? No, because P is
flipping both subsystems, this flips only one.

-rename conjugate states?
\end{comment}

\section{\label{sec:Special-Classes-of-States}Special Classes of States}

\subsection{\label{subsec:Maximally-Entangled-states}Maximally Entangled states}

In this section we find the Bloch matrix description for some important
classes of states. We begin by characterizing maximally entangled
states, a class that includes Bell states. A maximally entangled state
may be defined as being (i) pure, and (ii) locally maximally mixed
(LMM), i.e. once a partial trace eliminates one subsystem, the other
is left in a maximally mixed state. 

A pure density matrix $\rho$ has a single nonzero eigenvalue, equal
to unity. From the derivation of the in Sec. \ref{sec:The-Positivity-Condition},
it is clear that achieving purity is equivalent to all three positivity
inequalities achieving equality. The LMM condition is equivalent to
both local Bloch vectors being zero. Setting $\vec{g}=\vec{h}=0$
and equality in \prettyref{eq:svbounds}, we have
\begin{gather}
\|\Sigma\|^{2}=x_{1}^{2}+x_{2}^{2}+x_{3}^{2}=3,\nonumber \\
d\det\Sigma=dx_{1}x_{2}x_{3}=-1,\nonumber \\
\|\tilde{\Sigma}\|^{2}=x_{2}^{2}x_{3}^{2}+x_{3}^{2}x_{1}^{2}+x_{1}^{2}x_{2}^{2}=3.\label{eq:MaxEntangledCondition}
\end{gather}

Given that $x_{i}\ge0$, the only solution to the above is $d=-1$
and $x_{1}=x_{2}=x_{3}=1$. That is, $\Sigma=I$. Therefore
\begin{equation}
R=MN^{\dagger}\equiv O_{-},\label{eq:MaxEntangledRCondition}
\end{equation}
where $O_{-}$ is an orthogonal matrix with $\det O_{-}=d=-1$.

Maximally entangled states are characterized as those whose local
Bloch vectors $\vec{u},\vec{v}$ are zero, and whose correlation matrix
$R$ is orthogonal with determinant $-1$, conditions clearly satisfied
by the Bell states \prettyref{eq:bellBlochMatrices}.

The uniqueness of $\Sigma,\vec{g},\vec{h},$ and $d$ in the above
solution implies that\emph{ there exists a single maximally entangled
state, unique up to local unitary transformations}.

\subsection{\label{subsec:Pure-States}Pure states}

As mentioned above, requiring that the positivity inequalities \prettyref{eq:svbounds}
achieve equality suffices to characterize pure states. However, solving
the resulting equalities is in general algebraically involved. It
is easier to note that any pure state can be reached from another
by the action of an arbitrary unitary transformation, as the latter
do not affect purity.

There are $7$ degrees of freedom in bipartite pure states: $2$ for
each of the $4$ complex coefficients, less one for an irrelevant
global phase. Local unitary operations create $6$ of the $7$, and
so we expect basic nonlocal unitary operation $\mathring{U}$ to effect
the remaining degree of freedom. Since $\Sigma,\vec{g},\vec{h},d$
are invariant under local unitaries, we can start with their values
for a known pure state and then apply $\mathring{U}$, expecting it
to generate the final degree of freedom on these quantities.

Therefore we start with the pure state $\vec{g}=\vec{h}=\vec{u}=\vec{v}=(1,0,0),$
and $\Sigma=R=\diag(1,0,0)$. Applying $\mathring{U}\big(\theta_{1},\theta_{2},\theta_{3}\big)$,
whose effect on the Bloch components is shown in \prettyref{eq:UBasicEffect},
to this state:
\begin{align*}
\vec{u}^{\prime}=\vec{v}^{\prime} & =(\cos\theta,0,0),\\
R^{\prime} & =\begin{bmatrix}1 & 0 & 0\\
0 & 0 & \sin\theta\\
0 & \sin\theta & 0
\end{bmatrix},
\end{align*}
where $\theta=\theta_{2}-\theta_{3}$. As expected, the resulting
pure state has a single degree of freedom defined by $\theta$. The
above correlation matrix has the singular value decomposition $R^{\prime}=M^{\prime}\Sigma^{\prime}N^{\prime\dagger}$
where $M^{\prime}=\left[\begin{smallmatrix}1 & 0 & 0\\
0 & 0 & 1\\
0 & 1 & 0
\end{smallmatrix}\right]$, $N^{\prime}=I$ and $\Sigma^{\prime}=\diag(1,\sin\theta,\sin\theta)$.
We also find $d=\det(M^{\prime})\det(N^{\prime})=-1$, $\vec{g}^{\prime}=M^{\prime\dagger}\vec{u}^{\prime}=\vec{u}^{\prime}$,
and $\vec{h}^{\prime}=N^{\prime\dagger}\vec{v}^{\prime}=\vec{v}^{\prime}$.

Therefore pure states are characterized by
\begin{align}
\vec{g} & =\vec{h}=(\cos\theta,0,0),\nonumber \\
\Sigma & =\diag(1,\sin\theta,\sin\theta),\nonumber \\
d & =-1,\label{eq:PureStateSVDDef}
\end{align}
for some arbitrary $\theta$, up to an identical reordering of the
entries in $\Sigma,\vec{g},\vec{h}$. The single nonlocal degree of
freedom in \prettyref{eq:PureStateSVDDef} along with $6$ local ones
in the choice of $M,N$ (so long as they satisfy $d=-1$) make up
the $7$ degrees of freedom in pure states.

One can substitute \prettyref{eq:PureStateSVDDef} into the positivity
inequalities \prettyref{eq:svbounds} and verify they all satisfy
equality. The pure state defined in \prettyref{eq:PureStateSVDDef}
is a product state if $\theta=0$ (the state we started with) and
maximally entangled when $\theta=\frac{\pi}{2}$. Hence, the quantity
$\sin\theta$ may be taken as measure of entanglement for pure states.

\subsection{\label{subsec:Generalized-Isotropic-States}Generalized isotropic
states}

Werner states are defined as invariant under local unitary transformations
of the form $U\otimes U$ \citep{Werner1989}. In two-qubit systems,
their density matrix takes the well known form
\begin{equation}
\rho_{wer}(z)=\frac{1-z}{4}I+z\ket{\Psi^{-}}\bra{\Psi^{-}},\label{eq:wernerState}
\end{equation}
where $z$ is a scalar parameter. Similarly, isotropic state are defined
as invariant under local unitary transformations of the form $U\otimes U^{*}$,
with density matrix of the form
\begin{equation}
\rho_{iso}(z)=\frac{1-z}{4}I+z\ket{\Phi^{+}}\bra{\Phi^{+}}.\label{eq:isotropicState}
\end{equation}

It is known that both Werner and isotropic states are physical for
$-\frac{1}{3}\le z\le1$ and entangled for $\frac{1}{3}\le z$. More
inclusively, we\emph{ }define\emph{ generalized isotropic states}
as those invariant under local unitary transformations of the form
$U_{\hat{a},\alpha}\otimes U_{\hat{b},\beta}$, where the unitary
transforms are defined in \prettyref{eq:unitaryDef}, $\hat{a},\alpha$
vary freely, and $\hat{b},\beta$ are assumed to be one-to-one functions
of $\hat{a},\alpha$. The Bloch matrix components $\vec{u},\vec{v},R$,
of the invariant state should satisfy
\begin{align}
\vec{u}^{\prime} & =Q(\hat{a},\alpha)\vec{u}=\vec{u},\nonumber \\
\vec{v}^{\prime} & =Q(\hat{b},\beta)\vec{v}=\vec{v},\nonumber \\
R^{\prime} & =Q(\hat{a},\alpha)RQ^{\dagger}(\hat{b},\beta)=R,\label{eq:GenIsotropicTransform}
\end{align}
and $Q$ are rotations with the specified parameters. 

The above should hold for all $\hat{a},\alpha$, and all $\hat{b},\beta$,
with some relationship to be found between the two pairs. Therefore
$\vec{u}=\vec{v}=0$, as the zero vector is the only one invariant
under all rotations. Further, $R$ then satisfies
\begin{align*}
Q(\hat{a},\alpha)RR^{\dagger}Q^{\dagger}(\hat{a},\alpha) & =RR^{\dagger},\\
Q(\hat{b},\beta)R^{\dagger}RQ^{\dagger}(\hat{b},\beta) & =R^{\dagger}R.
\end{align*}

The two equalities above mean that $RR^{\dagger}$ and $R^{\dagger}R$
are invariant under any orthogonal change of basis. The only such
matrices are proportional to the identity. %
\begin{comment}
To see this, note that $RR^{\dagger}$ and $R^{\dagger}R$ are symmetric
and hence diagonalizable. Since they are invariant under orthogonal
transformations that take one eigenvector to another, and hence switch
positions of eigenvalues in the diagonalization, all the eigenvalues
must be equivalent.
\end{comment}
{} Given that $RR^{\dagger}$ and $R^{\dagger}R$ are positive with
the same magnitude, they both have the same positive proportionality
constant. Hence we can write
\begin{equation}
RR^{\dagger}=R^{\dagger}R=z^{2}I,\label{eq:RRdag=00003DzI}
\end{equation}
where $z$ is some real scalar. This implies that $R=zO$ for some
orthogonal $O$. Substituting this in the last equality in \prettyref{eq:GenIsotropicTransform}
and rearranging, we have
\[
Q(\hat{a},\alpha)=OQ(\hat{b},\beta)O^{\dagger}.
\]
Making use of the explicit expression for a rotation in \prettyref{eq:unitaryRotation2},
the last equation reduces to
\begin{multline}
\cos\alpha I+(1-\cos\alpha)\hat{a}\hat{a}^{\dagger}+\sin\alpha\lfloor\hat{a}\rfloor_{\times}\\
=\cos\beta I+(1-\cos\beta)O\hat{b}(O\hat{b})^{\dagger}+\sin\alpha O\lfloor\hat{b}\rfloor_{\times}O^{\dagger}.\label{eq:RotationsOrthogonallyEqual}
\end{multline}
Taking the trace of both sides, $\cos\alpha=\cos\beta\Rightarrow\alpha=\pm\beta.$
Without loss of generality, set $\alpha=\beta$. Then \prettyref{eq:RotationsOrthogonallyEqual}
implies
\begin{align}
\hat{a}\hat{a}^{\dagger} & =O\hat{b}(O\hat{b})^{\dagger},\label{eq:RotationsOrthogonallyEqual2}\\
\lfloor\hat{a}\rfloor_{\times} & =O\lfloor\hat{b}\rfloor_{\times}O^{\dagger}.\label{eq:RotationsOrthogonallyEqual3}
\end{align}

Multiplying \prettyref{eq:RotationsOrthogonallyEqual2} by $\hat{a}$
from both sides yields $(\hat{a}^{\dagger}O\hat{b})^{2}=1.$ Noting
that the indices of the cross product matrix satisfy $\big(\lfloor\hat{a}\rfloor_{\times}\big)_{ij}=\varepsilon_{jik}a_{k}$,
\prettyref{eq:RotationsOrthogonallyEqual3} then implies
\begin{align}
\varepsilon_{jik}a_{k} & =O_{im}\varepsilon_{nml}b_{l}O_{jn}\nonumber \\
 & =\varepsilon_{jik}\det(O)O_{kl}b_{l},\label{eq:CrossProdMatrixEq}
\end{align}
where we have used \prettyref{eq:orthogonalDetID} with $O^{\dagger}$
in the place of $O$. Canceling the Levi-Civita factor, \prettyref{eq:CrossProdMatrixEq}
is equivalent to $\hat{a}=(\det O)O\hat{b}$. More symmetrically,
\begin{equation}
\hat{a}^{\dagger}O\hat{b}=\det O.\label{eq:GenIsotropicabRelation}
\end{equation}

Therefore, a generalized isotropic state is defined as invariant under
$U_{\hat{a},\alpha}\otimes U_{\hat{b},\alpha}$ for any angle $\alpha,$
and any $\hat{a},\hat{b}$ satisfying \prettyref{eq:GenIsotropicabRelation}
for some fixed orthogonal $O$. Explicitly, the state has Bloch matrix
components
\begin{equation}
\vec{u}=\vec{v}=0,\,R=zO.\label{eq:GenIsotropicBlochMatrix}
\end{equation}
Without loss of generality, choose $\det O=-1$, the sign of $z$
offsetting our choice. Finally, we find the range of $z$ for which
the general isotropic state is positive or entangled. Note that \prettyref{eq:cofactorDet}
implies the cofactor matrix is $\tilde{R}=-z^{2}O$. Substituting
\prettyref{eq:GenIsotropicBlochMatrix} into the positivity equations
\prettyref{eq:bounds},
\begin{gather}
1-z^{2}\ge0,\nonumber \\
2z^{3}-3z^{2}+1\ge0,\nonumber \\
-3z^{4}+8z^{3}-6z^{2}+1\ge0.\label{eq:zInequalities}
\end{gather}
The polynomials on the left hand side appear in Ref. \citep{Byrd2003},
though apply more generally here. The first inequality simplifies
to $-1\le z\le1$ . The last two factor to 
\begin{gather}
(1-z)^{2}(2z+1)\ge0,\nonumber \\
(1-z)^{3}(3z+1)\ge0.\label{eq:zInequalities2}
\end{gather}
The last inequality is satisfied for 
\begin{equation}
-\frac{1}{3}\le z\le1,\label{eq:zdetORangePositivity}
\end{equation}
which is the range for $z$ common to all three inequalities. 

To check for entanglement, we apply the positivity criterion from
Sec. \ref{sec:Entanglement-Criteria}, which amounts to flipping the
sign of $z$ in the inequalities \prettyref{eq:zInequalities}. Reproducing
the steps above with sign reversal, we conclude that the state is
separable for $-1\le z\le\frac{1}{3}$ and entangled for 
\begin{equation}
\frac{1}{3}<z.\label{eq:zdetORangeEntanglement}
\end{equation}

Section \ref{subsec:Maximally-Entangled-states} showed that an orthogonal
matrix with negative determinant characterizes the correlation matrix
of all maximally entangled states. Therefore, a general isotropic
state takes the form 
\begin{equation}
\rho_{GI}(z)=\frac{1-z}{4}I+z\ket{\Psi}\bra{\Psi},\label{eq:GenIsotropicState}
\end{equation}
where $\ket{\Psi}$ is any maximally entangled state. Given \prettyref{eq:zdetORangePositivity}
and \prettyref{eq:zdetORangeEntanglement}, $\rho_{GI}(z)$ is positive
for $-\frac{1}{3}\le z\le1$, and entangled for $\frac{1}{3}<z$.
This successfully generalizes Werner and isotropic states, reproducing
their parameter ranges.

\section{\label{sec:Geometry-of-State-Space}Geometry of The State Space}

It is instructive to use the results thus far to visualize the quantum
state space. We draw the regions of physically allowable quantum states,
where the positivity inequalities \prettyref{eq:svbounds} hold. As
the latter are functions only of $\Sigma,\vec{g},\vec{h},$ $d$,
each point in our diagrams will represent a family of states equivalent
up to local unitary operations.

There are $9$ continuous degrees of freedom in the aforementioned
variables, we hold constant $6$ and plot the physical regions for
the remaining $3$. We create two types of diagrams, \emph{singular
value diagrams}, with $\vec{g}$ and $\vec{h}$ constant the singular
values $x_{1},x_{2},x_{3}$ along the diagonal of $\Sigma$ varying
on the axes, and \emph{relative Bloch vector diagrams}, with $\Sigma$
and $\vec{h}$ constant the components of $\vec{g}$ varying on the
axes.

In each case, regions are plotted twice; once for each value of the
orientation $d$. Regions with $d=1$ are colored in blue and $d=-1$
in red. As per the entanglement criterion in Sec. \ref{sec:Entanglement-Criteria},
states in the the intersection of the two regions are separable, and
states in one region but not the other are entangled.

It can be shown that all three positivity inequalities are needed,
in the sense that no two among them imply the third, in general. However,
it is the third inequality that determines the surface of the convex
allowable region; while the other two eliminate superfluous disconnected
regions. Since the third positivity inequality contains terms up to
the fourth power, the allowable regions are bordered by a family of
quartic surfaces \citep{Jessop1916}. 

\begin{figure}
\subfloat[\label{fig:svSpace1LMM}$\vec{g}=\vec{h}=(0,0,0)$]{\includegraphics[width=0.47\columnwidth]{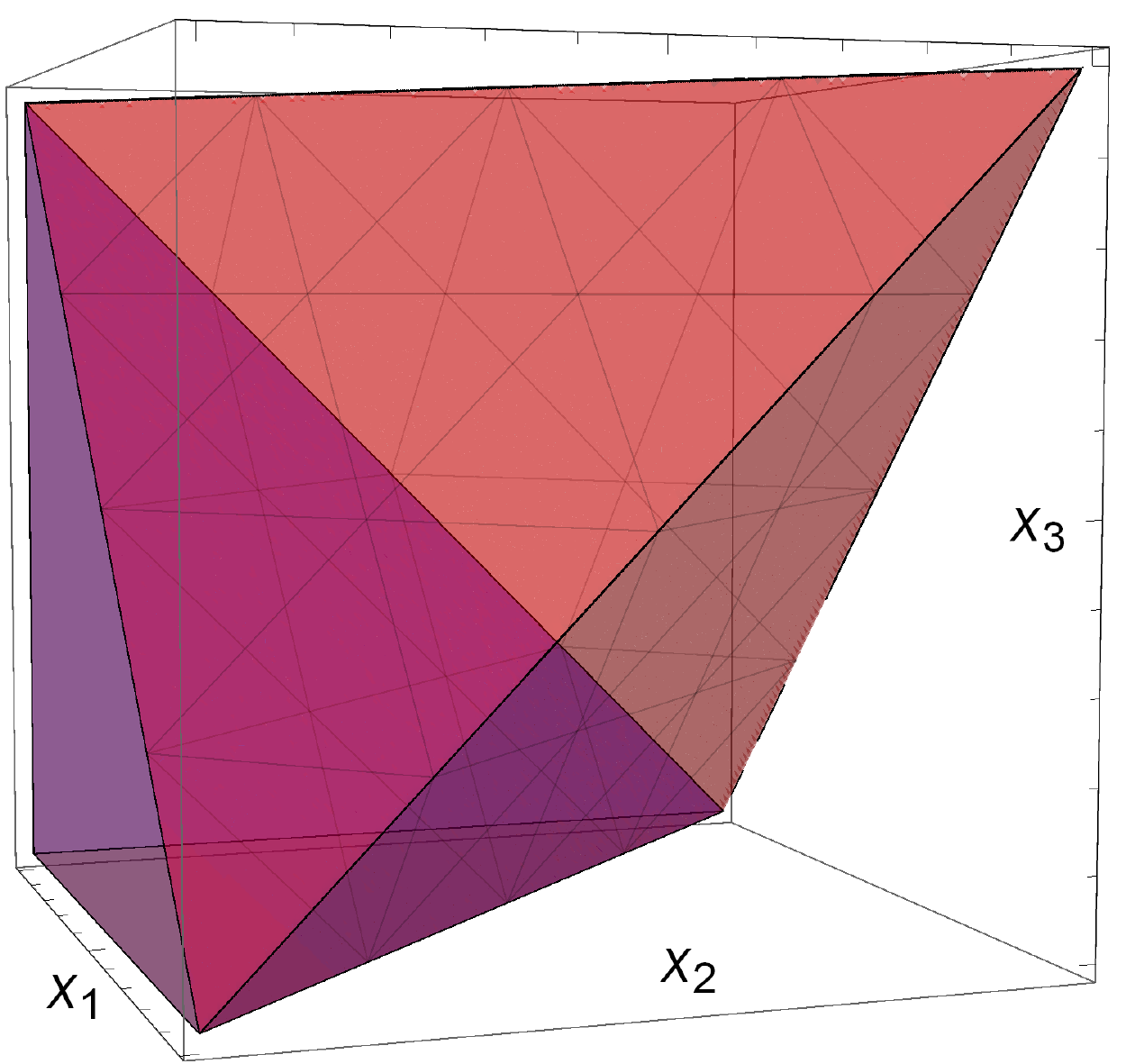}

} \subfloat[\label{fig:svSpace2}$\vec{g}=\vec{h}=(0,0,0.5)$]{\includegraphics[width=0.47\columnwidth]{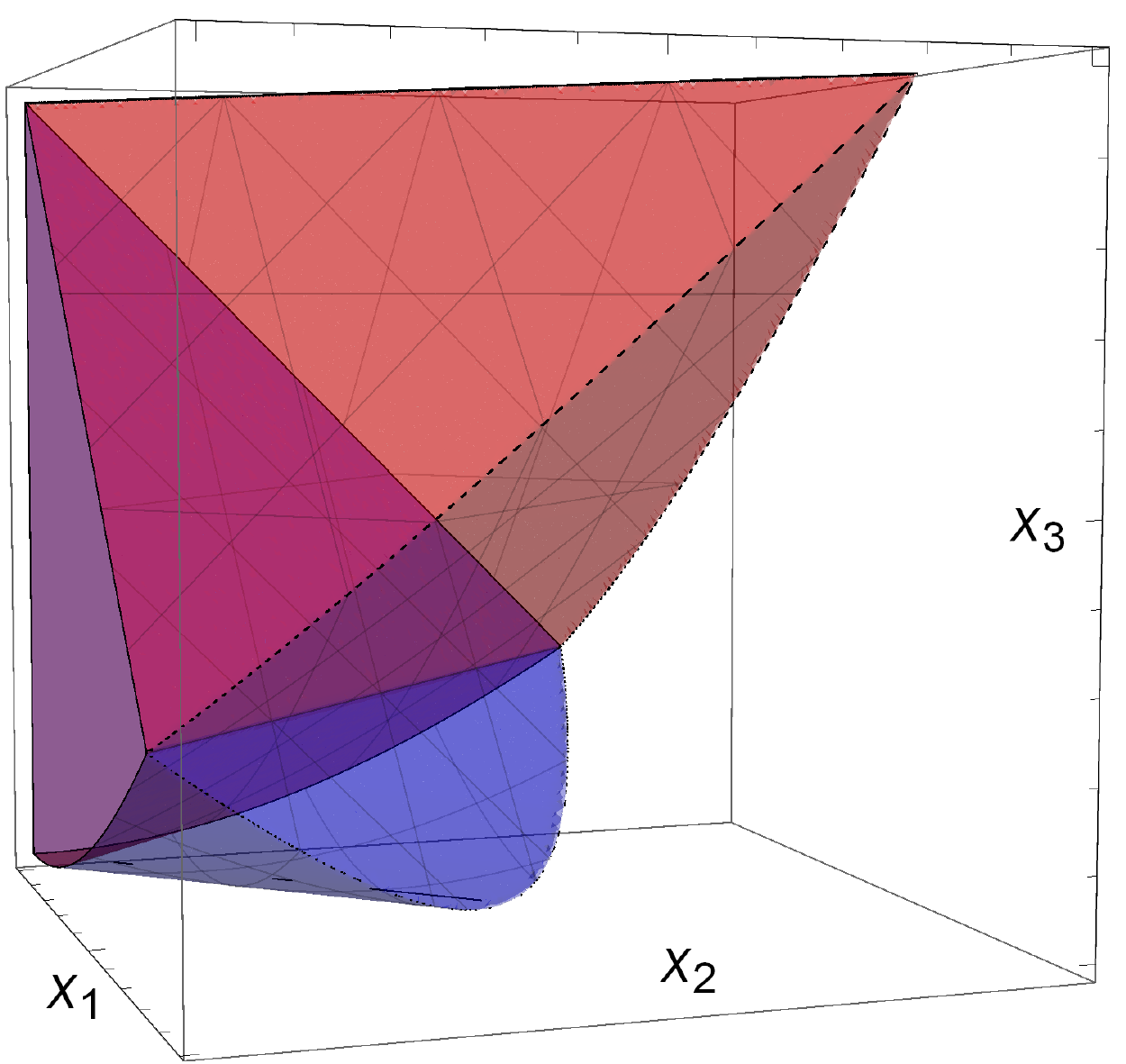}}\hfill{}\subfloat[\label{fig:svSpace3}$\vec{g}=(0,0,0.5)$ \qquad{}$\:\vec{h}=(0,0,0)$]{\includegraphics[width=0.47\columnwidth]{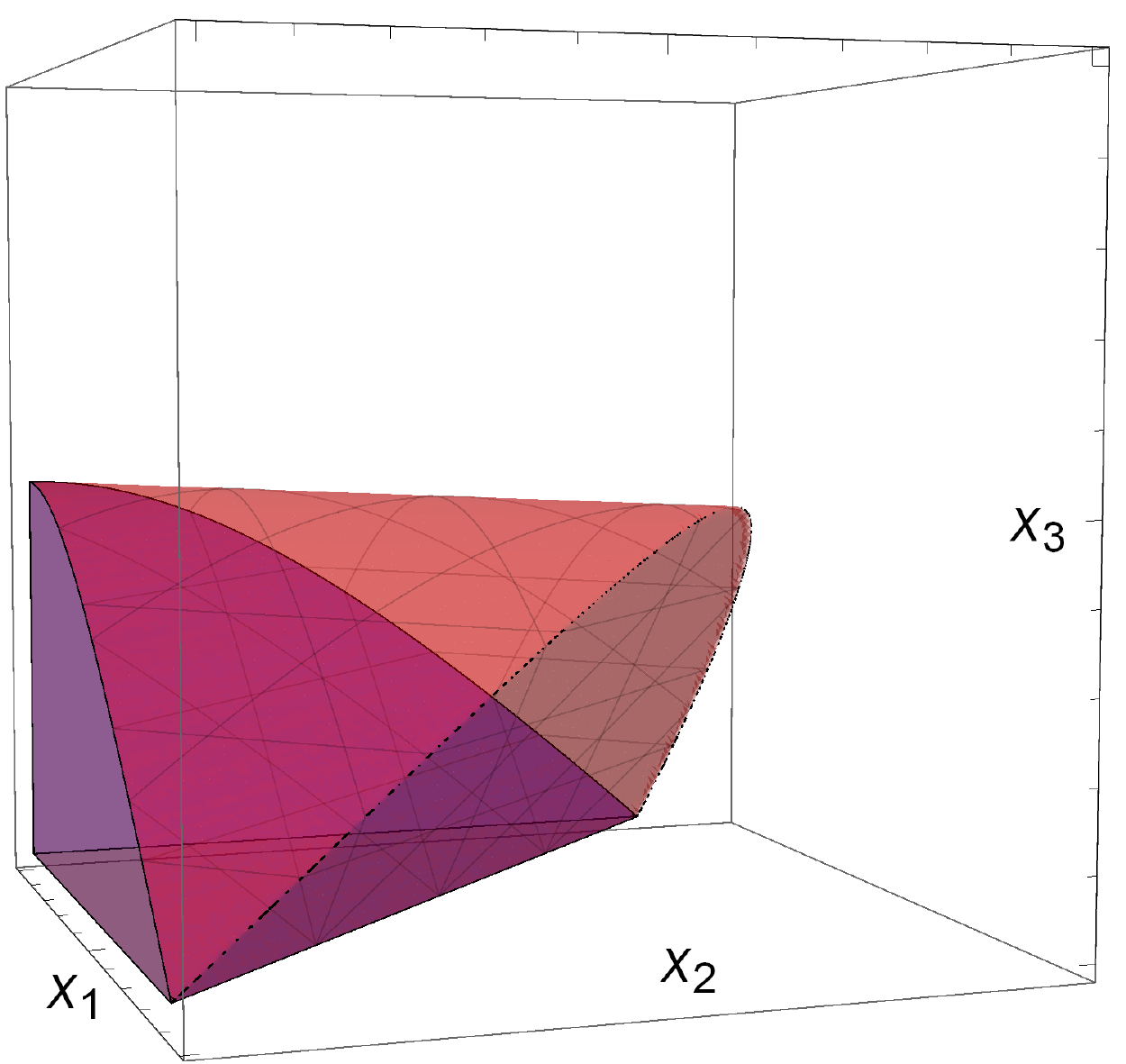}

} \subfloat[\label{fig:svSpace4}$\vec{g}=\vec{h}=(0.3,0.3,0.3)$]{\includegraphics[width=0.47\columnwidth]{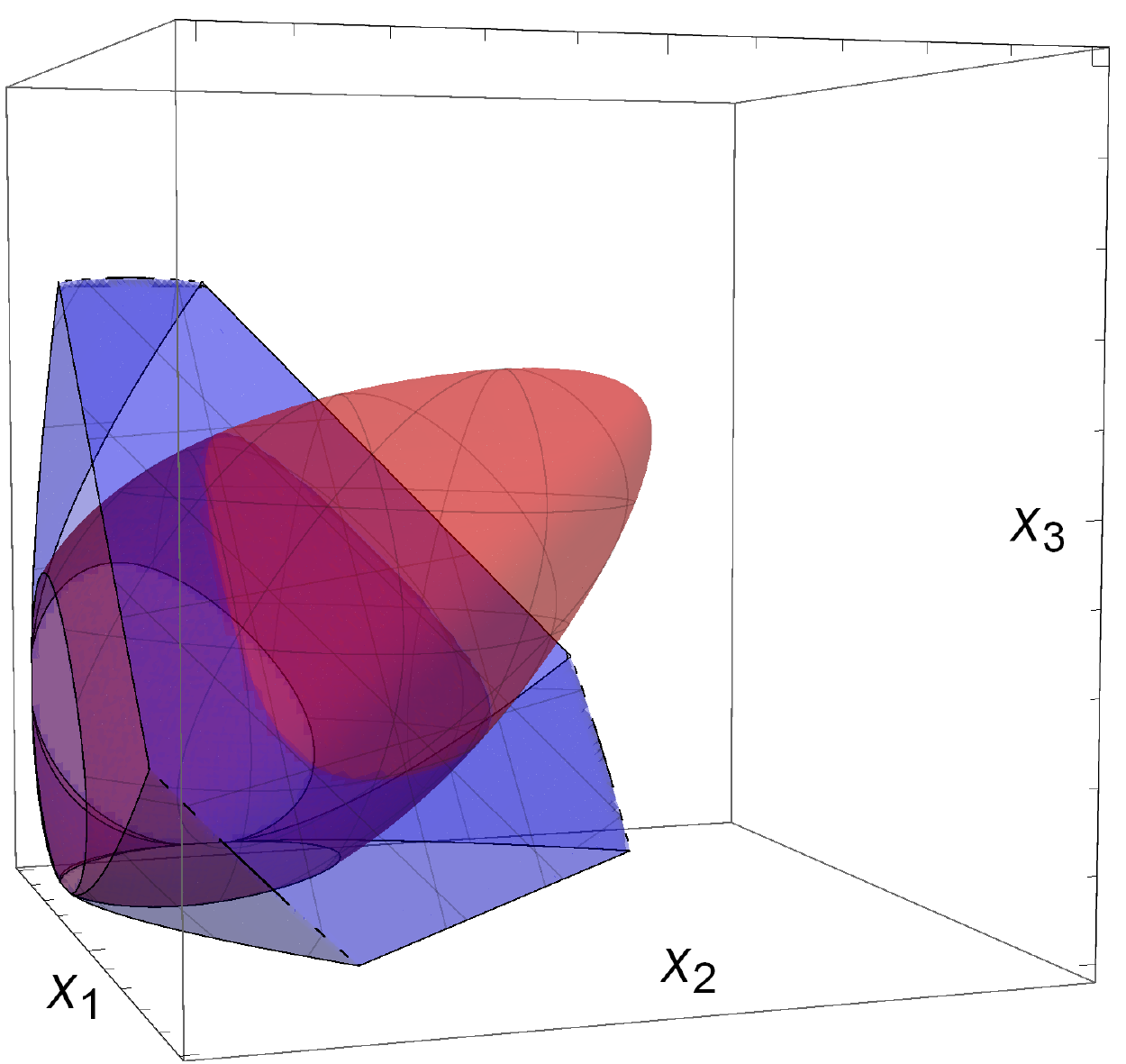}}\hfill{}\subfloat[\label{fig:svSpace5}$\vec{g}=(0.4,0.4,0.4)$\qquad{}$\:\vec{h}=(0.3,0.3,0.3)$]{\includegraphics[width=0.47\columnwidth]{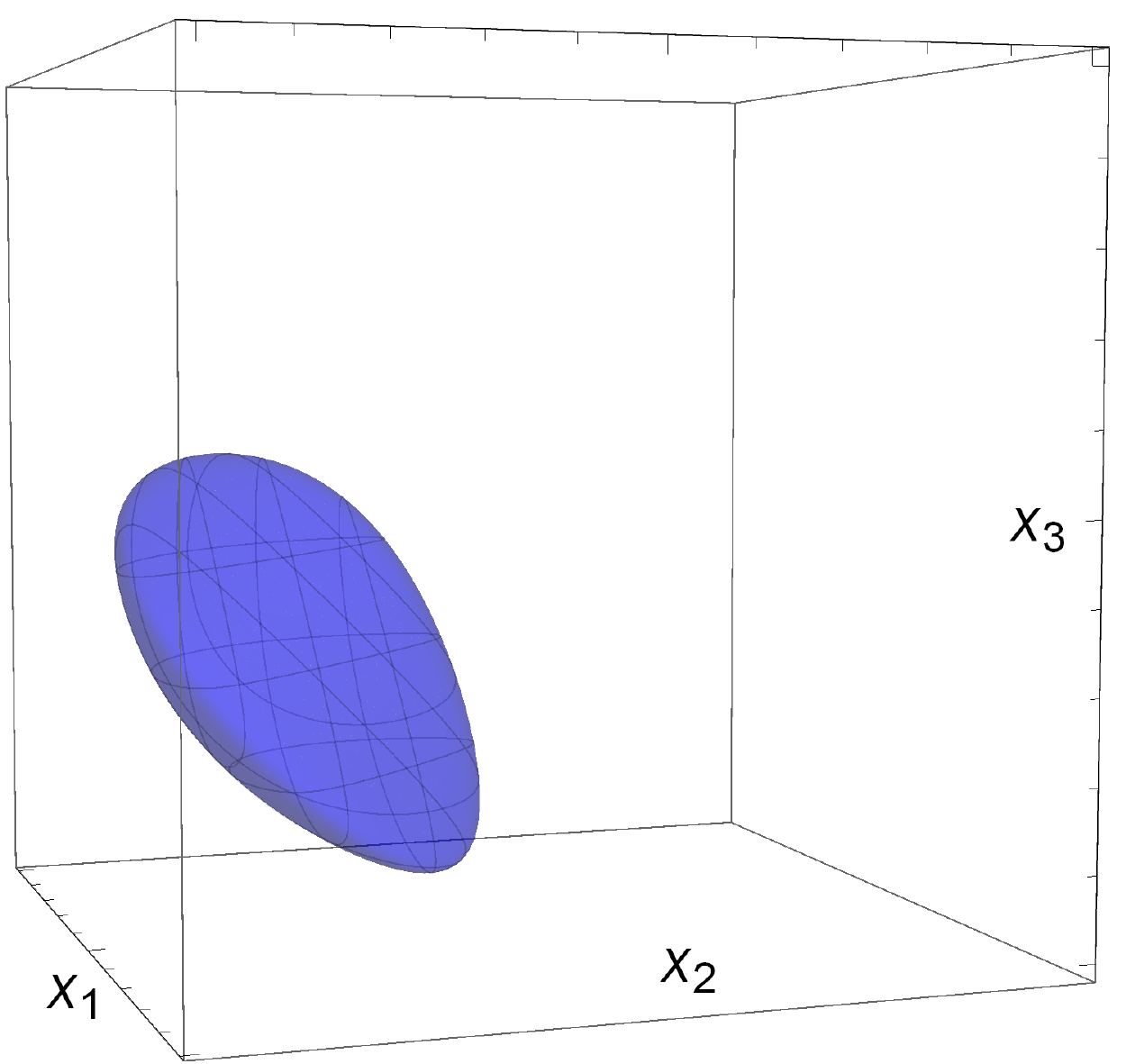}

} \subfloat[\label{fig:svSpace6}$\vec{g}=(-0.4,-0.5,0.2)$\qquad{}$\vec{h}=(-0.33,-0.33,0.1)$]{\includegraphics[width=0.47\columnwidth]{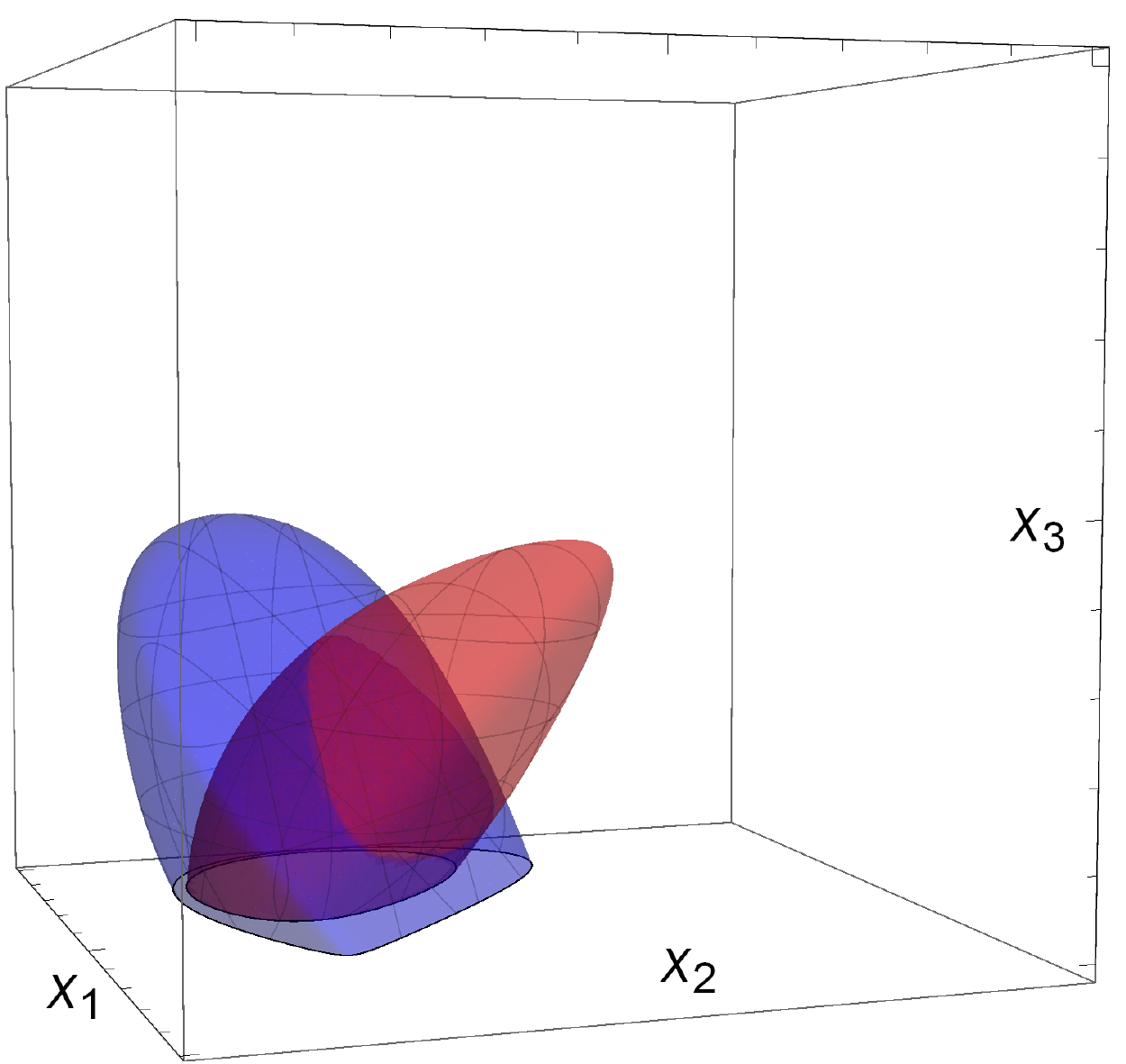}}

\caption{\label{fig:Singular-value-diagrams}Singular value diagrams: The regions
in singular value space $x_{1},x_{2},x_{3}$ where positivity is satisfied,
for fixed values of $\vec{g}$ and $\vec{h}$. Bounded by the unit
cube, with the origin in the rear bottom left. Regions with $d=1$
in blue and $d=-1$ in red.}
\end{figure}

Figure \ref{fig:Singular-value-diagrams} contains the singular value
diagrams for several values of $\vec{g}$ and $\vec{h}$. The most
interesting is Fig. \ref{fig:svSpace1LMM}, where both local Bloch
vectors are zero, i.e. LMM states. In this case, the last positivity
inequality \prettyref{eq:svbound3} factors to
\begin{gather}
(d{-}x_{1}{+}x_{2}{+}x_{3})(d{+}x_{1}{-}x_{2}{+}x_{3})\times\qquad\nonumber \\
\qquad(d{+}x_{1}{+}x_{2}{-}x_{3})(d{-}x_{1}{-}x_{2}{-}x_{3})\ge0,\label{eq:LMMplanesIneq}
\end{gather}
which describes a tetrahedral region bounded by four planes for $d=\pm1$.
If $x_{i}$ were allowed to go negative, the vertices would be $(-d,d,d)$,
$(d,-d,d)$, $(d,d,-d)$, and $(-d,-d,-d)$. This ``large'' tetrahedron
is analogous to the one usually representing linear combinations of
Bell states, with a Bell state at each vertex \citep{Horodecki1996a,Bertlmann2001,Spengler2011}.
One can see this if $d=1$, whence the vertex coordinates are the
diagonals of Bell state correlation matrices in \prettyref{eq:bellBlochMatrices}.

However, since $x_{i}\ge0$, only the octant in Fig. \ref{fig:svSpace1LMM}
is physical. The wedge bounded by points $(1,0,0)$, $(0,1,0)$, $(0,0,1)$,
and $(0,0,0)$ gives the set of separable states (for both values
of $d$). The ``small'' tetrahedron in the figure bounded by points
$(1,0,0)$, $(0,1,0)$, $(0,0,1)$, and $(1,1,1)$ contains entangled
states (with $d=-1$). The origin corresponds to the maximally mixed
state and the point $(1,1,1)$ is the maximally entangled state, unique
up to local operations. This graphical representation is more powerful
than the usual one as all maximally entangled states are included
in a single point. 

The straight line from the origin to $(1,1,1)$ represents the generalized
isotropic states of Sec. \ref{subsec:Generalized-Isotropic-States}.
As expected, $\frac{1}{3}$ of this line lies in the separable region,
and the rest in the entangled. The volume occupied by entangled states
is double that of separable states, so by a natural measure, there
are twice as many entangled as there are separable LMM states.

As the relative Bloch vectors $\vec{g}$ and $\vec{h}$ change, they
continuously deform the blue and red regions as shown in the figures.
Either the blue or red regions may vanish entirely, as is the case
with Fig. \ref{fig:svSpace5}, in which case the states are all entangled.

Given the result in Sec. \ref{subsec:Pure-States}, pure states must
lie along the diagonal of the outer surfaces of the unit cube, and
there is only a single pure state for a suitable choice of $\vec{g}=\vec{h}$.
For LMM states, the pure state is the maximally entangled state. In
Fig. \ref{fig:svSpace2}, the pure state is at the vertex of the red
deformed tetrahedron at $(\frac{\sqrt{3}}{2},\frac{\sqrt{3}}{2},1)$.
Product states must lie on one of the Cartesian axes.

For degenerate choices of $\vec{g}$ and $\vec{h}$, i.e., $g_{i}=g_{j}$
and $h_{i}=h_{j}$, $i\neq j$, local operations may switch the ordering
of $x_{i}$ and $x_{j}$. There is threefold degeneracy in Figs. \ref{fig:svSpace1LMM},
\ref{fig:svSpace4} and \ref{fig:svSpace5}, and twofold degeneracy
in Figs. \ref{fig:svSpace2} and \ref{fig:svSpace3}. One may eliminate
the degeneracy by restricting the singular values to a subset of the
space, e.g. the region $x_{1}\ge x_{2}\ge x_{3}$ for threefold degeneracy.

\begin{figure}[h]
\subfloat[\label{fig:rbvSpace1}$\Sigma=\protect\diag(0,0,0)$\qquad{}$\vec{h}=(0,0,0)$]{\includegraphics[width=0.47\columnwidth]{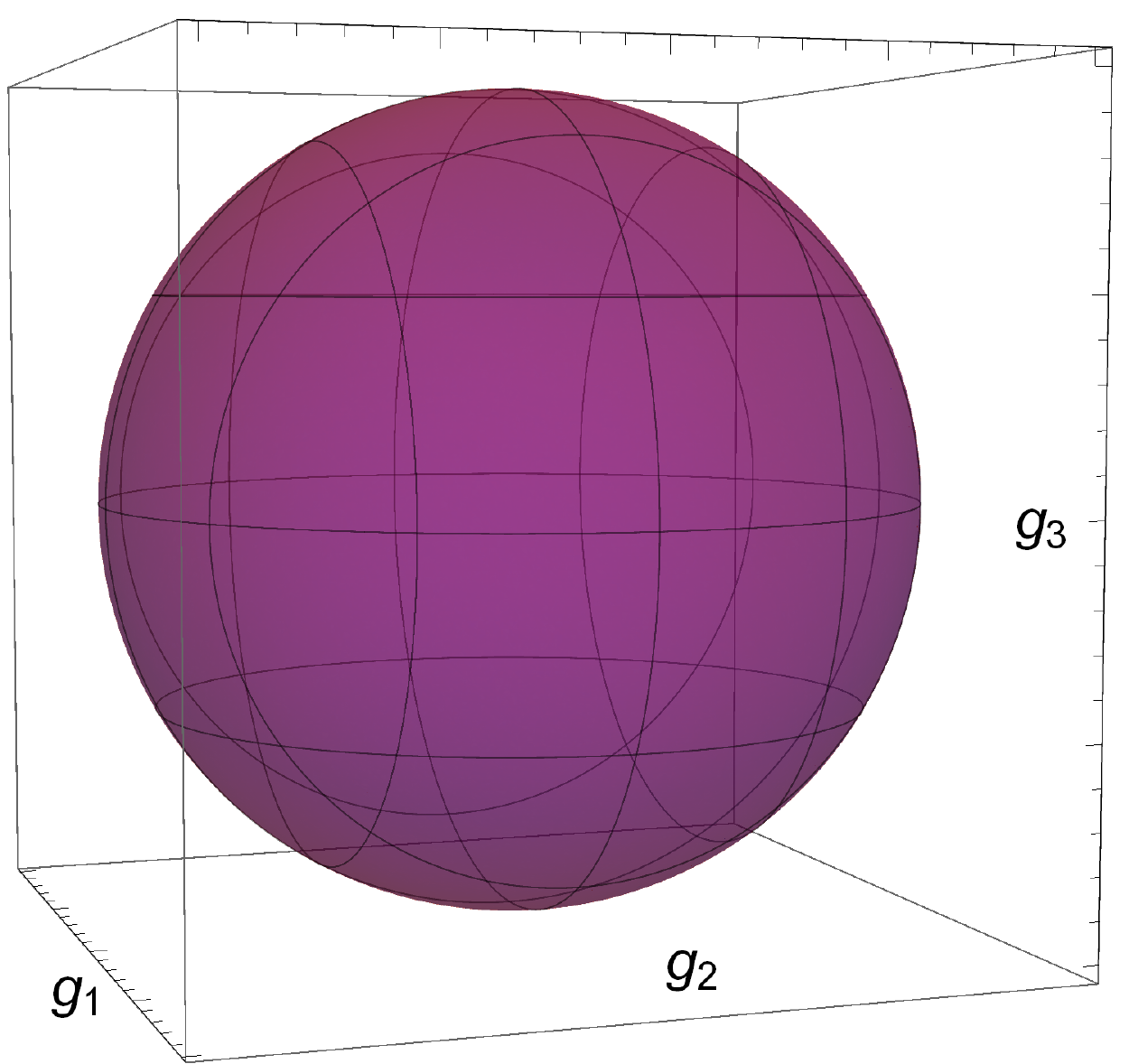}

} \subfloat[\label{fig:rbvSpace2}$\Sigma=\protect\diag(0.3,0.3,0.3)$\qquad{}$\vec{h}=(0,0,0)$]{\includegraphics[width=0.47\columnwidth]{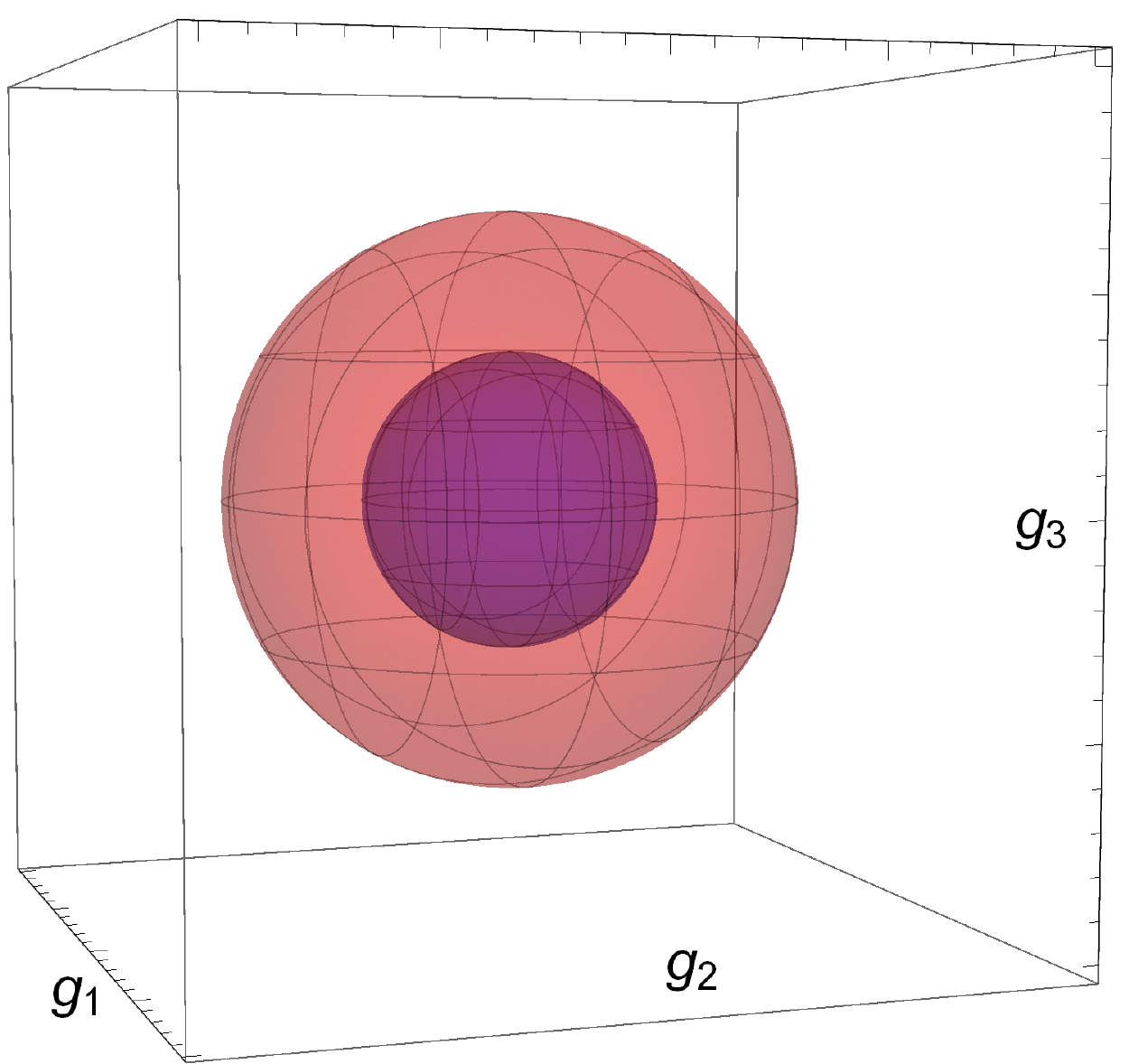}}\hfill{}\subfloat[\label{fig:rbvSpace3}$\Sigma=\protect\diag(0.5,0.5,0.3)$\qquad{}$\vec{h}=(0,0,0)$]{\includegraphics[width=0.47\columnwidth]{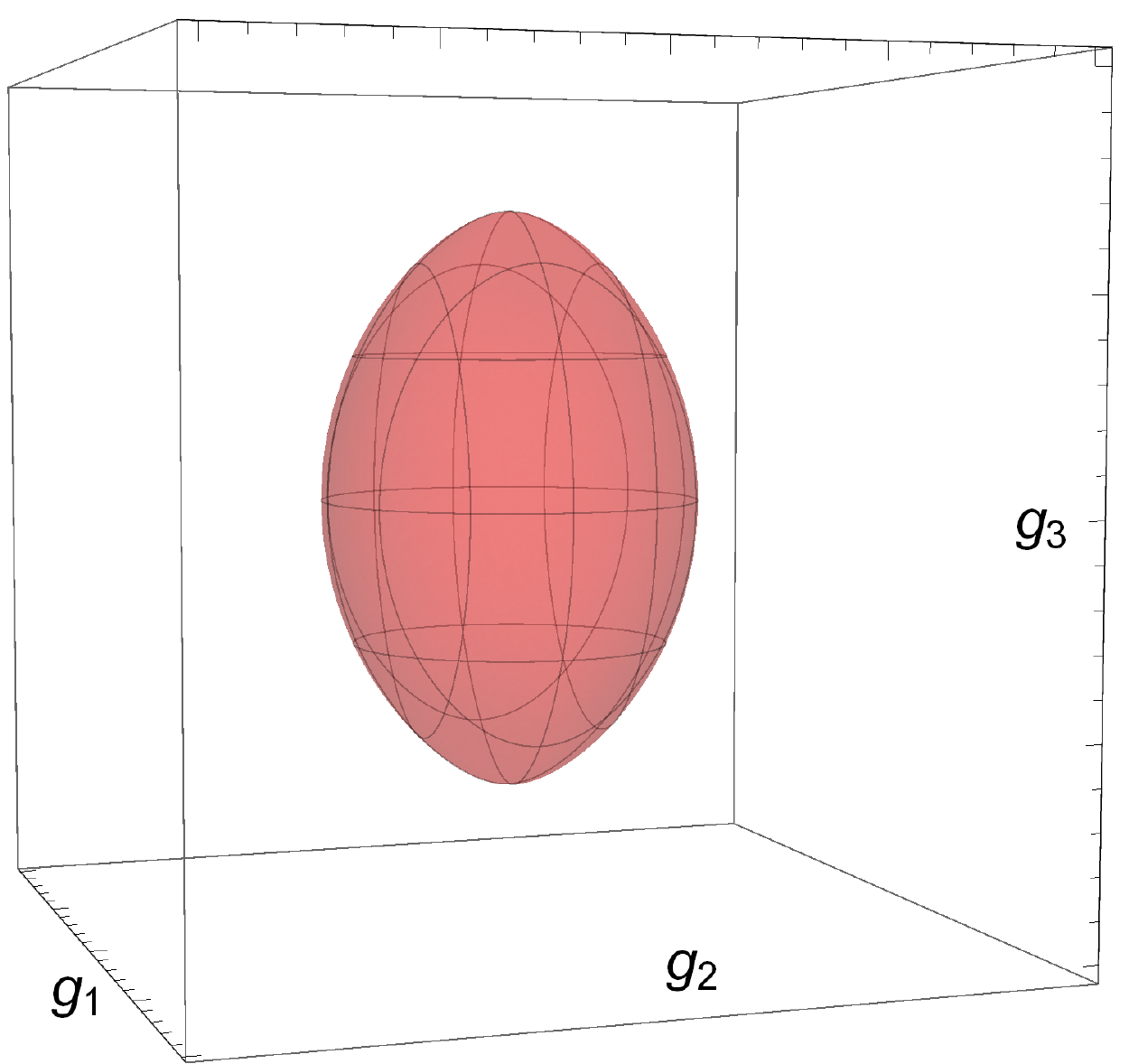}

} \subfloat[\label{fig:rbvSpace4}$\Sigma=\protect\diag(0.3,0,0.3)$\qquad{}$\vec{h}=(0,0.3,0)$]{\includegraphics[width=0.47\columnwidth]{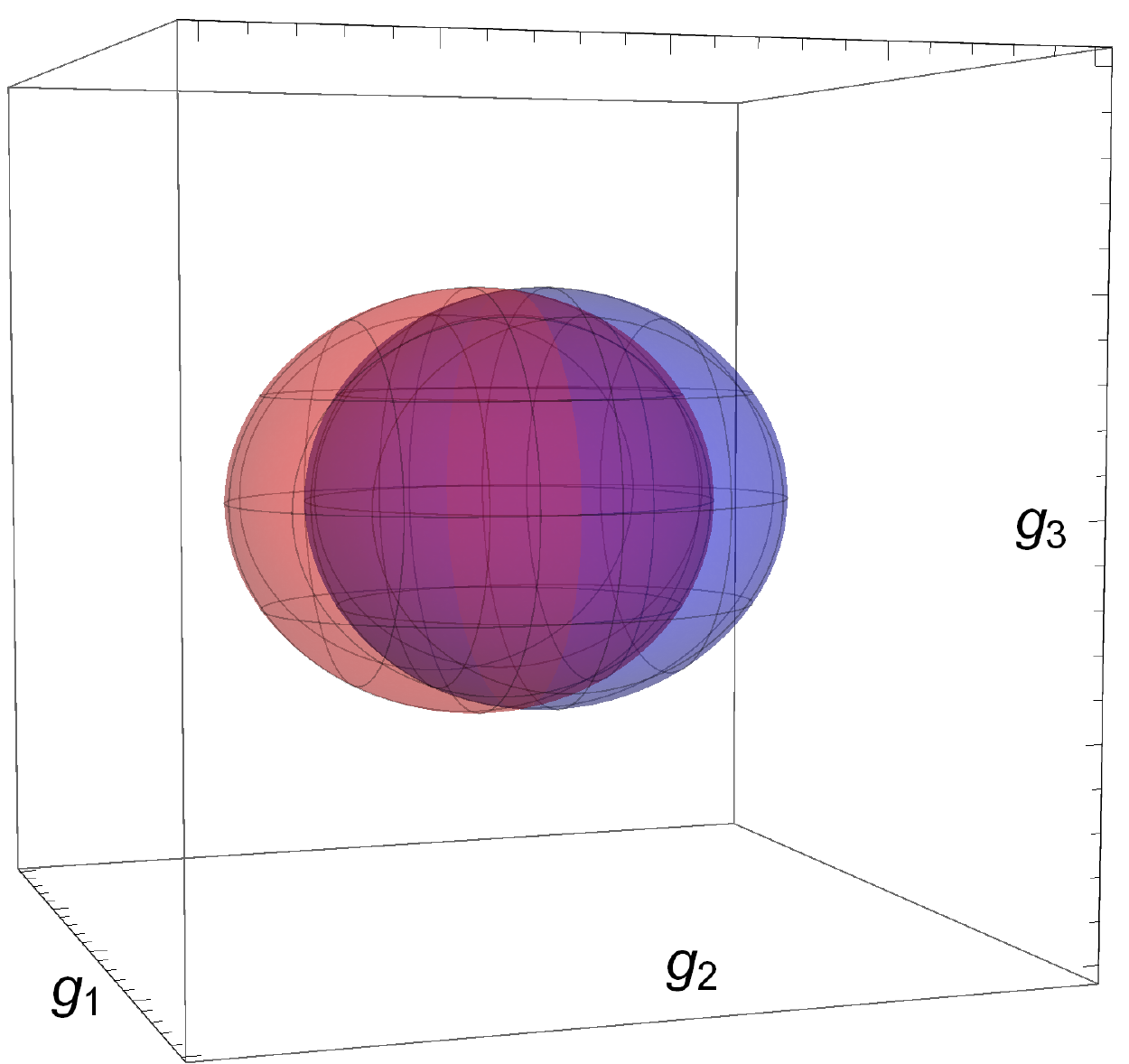}}\hfill{}\subfloat[\label{fig:rbvSpace5}$\Sigma=\protect\diag(0.4,0.3,0.2)$\qquad{}$\vec{h}=(0,0,0.5)$]{\includegraphics[width=0.47\columnwidth]{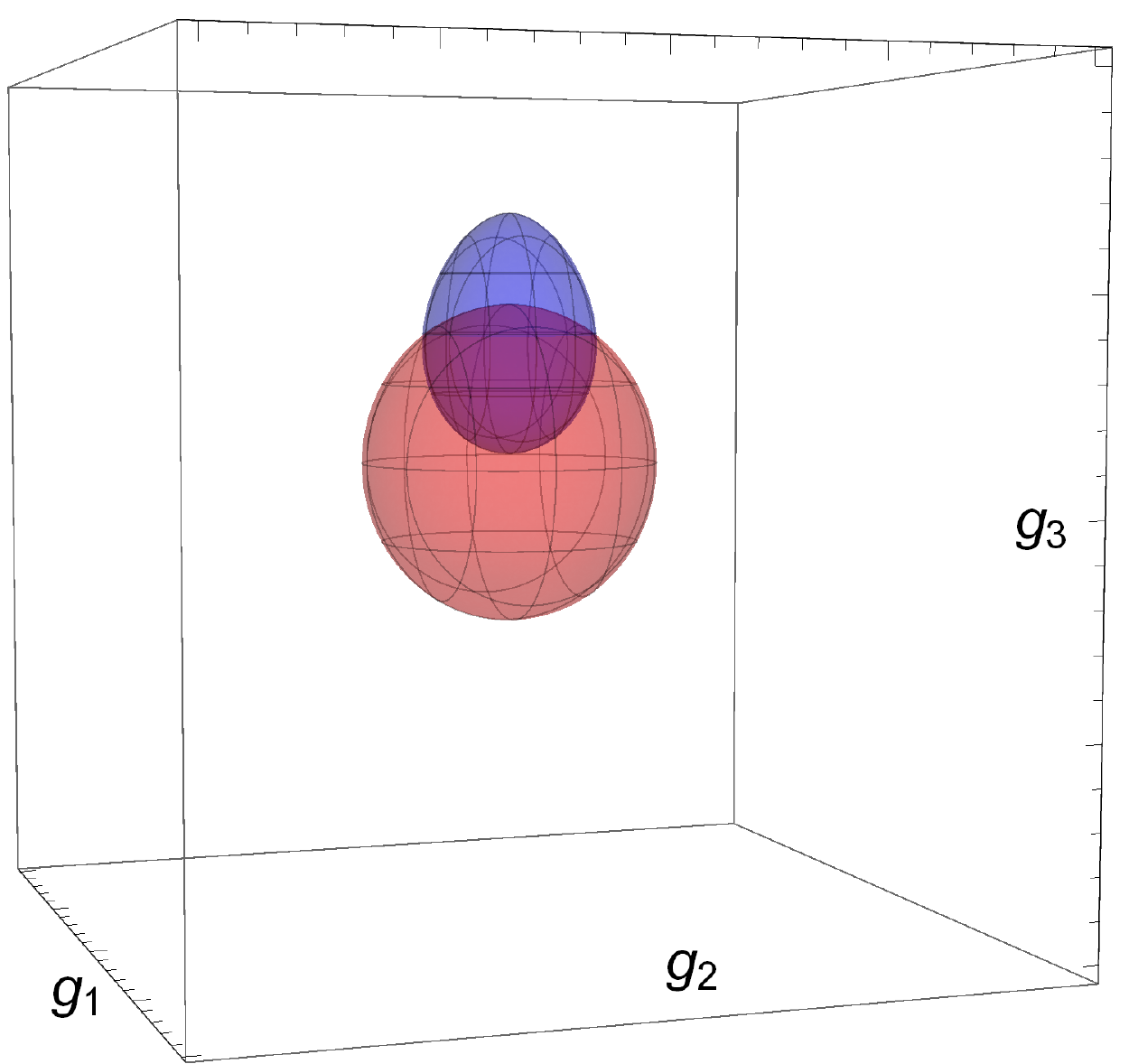}

} \subfloat[\label{fig:rbvSpace6}$\Sigma=\protect\diag(0.25,0.3,0.3)$\qquad{}$\vec{h}=(0.4,0.5,0.5)$]{\includegraphics[width=0.47\columnwidth]{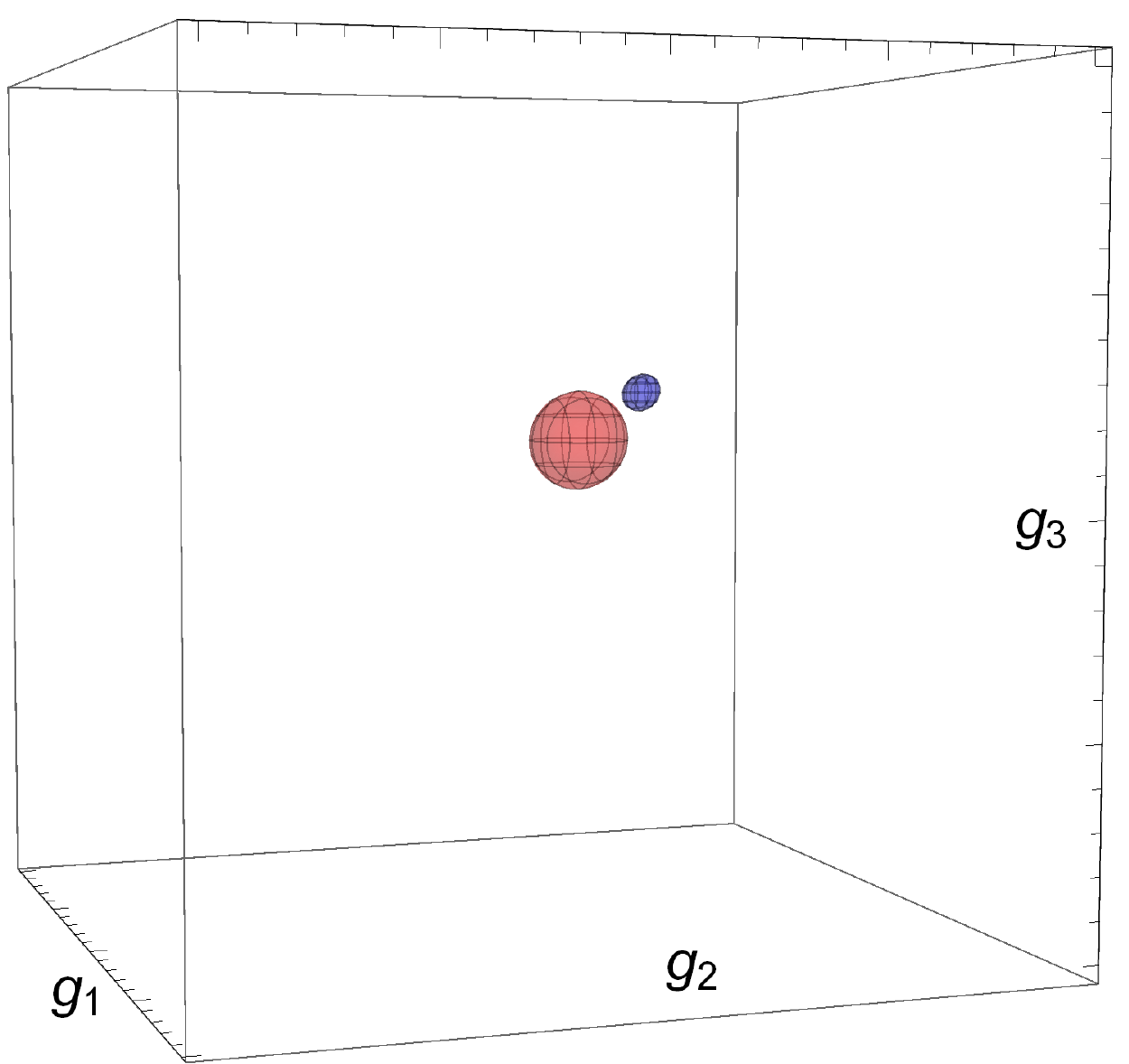}}

\caption{\label{fig:Relative-Bloch-Vector-diagrams}Relative Bloch vector diagrams.
The coordinates of $\vec{g}=(g_{1},g_{2},g_{3})$ where positivity
is satisfied, for fixed values of $\Sigma$ and $\vec{h}$. Axes in
the range $[-1,1]$, with the origin at the center of the cube. Regions
with $d=1$ in blue and $d=-1$ in red.}
\end{figure}

Figure \ref{fig:Relative-Bloch-Vector-diagrams} contains the relative
Bloch vector diagrams, with allowed regions of the vector $\vec{g}$
for several fixed values of $\Sigma,\vec{h}$. Figure \ref{fig:rbvSpace1}
shows the simplest case when the singular values and the second subsystem's
Bloch vector are zero. The allowed $\vec{g}$ region is a complete
Bloch sphere for both values of $d$, with all the states separable.
The quartic \prettyref{eq:svbound3} reduces to a sphere via $(\|\vec{g}\|^{2}-1)^{2}=0$.

Figure \ref{fig:rbvSpace2} shows concentric spheres, with the smaller
sphere containing separable states and spherical shell between the
two containing entangled states. Figure \ref{fig:rbvSpace3} shows
a ``football'' for $d=-1$ that is entirely entangled. Figures \ref{fig:rbvSpace4}
and \ref{fig:rbvSpace5} demonstrate a partial overlap between the
regions for the two values of $d$. In Fig. \ref{fig:rbvSpace6} they
are disjoint, meaning all the states are entangled.

\section{\label{sec:Summary}Summary}

With the goal of generalizing the Bloch sphere, we have examined two-qubit
systems in much detail. Representing the density matrix $\rho$ in
the Dirac basis yields the Bloch matrix $\dvec r$ with real entries.
The latter was split to three components, the local Bloch vectors
$\vec{u},\vec{v}$ and correlation matrix $R$. We then derived the
positivity condition of the quantum state on $\vec{u},\vec{v},R$,
in the form of three important inequalities in \prettyref{eq:bounds},
allowing us to parametrize and visualize the quantum state space.

The form of the positivity inequalities suggested the singular value
decomposition of $R$, and redefining the degrees of freedom in terms
of singular value matrix $\Sigma$, singular vector matrices $M,N$
and relative Bloch vectors $\vec{g},\vec{h}$. It was found that positivity
only depends on $9$ continuous degrees of freedom in $\Sigma,\vec{g},\vec{h}$
and the discrete orientation $d\equiv\det(M)\det(N)=\pm1$, all invariant
under local unitary transformations. The SVD also allowed us to visualize
a quantum state as two Bloch spheres with local Bloch vectors and
scaled correlation axes.

We showed that nonlocal unitary transformations have a mixing effect
on the Bloch matrix components $\vec{u},\vec{v},R$. The SVD components
are affected in complicated nonlocal unitaries, and the singular values
can experience what resembles avoided crossings. 

The three unitary invariants of the quantum state were found in terms
of $\vec{u},\vec{v},R$, and in terms of $\Sigma,\vec{g},\vec{h},d$.
The latter representation in particular is significant in that it
represents the general unitary invariants of a state in terms of its
local unitary invariants. The positive partial transpose criterion
was generalized, and entanglement of a state was found equivalent
to the positivity of its conjugate state, defined as the state with
$\Sigma,\vec{g},\vec{h}$ unchanged and $d$ reversed in sign. We
also characterized maximally entangled, pure, and generalized isotropic
states.

Finally, the positivity conditions were used to visualize the quantum
state space, by holding $6$ degrees of freedom in $\Sigma,\vec{g},\vec{h}$
constant, and drawing the physicality region for the other $3$. The
regions were drawn for both values of orientation $d$, with the intersection
indicating separable states, and the symmetric difference entangled
states.

This investigation deepens our understanding of two-qubit states and
aids intuition when dealing with them. Looking ahead, there are several
potential extensions to this work. We may examine the effect of dissipative
and open system evolution on the Bloch components, the SVD, and the
unitary invariants. 

One may consider the case of more qubits. Though there is no simple
singular value decomposition in higher dimensions, it may prove fruitful
in understanding entanglement. For instance, shedding light on the
different orders of multipartite entanglement. If there turn out to
be several orientation signs similar to $d$, this approach may yield
an entanglement criterion that is both necessary and sufficient in
higher dimensions. 
\begin{acknowledgments}
My heartfelt thanks to Prof. Graham Fleming for his support, mentorship,
and for proposing research problems that motivated this manuscript.
I also thank Prof. Birgitta Whaley for bringing to my attention several
important references, and the anonymous referees for valuable suggestions.

This work was supported by the Director, Office of Science, Office
of Basic Energy Sciences, of the USA Department of Energy under Contract
No. DE-AC02-05CH11231 and the Division of Chemical Sciences, Geosciences
and Biosciences Division, Office of Basic Energy Sciences through
Grant No. DE-AC03-76F000098 (at LBNL and UC Berkeley).
\end{acknowledgments}

\appendix

\section{Hermitian matrix basis sets}

\subsection{\label{subsec:Gell-Mann-Matrices}Gell-Mann matrices (3$\times$3)}

The Gell-Mann matrices, $G_{m}$, are the most widely used set of
generators for the group of special unitary 3$\times$3 matrices,
$SU(3)$ \citep{GellMann1964}. With the identity matrix ($G_{0}$),
they form a basis for the space of 3$\times$3 Hermitian matrices:

\noindent\begin{minipage}[t]{1\columnwidth}%
\begin{align*}
\begin{aligned}G_{1}=\left(\begin{array}{ccc}
0 & 1 & 0\\
1 & 0 & 0\\
0 & 0 & 0
\end{array}\right), &  & G_{2}=\left(\begin{array}{ccc}
0 & -i & 0\\
i & 0 & 0\\
0 & 0 & 0
\end{array}\right),\\
G_{3}=\left(\begin{array}{ccc}
1 & 0 & 0\\
0 & -1 & 0\\
0 & 0 & 0
\end{array}\right), &  & G_{4}=\left(\begin{array}{ccc}
0 & 0 & 1\\
0 & 0 & 0\\
1 & 0 & 0
\end{array}\right),\\
G_{5}=\left(\begin{array}{ccc}
0 & 0 & -i\\
0 & 0 & 0\\
i & 0 & 0
\end{array}\right), &  & G_{6}=\left(\begin{array}{ccc}
0 & 0 & 0\\
0 & 0 & 1\\
0 & 1 & 0
\end{array}\right),\\
G_{7}=\left(\begin{array}{ccc}
0 & 0 & 0\\
0 & 0 & -i\\
0 & i & 0
\end{array}\right), &  & G_{8}=\frac{1}{\sqrt{3}}\left(\begin{array}{ccc}
1 & 0 & 0\\
0 & 1 & 0\\
0 & 0 & -2
\end{array}\right).
\end{aligned}
\end{align*}
\end{minipage}

\subsection{\label{subsec:Dirac-Matrices}Dirac matrices (4$\times$4)}

The Dirac matrices are defined by $D_{\mu\nu}=\sigma_{\mu}\otimes\sigma_{\nu}$,
$\mu,\nu=0,1,2,3$, and form a 16-element basis for the space of 4$\times$4
Hermitian matrices. Excluding the identity $D_{00}$, the remaining
15 matrices constitute a set of generators for the group of special
unitary 4$\times$4 matrices, $SU(4)$. The matrices are explicitly
given in Table \ref{tab:Dirac-Matrices}.

\begin{table}[h]
\begin{tabular}{l|cccc}
\noalign{\vskip5pt}
\backslashbox{$\mu$}{$\nu$} & $0$ & $1$ & $2$ & $3$\tabularnewline[5pt]
\hline 
\noalign{\vskip5pt}
$0$ & $\scaledmatrix{\left(\begin{array}{cccc}
1 & 0 & 0 & 0\\
0 & 1 & 0 & 0\\
0 & 0 & 1 & 0\\
0 & 0 & 0 & 1
\end{array}\right)}$ & $\scaledmatrix{\left(\begin{array}{cccc}
0 & 1 & 0 & 0\\
1 & 0 & 0 & 0\\
0 & 0 & 0 & 1\\
0 & 0 & 1 & 0
\end{array}\right)}$ & $\scaledmatrix{\left(\begin{array}{cccc}
0 & -i & 0 & 0\\
i & 0 & 0 & 0\\
0 & 0 & 0 & -i\\
0 & 0 & i & 0
\end{array}\right)}$ & $\scaledmatrix{\left(\begin{array}{cccc}
1 & 0 & 0 & 0\\
0 & -1 & 0 & 0\\
0 & 0 & 1 & 0\\
0 & 0 & 0 & -1
\end{array}\right)}$\tabularnewline[5pt]
\noalign{\vskip5pt}
$1$ & $\scaledmatrix{\left(\begin{array}{cccc}
0 & 0 & 1 & 0\\
0 & 0 & 0 & 1\\
1 & 0 & 0 & 0\\
0 & 1 & 0 & 0
\end{array}\right)}$ & $\scaledmatrix{\left(\begin{array}{cccc}
0 & 0 & 0 & 1\\
0 & 0 & 1 & 0\\
0 & 1 & 0 & 0\\
1 & 0 & 0 & 0
\end{array}\right)}$ & $\scaledmatrix{\left(\begin{array}{cccc}
0 & 0 & 0 & -i\\
0 & 0 & i & 0\\
0 & -i & 0 & 0\\
i & 0 & 0 & 0
\end{array}\right)}$ & $\scaledmatrix{\left(\begin{array}{cccc}
0 & 0 & 1 & 0\\
0 & 0 & 0 & -1\\
1 & 0 & 0 & 0\\
0 & -1 & 0 & 0
\end{array}\right)}$\tabularnewline[5pt]
\noalign{\vskip5pt}
$2$ & $\scaledmatrix{\left(\begin{array}{cccc}
0 & 0 & -i & 0\\
0 & 0 & 0 & -i\\
i & 0 & 0 & 0\\
0 & i & 0 & 0
\end{array}\right)}$ & $\scaledmatrix{\left(\begin{array}{cccc}
0 & 0 & 0 & -i\\
0 & 0 & -i & 0\\
0 & i & 0 & 0\\
i & 0 & 0 & 0
\end{array}\right)}$ & $\scaledmatrix{\left(\begin{array}{cccc}
0 & 0 & 0 & -1\\
0 & 0 & 1 & 0\\
0 & 1 & 0 & 0\\
-1 & 0 & 0 & 0
\end{array}\right)}$ & $\scaledmatrix{\left(\begin{array}{cccc}
0 & 0 & -i & 0\\
0 & 0 & 0 & i\\
i & 0 & 0 & 0\\
0 & -i & 0 & 0
\end{array}\right)}$\tabularnewline[5pt]
\noalign{\vskip5pt}
$3$ & $\scaledmatrix{\left(\begin{array}{cccc}
1 & 0 & 0 & 0\\
0 & 1 & 0 & 0\\
0 & 0 & -1 & 0\\
0 & 0 & 0 & -1
\end{array}\right)}$ & $\scaledmatrix{\left(\begin{array}{cccc}
0 & 1 & 0 & 0\\
1 & 0 & 0 & 0\\
0 & 0 & 0 & -1\\
0 & 0 & -1 & 0
\end{array}\right)}$ & $\scaledmatrix{\left(\begin{array}{cccc}
0 & -i & 0 & 0\\
i & 0 & 0 & 0\\
0 & 0 & 0 & i\\
0 & 0 & -i & 0
\end{array}\right)}$ & $\scaledmatrix{\left(\begin{array}{cccc}
1 & 0 & 0 & 0\\
0 & -1 & 0 & 0\\
0 & 0 & -1 & 0\\
0 & 0 & 0 & 1
\end{array}\right)}$\tabularnewline[5pt]
\end{tabular}

\caption{\label{tab:Dirac-Matrices}The Dirac Matrices, $D_{\mu\nu}=\sigma_{\mu}\otimes\sigma_{\nu}$.}
\end{table}

Aside, the gamma matrices, standard in modern treatments of the Dirac
equation \citep{Peskin1995}, are given by $\gamma^{0}=D_{30},\,\gamma^{1}=iD_{21},\,\gamma^{2}=iD_{22},\,\gamma^{3}=iD_{23},\,\gamma^{5}=D_{10}$. 

\section{\label{sec:svd-Cofactor} Cofactor matrix singular value decomposition }

We show that for any $R=M\Sigma N^{\dagger},$ its cofactor matrix
satisfies $\tilde{R}=\det(M)\det(N)M\tilde{\Sigma}N^{\dagger}$, where
$\tilde{\Sigma}=\diag(x_{2}x_{3},x_{3}x_{1},x_{1}x_{2})$ is the cofactor
matrix of the singular value matrix $\Sigma=\diag(x_{1},x_{2},x_{3})$. 

In this appendix, we use extended Einstein notation, in which any
index that is repeated\emph{ twice or more }is summed over. 

The cross product of two columns of a $3\times3$ orthogonal matrix
$O$ yields the remaining column, up to a sign determined by $\det O$
and the column indices;
\begin{eqnarray}
\varepsilon_{i_{1}i_{2}i}O_{i_{1}k}O_{i_{2}l} & = & \varepsilon_{i_{1}i_{2}j}\delta_{ij}O_{i_{1}k}O_{i_{2}l}\nonumber \\
 & = & \big(\varepsilon_{i_{1}i_{2}j}O_{jm}O_{i_{1}k}O_{i_{2}l}\big)O_{im}\nonumber \\
 & = & \det(O)\varepsilon_{klm}O_{im},\label{eq:orthogonalDetID}
\end{eqnarray}
where in the second line we used $O_{im}O_{jm}=\delta_{ij}$, and
in the last line we used a determinant identity. Proceeding from the
cofactor matrix definition \prettyref{eq:cofactor}, we have
\begin{eqnarray}
\tilde{R}_{ij} & = & \frac{1}{2}\varepsilon_{i_{1}i_{2}i}\varepsilon_{j_{1}j_{2}j}r_{i_{1}j_{1}}r_{i_{2}j_{2}}\nonumber \\
 & = & \frac{1}{2}\varepsilon_{i_{1}i_{2}i}\varepsilon_{j_{1}j_{2}j}M_{i_{1}k}x_{k}N_{j_{1}k}M_{i_{2}l}x_{l}N_{j_{2}l}\nonumber \\
 & = & \frac{1}{2}\det(M)\det(N)M_{im}\varepsilon_{klm}\varepsilon_{kln}x_{k}x_{l}N_{jn}\nonumber \\
 & = & \det(M)\det(N)M_{im}\tilde{\Sigma}_{mn}N_{jn},\label{eq:cofactorsvdProof}
\end{eqnarray}
where in the third line we twice applied \prettyref{eq:orthogonalDetID}
and in the last line we noted that $\varepsilon_{klm}\varepsilon_{kln}x_{k}x_{l}=2\tilde{\Sigma}_{mn}$. 

\section{\label{sec:BasicNonlocal}Nonlocal operators on the Bloch matrix}

We first derive the effect of the irreducible nonlocal operator $\mathring{U}_{j}(\theta_{j})$,
defined in \prettyref{eq:UjIrreducible}, on the Bloch matrix entries
$r_{\mu\nu}$. In the derivation \prettyref{eq:longUjDerive} below,
we suppress the subscript on $\theta$, and repeated indices are summed
over except $j,$ which is fixed. We have,
\begin{widetext}
\begin{align}
r_{\mu\nu}^{\prime}\sigma_{\mu}{\otimes}\sigma_{\nu}= & \exp\left[\frac{i}{2}\theta\sigma_{j}{\otimes}\sigma_{j}\right]r_{\mu\nu}\sigma_{\mu}{\otimes}\sigma_{\nu}\exp\left[-\frac{i}{2}\theta\sigma_{j}{\otimes}\sigma_{j}\right]\nonumber \\
= & r_{\mu\nu}\left(\cos\frac{\theta}{2}I{\otimes}I+i\sin\frac{\theta}{2}\sigma_{j}{\otimes}\sigma_{j}\right)\sigma_{\mu}{\otimes}\sigma_{\nu}\left(\cos\frac{\theta}{2}I{\otimes}I-i\sin\frac{\theta}{2}\sigma_{j}{\otimes}\sigma_{j}\right)\nonumber \\
= & r_{\mu\nu}\left(\cos^{2}\frac{\theta}{2}\sigma_{\mu}{\otimes}\sigma_{\nu}-i\cos\frac{\theta}{2}\sin\frac{\theta}{2}\commutator{\sigma_{\mu}{\otimes}\sigma_{\nu}}{\sigma_{j}{\otimes}\sigma_{j}}+\sin^{2}\frac{\theta}{2}\sigma_{j}\sigma_{\mu}\sigma_{j}{\otimes}\sigma_{j}\sigma_{\nu}\sigma_{j}\right)\nonumber \\
= & r_{\mu\nu}\left(\cos^{2}\frac{\theta}{2}\sigma_{\mu}{\otimes}\sigma_{\nu}+\sin\theta\big(\theta_{\mu j\alpha}\varepsilon_{\nu j\beta}{+}\varepsilon_{\mu j\alpha}\theta_{\nu j\beta}\big)\sigma_{\alpha}{\otimes}\sigma_{\beta}+\sin^{2}\frac{\theta}{2}(2\delta_{\mu0}I{+}2\delta_{\mu j}\sigma_{j}{-}\sigma_{\mu}){\otimes}(2\delta_{\nu0}I{+}2\delta_{\nu j}\sigma_{j}{-}\sigma_{\nu})\right)\nonumber \\
= & r_{\mu\nu}\sigma_{\mu}{\otimes}\sigma_{\nu}+\sin\theta\big(R_{jk}\varepsilon_{kjn}I{\otimes}\sigma_{n}{+}u_{k}\varepsilon_{kjn}\sigma_{j}{\otimes}\sigma_{n}{+}R_{kj}\varepsilon_{kjm}\sigma_{m}{\otimes}I{+}v_{k}\varepsilon_{kjm}\sigma_{m}{\otimes}\sigma_{j}\big)\nonumber \\
 & \hphantom{r_{\mu\nu}\sigma_{\mu}{\otimes}\sigma_{\nu}}+(\cos\theta-1)\sum_{k\neq j}\big(u_{k}\sigma_{k}{\otimes}I+v_{k}I{\otimes}\sigma_{k}+R_{kj}\sigma_{k}{\otimes}\sigma_{j}+R_{jk}\sigma_{j}{\otimes}\sigma_{k}\big).\label{eq:longUjDerive}
\end{align}
\end{widetext}

Gathering like terms and comparing the coefficients of $\sigma_{\mu}{\otimes}\sigma_{\nu}$
on both sides yields the transformed $\dvec r^{\prime}$ shown in
\prettyref{eq:U1effect} for $j=1$.

\begin{comment}
\begin{align*}
= & r_{\mu\nu}\sigma_{\mu}{\otimes}\sigma_{\nu}+\sin\theta\big(R_{jk}\varepsilon_{kjn}I{\otimes}\sigma_{n}{+}u_{k}\varepsilon_{kjn}\sigma_{j}{\otimes}\sigma_{n}{+}R_{kj}\varepsilon_{kjm}\sigma_{m}{\otimes}I{+}v_{k}\varepsilon_{kjm}\sigma_{m}{\otimes}\sigma_{j}\big)\\
 & +\sin^{2}\frac{\theta}{2}\Big[4\big(I{\otimes}I+u_{j}I{\otimes}\sigma_{j}+v_{j}\sigma_{j}{\otimes}I+R_{jj}\sigma_{j}{\otimes}\sigma_{j}\big)\\
 & -2\big(I{\otimes}I+u_{k}\sigma_{k}{\otimes}I+I{\otimes}I+v_{k}I{\otimes}\sigma_{k}+v_{j}I{\otimes}\sigma_{j}+R_{kj}\sigma_{k}{\otimes}\sigma_{j}+u_{j}\sigma_{j}{\otimes}I+R_{jk}\sigma_{j}{\otimes}\sigma_{k}\big)\Big]
\end{align*}
\end{comment}

We now combine the three irreducible nonlocal operators to find the
full effect of the basic nonlocal operator $\mathring{U}\big(\theta_{1},\theta_{2},\theta_{3}\big)=\mathring{U}_{1}\big(\theta_{1}\big)\mathring{U}_{2}\big(\theta_{2}\big)\mathring{U}_{3}\big(\theta_{3}\big)$.
Rather that write the modified Bloch matrix $\dvec r$ explicitly,
we use more compact index notation. In what follows, \emph{repeated
indices do not indicate a sum}, and in the first two equations we
implicitly assume the indices $i,j,k$ are distinct. 

Combining the effects of $\mathring{U}_{1},\mathring{U}_{2},\mathring{U}_{3}$,
we find $\mathring{U}$ transforms the Bloch matrix components as
\begin{widetext}
\begin{subequations} \label{eq:UBasicEffect}

\begin{align}
u_{k}^{\prime} & =u_{k}\cos\theta_{i}\cos\theta_{j}+v_{k}\sin\theta_{i}\sin\theta_{j}+\varepsilon_{ijk}\big(R_{ij}\cos\theta_{i}\sin\theta_{j}-R_{ji}\sin\theta_{i}\cos\theta_{j}\big),\label{eq:UBasicEffectu}\\
v_{k}^{\prime} & =v_{k}\cos\theta_{i}\cos\theta_{j}+u_{k}\sin\theta_{i}\sin\theta_{j}+\varepsilon_{ijk}\big(R_{ji}\cos\theta_{i}\sin\theta_{j}-R_{ij}\sin\theta_{i}\cos\theta_{j}\big),\label{eq:UBasicEffectv}\\
R_{ij}^{\prime} & =R_{ij}\cos\theta_{i}\cos\theta_{j}+R_{ji}\sin\theta_{i}\sin\theta_{j}-\varepsilon_{ijk}\big(u_{k}\cos\theta_{i}\sin\theta_{j}-v_{k}\sin\theta_{i}\cos\theta_{j}\big).\label{eq:UBasicEffectR}
\end{align}

\end{subequations} %
\begin{comment}
$R_{ji}^{\prime}=R_{ji}\cos\theta_{i}\cos\theta_{j}+R_{ij}\sin\theta_{i}\sin\theta_{j}-\varepsilon_{ijk}\big(v_{k}\cos\theta_{i}\sin\theta_{j}-u_{k}\sin\theta_{i}\cos\theta_{j}\big).$
\end{comment}

With suitable sums, differences and trigonometric identities, the
above can be written as a single two-dimensional rotation matrix acting
on an artificial $2$-vector, mixing $\vec{u}\pm\vec{v}$ with $R\pm R^{\dagger}$
to generate entanglement: 
\begin{equation}
\begin{bmatrix}(\vec{u}\pm\vec{v})_{k}\\
\varepsilon_{ijk}(R\pm R^{\dagger})_{ij}
\end{bmatrix}^{\prime}=\begin{bmatrix}\cos(\theta_{i}\mp\theta_{j}) & \mp\sin(\theta_{i}\mp\theta_{j})\\
\pm\sin(\theta_{i}\mp\theta_{j}) & \cos(\theta_{i}\mp\theta_{j})
\end{bmatrix}\begin{bmatrix}(\vec{u}\pm\vec{v})_{k}\\
\varepsilon_{ijk}(R\pm R^{\dagger})_{ij}
\end{bmatrix}.
\end{equation}
\end{widetext}

\end{document}